# THE CONCORDANCE MODEL - A HEURISTIC APPROACH FROM A STATIONARY UNIVERSE

PETER OSTERMANN*

ABSTRACT

Given there has been something where a big-bang origin of our evolutionary cosmos took place: What is the relativistic line element describing the energy density and pressure of such a pre-existing universal background? The simplest conceivable ansatz leads to a Stationary-Universe Model (SUM), which instead of the 'Steady-state Theory' is shown to be an arguable alternative to the Cosmological Concordance Model (CCM) commonly accepted today. The SUM stands out with redshift values statistically independent of time, a significant Hubble parameter is proved in contrast to the conventional one. It requires a negative gravitational 'dark' pressure of –1/3 the critical density. Intrinsic limitations of proper length and time are derived, which cause a struggle of local SRT (quantum mechanics) and universal GRT (gravitation). Using one macroscopic constant $H$ in addition to $c$ and $G$ only, the model describes a background free of coincidences or horizon problems. While the CCM's key parameter $\Omega_\Lambda$ seems determined by SUM 'boundary' conditions, there is a chance of having already observed parts of a stationary universe: With no need for 'dark energy', this alternative explains straightforwardly the SNe-Ia data on universal scales. In addition to its currently assumed parts, a non-lensing homogeneous background of matter might fill the gap to critical density. A mathematical solution for a perfect black-body spectrum composed of redshifted microwave radiation emitted from 'dark' sources within the universe is derived, thus the CMB might exist as a special part of the extragalactic background light. Given the law of entropy restricted to evolutionary processes, an open concept is revealed to imply a 'chaotic' quasi-inflation background, embedding 'local-bang' cosmoses therein. – The SUM is shown to be the only arguable solution of Einstein's original equations without cosmological constant.

*Subject headings:* gravitation – cosmology: Hubble parameter, SNe-Ia, CMB, dark matter, entropy
*Online material:* HZT and SCP SNe-Ia data as used here; list of symbols, notations, and coordinates

## 1. INTRODUCTION

It is well-known that the 'Steady-state Theory' (SST) as developed by [Bondi & Gold 1948] and [Hoyle 1948/49] has proved obsolete. This statement certainly applies to its mathematical model as well as to several of its concepts. It is hard to believe, however, that Einstein's equations should definitely fail to describe a stationary background while, on the other hand, it is widely assumed that something like quantum fluctuations existed when according to the Cosmological Concordance Model (CCM) the 'big bang' took place. Such a 'false vacuum', however, would have been anything but empty space, thus requiring a solution in the framework of Einstein's gravitational equations.

Going beyond the first beginnings of relativistic cosmology by [Einstein 1917], [de Sitter 1917], [Friedman(n) 1922/24], and trying to model an eternal infinite universe, it is thought since more than half a century that the only reasonable alternative to a 'big bang' solution as essentially suggested by [Lemaître 1931b/c] would be the SST. Here it will be shown, however, that for several reasons an essentially different Stationary-Universe Model (SUM) – though of corresponding intention – might prove an actual alternative instead.

* Independent Research, Valpichlerstr. 150, 80689 Munich, GER
  Electronic address: peos@independent-research.org

Undoubtedly there has been an origin of our evolutionary cosmos billions of years ago, there is no need to emphasize this feature at large. The CCM as developed in recent years is representing nearly all relevant observational facts at least numerically well. The underlying concept of a hot 'big bang' singularity, followed by an assumed phase of 'inflation' leading to a Lambda-Cold-Dark-Matter (ΛCDM) universe, has inspired the overwhelming cosmological discoveries of the last decades.

Besides those physicists who believe that the entire universe once had originated together with space and time from one singular 'big bang' out of nothing, however, there may be an increasing number of others preferring alternatives. A 'multiverse', though, may be just another word for the one and only universe since any 'parallel-universes' – if never causally connected – would physically not exist.

In the following, to distinguish our *cosmos* from a pre-existing background – possibly allowing for other cosmoses as well – only this all-embracing background may be named *universe*. Remarkably unlike the word 'cosmos', meaning *order* of our world, the original meaning of 'universe' is *all* of the world.

The question of an eternal universe behind our evolutionary cosmos leads immediately to the idea of stationarity, though any such attempt seemed blocked by the failure of the SST. Actually that theory, however, did not describe a steady state. Its redshift parameters – together with all observable quantities depending on *z* – are functions of time. There would be no need to mention this, if not for the sake of clarity in the sense of a dissociation from SUM now.

Though of unique mathematical simplicity, the new model's line element based on both General Relativity Theory (GRT) and Special Relativity Theory (SRT) has not been recognized to stand for a stationary background universe so far. A reason may be that it reveals this feature best in universal coordinates instead in its FLRW form (as developed in general by [Friedman(n) 1922/24], [Lemaître 1927/31], [Robertson 1935/36], [Walker 1936]).

Looking back from SUM, [Einstein 1917] was right to assume a universe without peculiar history. It seems an unnecessary assumption, however, that such a universe had to be static (at that times it has been correspondingly assumed that stable radiationless atoms should be static, while the characteristic feature in both cases is stationarity instead).

Now, though the model presented here proves stationary, it yet implies a maximum age of macroscopic structures, for example. This raises the question whether the SUM can keep unquestionable achievements of the CCM without suffering from its various problems. No external hypotheses are needed to show some basic features, otherwise rather speculatively developed in the CCM framework before. Examples are spatial flatness or an 'age of the universe' just equaling the Hubble time $T_{H(o)} \equiv 1/H_{(0)}$, what will be shown to determine the numerical value of a 'cosmological constant' heuristically.

As is well known, the CCM is governed by a spatially flat line element of FLRW form, with a matter density $\rho_M \approx 0.27\rho_c$ including 'dark matter' (DM), and an amount of 'dark energy' $\Lambda/(8\pi G/c^4) \equiv \varepsilon_\Lambda = (\rho_0 - \rho_M)c^2 \approx 0.73\rho_c$ due to a cosmological constant $\Lambda$, see e.g. from [Bennett et al. 2003] to [Jarosik et al. 2011] and references therein. Here is $\rho_0 \equiv \rho_{total} \approx \rho_c$ with $\rho_c \equiv 3H_0^2/(8\pi G)$ the critical mass density, $G$ Newton's gravitational constant, and $H_0$ the conventional Hubble parameter $H_c(t'=0)$ today, s. below. The present 'deceleration' parameter is $q_0$, and $T_0$ the 'age of the universe'.

Well-known pillars are supporting the CCM. Besides the 'predictions' concerning the magnitude-redshift relation of SNe-Ia or the primordial nucleosynthesis, in particular the Cosmic Microwave Background (CMB) black-body radiation together with the almost perfect description of its anisotropies are the strongest arguments for a hot 'big bang' in the framework of the $\Lambda$CDM model, see e.g. [Durrer 2008, where also fundamental unproven hypotheses underlying the CCM and its mathematical treatment are explicitly addressed].

The indispensable CCM-paradigm of inflation, however, is raising serious doubts [Steinhardt 2011], since there is neither a clear theory of such a scenario nor any detection of a corresponding scalar field needed there to solve the problems of universal horizons or approximate flatness. Not least the baryon asymmetry has to be mentioned in this context as well as several other questions concerning the initial singularity, coincidences, and fine tuning, for example.

It is clear from the beginning that in any 'big-bang' cosmology there will remain purely coincidental aspects which seem particularly difficult to accept, as long as they concern the universe as a whole. Therefore it is a natural question whether instead of inflation there might be an alternative to reconcile relativistic cosmology with those observational facts which otherwise mean a fundamental dilemma each.

There are three main reasons to take the SUM into consideration. At first, if relativistic cosmology shall be more than a tale of creation there must have been some physical background behind that big-bang cosmos described by today's CCM. At second, it is of interest in its own right that – essentially different from various versions of the SST – there is a stationary solution of Einstein's equations implying some unexpected features concerning Relativity Theory (RT). At third, without any unproven physics, the SUM line element yields magnitude-redshift relations which obviously describe the Supernovae Type Ia (SNe-Ia) data on universal scales $z > 0.1$ as well as those derived from the CCM.

Together with the prove that there is at least a mathematical solution for a perfect CMB black-body radiation composed of redshifted contributions, and a law of entropy restricted to evolutionary processes outside Supermassive Gravitational Centers (SGCs), the magnitude-redshift accordance indicates the chance for even a *straight* SUM. This term may occasionally differentiate from the otherwise much more adaptable *open* SUM, the latter with the option of embedding a modified CCM cosmos directly.

Historically, in addition to today's CCM, there has been a chaotic inflationary approach where some early papers once also referred to a 'stationary universe model' [Mezhlumian 1993/94], [Linde & Mezhlumian 1993], [Linde, Linde, & Mezhlumian 1994]. Besides the heading expressing a corresponding intention, however, that approach is quite different from the SUM proposed here. Instead, it seems to give rise to those completely separated 'parallel universes' of inflationary scalar fields mentioned above. Each of them would have to be described by a variant of the concordance model respectively, but the one decisive line element of general relativity to describe a coherent background universe is missing there. On the other hand, in view of the SUM, it is a large improvement of that 'chaotic inflation' concept to have established a universal background, though only in form of mere quantum fluctuations so far. Nevertheless the concept of a singular 'big bang' has been effectively overcome there.

Following another track, a "Coasting Cosmology" [Kolb 1989] has been discussed – closely related to later [Melia & Shevchuk 2012] – whose general line element in the special case of flat space would be the same as the FLRW form of the SUM. In view of such an assumed coasting expansion, however, most stationary features – as developed in the following sections – have been left unnoticed there. In particular, the fundamental consequence seems to be missing, that corresponding redshift parameters are independent of time. – More remarks on the origin of the new SUM concept as well as on those earlier attempts are in Section 2.13 below.

Since the SST has turned out to conflict with cosmological facts, there may be little interest in another attempt to overcome a singular origin of all as long as there is no observational indication. Therefore, the reader may take a preliminary glance at ten figures below which straightforwardly result from the SUM as the new concept presented here.

### *1.1 Concepts and Notation*

Using the [Landau & Lifschitz 1992] notation, the signature of the GRT fundamental tensor $g_{ik}$ determining the line element $d\sigma^2 = g_{ik}dx^i dx^k$ is always assigned according to $\eta_{ik} = (+1,-1,-1,-1)$ of SRT. Latin indices $i, k, l \ldots = 0, 1, 2, 3$ represent four-dimensional quantities, whereas Greek indices $\alpha, \beta ... = 1, 2, 3$ represent spatial quantities only. As usual, all symbols are explained at first occurrence.

Since 'relativity' – originally based on the principle of no preferred system – has effectively established a universal CMB restframe after all, something might have been mistaken there. Consequently in contrast to unambiguously testable physical concepts, historical terms like 'relativistic' or 'spatial curvature', for example, may not be taken literally in the context of Einstein's fundamental equations

$$E_{ik} \equiv R_{ik} - \tfrac{1}{2} R\, g_{ik} = \kappa T_{ik} \;, \qquad (1)$$

where $E_{ik}$ is the Einstein tensor, $R_{ik}$, $R$ are the Ricci tensor and its scalar, $g_{ik}$ the fundamental tensor and $\kappa$ means $8\pi G/c^4$ (not to be confused with an absorption coefficient below). In Einstein's 'extended' equations there would be an additional term $\Lambda g_{ik}$ with a cosmological constant on the right hand side.

The fact that there is no need for the still prevailing historical interpretation has already been explicitly accepted by [Einstein 1921] himself who in *'Geometrie und Erfahrung'* – six years after his final formulation of GRT – agreed to [Poincaré's 1902] understanding (*'La Science et l'Hypothèse'*).

According to the SUM concept of space and time, the principle of relativity actually means that freely falling *local inertial systems* allow for the existence of stable objects in spite of their accelerated motions, yet locally implying uniform velocities relative to each other. Furthermore, non-Euclidean geometry is understood here to be nothing but the mathematical tool to deal with rods and clocks systematically affected by gravitation and motion relative to the universal frame, whose coordinates are otherwise denoted as 'comoving' or 'conformal' ones. In addition to all excellent agreement in local gravitational fields, relativity theory then seems to be an appropriate tool also to describe a stationary universe 'embedding' our evolutionary cosmos therein.

It is widely believed that at least on Planck scales General Relativity (GR) and Quantum Mechanics (QM) prove incompatible. Such a statement, however, seems premature as long as Einstein's equations are not solved for a detailed quantum Energy-Momentum-Stress (EMS) tensor on the right hand side of (1) but only for Einstein's phenomenological substitute describing a perfect fluid, whose provisional nature once let him write of 'lumber instead of marble' [Einstein 1936]. A first step to a quantized EMS tensor has been proposed in an approach based on a still preliminary but consistent variational principle to a unified theory of electrodynamics, quantum mechanics, and gravitation [Ostermann 2008a, 2008b]. While regarding the Klein-Gordon equation, mathematical consistency seems already achieved there, this feature may be also established dealing with the Dirac equation on base of the relations below in this section (what will be shown elsewhere).

In spite of the fact, that detailed quantum solutions of (1) may be found rarely if at all, a resignation in view of the assumed incompatibility of GR and QM seems unjustified. As soon as one discards the strictly geometric interpretation of GR, most of the fundamental problems rather vanish into new chances – from particle physics up to cosmology. There is simply no need for geometric properties of space and time instead physical properties of material objects to recover the immense plenty of experimentally verified results derived from Einstein's wonderful equations.

To demonstrate the evidence of this claim I now give a simple derivation of Riemann's non-Euclidean line element without referring to any properties of space and time, before applying it to actual cosmology in the framework of Einstein's equations.

Retrospectively, his theories are mathematically based on the well-known fundamental tensors $\eta_{ab}$ of SRT and $g_{ik}$ of GRT, where for the rest of this paragraph the indices $a, b.. = 1..4$ refer to the first, while the indices $i, k.. = 1..4$ refer to the second. Now the 'non-Euclidean' $g_{ik}$ will be derived in flat space and with respect to a uniform time, both taken together in *universal coordinates* $x^a$ where at large e.g. galaxies are statistically at rest. In contrast, the arbitrary coordinates $x^i$ may refer to any mathematically acceptable system.

Given two neighboring points $P(x^a)$ and $Q(x^a + dx^a)$ in a quasi-Euclidean space-time of SRT as represented by the Poincaré group, their distance from an arbitrarily chosen origin measured with physical rods and clocks *affectable by gravitation or motion* will be

$$\sigma_P^a = x^a + \xi^a(x^a), \qquad (2)$$

$$\sigma_Q^a = (x^a + dx^a) + \xi^a(x^a + dx^a), \qquad (3)$$

where the function $\xi^a$ is describing the respective deviation from the Euclidean value $x^a$ due to physical deformation of the measuring tools. Now the second summand of $s_Q{}^a$ may be expanded according to

$$\xi^a(x^a + dx^a) = \xi^a(x^a) + \left[\partial_b \xi^a\right](x^a)\, dx^b + \ldots \quad (4)$$

with $\partial_b \equiv \partial/\partial x^b$ and, for the sake of readability, the designator $(x^a)$ hereafter omitted. The expansion (4) yields the 'properly' measurable infinitesimal interval

$$d\sigma^a \equiv \sigma_Q^a - \sigma_P^a = dx^a + \left(\partial_b \xi^a\right) dx^b + \ldots \quad (5)$$

between the two neighboring points Q and P. Here it is decisive to assign exactly *by definition* a mixed tensor $\xi^a{}_i$ according to the second identity of the following expression

$$d\xi^a \equiv \left(\partial_b \xi^a\right) dx^b + \ldots \equiv \xi^a{}_i\, dx^i, \qquad (6)$$

where $\xi^a{}_i$ and $dx^i$ may be not applied only with respect to the quasi-Euclidean coordinate system above, but with respect to any additional set of arbitrary coordinates $x^i$ as well. Now because of the '…'-symbol *in general* (GRT) it is

$$\xi^a{}_i \not\equiv \partial_i \xi^a, \qquad (7)$$

while otherwise this would lead to a *special* case (SRT). According to (6), relation (5) may be written as

$$d\sigma^a \equiv dx^a + d\xi^a \equiv e^a{}_i\, dx^i, \qquad (8)$$

where

$$e^a{}_i \equiv \delta^a_i + \xi^a{}_i. \quad (9)$$

As usual, there may be defined the covariant SRT 4-vector $d\sigma_a$ by lowering an index $b$ using the $\eta_{ab}$, what is equivalent *by definition* again to the second identity in

$$d\sigma_a \equiv \eta_{ab} d\sigma^b \equiv e_{ai} dx^i. \quad (10)$$

The square of the line element, $d\sigma^2 \equiv d\sigma_a d\sigma^a$, follows by direct multiplication from (10) and (8) in the form underlying the mathematics of GRT

$$d\sigma^2 \equiv g_{ik} dx^i dx^k, \quad (11)$$

where finally, as easily verified

$$g_{ik} \equiv e_{ai} e^a{}_k. \quad (12)$$

No property of space and time is used in this derivation but merely a 'deformability' by gravitation and motion of physical rods and clocks.

This deduction yields not only Einstein's GR fundamental tensor $g_{ik}$ itself which enables to effectively establish a non-Euclidean geometry of affected rods and clocks – but in particular, this immediately leads to the only appropriate form (12) to apply GRT also to half-integer spin particles governed by the Dirac equation. This form and its mathematical features are well-known as *vierbein* or *tetrad* representation, s. [Einstein 1928] or e.g. [Landau & Lifschitz 1992]. In addition [Rosen 1963] has pointed out an assumed link between his bi-metric formulation of GRT and the vierbein representation, while in view of straight SUM the underlying principle has been briefly elaborated now.

It is anything but coincidental that the mathematical description of spinning objects need mathematics going beyond pure Riemannian geometry, since it is even impossible to define exactly any angular momentum within the original GRT framework. In that the latter refers exclusively to what is called *proper* quantities it is dogmatically adhering to a pure geometric conception presupposing non-affectable standard rods. The reason for the failure to define GR angular momentum straightforwardly is that in the well-known SRT definition if transferred to GR, spatial non-proper coordinates will be necessarily involved. Otherwise there could not apply any non-local angular momentum conservation law. Strictly speaking, the validity of this law is already sufficient to disprove the claim to absoluteness of the historical geometric approach, which thus evidently fails in reducing physics to exclusively Riemannian properties of space and time.

A feature immediately stated by [Einstein 1928] may support this claim: In general 16 components of $e^a{}_i$ determine the 10 components of $g_{ik}$ uniquely, while the other way round the latter (fundamental tensor) do not determine the first (tetrad). Therefore Einstein tried to find field equations to fully determine the tetrads, too. In view of the SUM concept, however, such an attempt seems pointless. The existence of remaining 6 free parameters is necessary to allow for 4-dimensional rotation of particles within the quasi-Euclidean universal frame without changing the non-Euclidean metric $g_{ik}$.

The completion of what is called 'general relativistic space-time' by the quasi-Euclidean universal frame implied in the tetrad concept above – and reflected in Rosen's bi-metric approach below – seems to offer a solution in principle of two main problems of 20th century physics: the alleged incompatibility of GR with QM as well as an assumed 'big bang' creation of space and time.

In addition, it might be anything but coincidental again that the concept of angular momentum going beyond the strict general relativistic approach is closely related to the indirect observation of gravitational waves from decreasing periods of binaries, as well as in another context to the [Einstein, Podolsky, & Rosen 1935] paradox concerning the spin of entangled particles, too.

Now that the legitimacy has been explicitly shown here to understand spatial 'curvature' a gravitational effect on measuring rods instead on mathematical space, the latter therefore can be taken Euclidean at all events. Mathematically, the universal coordinates are only a special representation of what is called 'system coordinates' in general.

On the other hand, to understand the concepts of 'proper' length and 'proper' time as cool as possible – in fact without any loss of physical content – it is sufficient to accept the existence of a 'preferred' universal frame as presupposed above. This is not only possible, but in view of various well-known observations even realistic.

The assumed absence of a universal restframe has been the essential reason for Weyl [*Raum - Zeit - Materie*] to keep adhering to the literally geometric interpretation in spite of the mathematically equivalent alternative also stated there. With regard to such a unique universal frame, however, there is no longer a need to speak of 'pseudo'-tensors and 'pseudo'-tensor densities of the gravitational field, but rather of true bi-tensors and bi-tensor densities instead. The transformation properties of such quantities and the mathematical foundations for the transition from a preferred frame to an arbitrary other one is provided by the bi-metric formulation of GRT which has been established in [Rosen 1940 a/b, 1963, with references therein] on basis of a mathematical ansatz made by Levi-Civita [*The Absolute Differential Calculus*]. Rosen's reformulation called 'bi-metric relativity', however, must not be confused with his deviating 'bi-metric theory' later on (see [Will 1993] with references therein).

According to Rosen's approach, in view of the SUM it is sufficient at first to apply Einstein's equations as well as all tensors or 'pseudo'-tensors with respect to the universal frame in their familiar form. Then, for a transition to any other coordinates, all ordinary derivations – even occurring as parts of the Christoffel symbols or of any covariant derivatives in the original GRT framework – have to be afterwards replaced by a second kind of covariant derivations, now with respect to the new system. In addition, the negative determinant $g$ of the fundamental tensor $g_{ik}$ has to be replaced by $g/\gamma$ where here $\gamma$ is the negative determinant of $\eta_{ik}$ after both tensors are transformed to the new coordinates. On this base, the energy content of the gravitational field does no longer depend on the coordinate system. It is only this feature that would guarantee an objective reality of any energy transport within gravitational fields – in particular that of gravitational waves, whether these are directly observable or not.

While Rosen has convincingly shown that applying GRT, it is possible and of important advantage to refer to a second

'flat space' (bi-)metric, it may be emphasized here, that such a treatment is not only a chance but even a need, because: From all claims in the framework of GRT it is exactly that of a general covariance in choosing arbitrary coordinate systems, which forces to treat the so called pseudo-tensor as a true bi-tensor with respect to the universal frame. Only in this way it is possible to describe the processes leading to decreasing orbital periods of binary pulsars independently of the coordinates used there. This procedure even works if one might chose an appropriately rotating flexible coordinate system where the binaries are at rest all the time. There are other arguments as concerning the very definition of angular momentum mentioned above, s. also text V of [Ostermann 2008b]. These may be worked out elsewhere.

GR by itself cannot work without QM if applied to processes going beyond the 'geodesic' equations of motion, which attribute actually reflects an important geometric analogy only (s. also [Weinberg 1972]). Gravitation regarded as an isolated physical agent, however, would be unable to explain in particular how there can be explosions of gravitationally bound systems like SNe, for example.

Furthermore, a 'black hole' if taken literally may be only an old concept of GRT, while its phenomenological applicability breaks down as already assumed by Einstein [*Grundzüge der Relativitätstheorie*] himself. Quantum mechanics instead may set essential limits to GRT-applicability there in retaining matter from vanishing forever. What astronomers really see may be only Super-Massive-Objects (SMOs) instead. In jets of Active Galactic Nuclei (AGNi) or in close vicinity even of quiet SMOs there is not observed any inevitable disappearance of matter and radiation but rather the contrary all over the universe. This understanding is only a consequent extension of the idea which has led to the SUM as the stationary cosmological solution of Einstein's equations.

The possibility for the law of entropy to be restricted to evolutionary processes outside SGCs mentioned above, is supported by a well-known, otherwise puzzling, microscopic reversibility of elementary interactions implying the principle of detailed balance. Together with gravitationally disabled diffusion, this balance may turn to a reversal from increasing to decreasing entropy where the densities of matter and energy would approximate those of the Schwarzschild radius.

Only as long as the redshift of galaxies is understood to originate from real motion completely, this seems to imply a peculiar history of the entire universe. The associated Doppler approach actually underlying there, however, is questionable as already mentioned by [Hubble 1929] indirectly.

Ordinary gravitational redshift in local fields was predicted by Einstein and has nothing to do with any increasing distances. The SUM – as well as nearly every approach to cosmology today – is based on Einstein's equations. The redshift of starlight from extragalactic objects may therefore essentially be interpreted as a particular extension of ordinary gravitational redshift only. There is actually no need for a universal expansion, though the interpretation as Doppler effect is suggestive because time is involved in all cosmological solutions since Friedman(n)'s work. These solutions are not static of course. In the SUM framework, however, this means only a plenty of evolutionary processes, well compatible to stationarity with respect to sufficiently large scales of space and time.

Independently whether the phenomenon of gravitational redshift may be caused by local potentials as commonly accepted, or by the potential of the background universe, in both cases one is dealing with previously unknown effects of gravity derivable from the Einsteinian equations of RT, what for the latter effect will be shown in the following deductive sections. Apart from the historical view, there seem to be neither any reproducible facts nor any testable physical reasons which – applying Occam's razor – make a model of receding galaxies necessary for cosmology.

Apparently related to the well-known phenomenon called 'reduction of wave packets', GR may apply analogously to the universe in processes where QM is essentially involved. While in quantum leaps various physical possibilities are reduced to one single reality respectively, there is an analogy in the self-restoring aspects of SRT shown below. Therefore the description of physical reality by both RT and QM might be effectively 'quantized' itself, thus corresponding to a sequence of single snapshots making a movie. According to the SUM it does make no sense to search for a complete continuous history of the lively universe – in contrast to the natural search for the vital history of our cosmos instead.

Throughout this paper 'stationarity' means rather a revolutionary interplay than a 'steady state'.

### *1.2 Organization of the Paper*

For the sake of clarity, the deductive foundations of the SUM in part I are largely separated from the confrontation with observational facts in part II, which of course will require further development and ultimately decide.

I. The deductive part – In *SECTION* 2 the SUM is developed as a new stationary background cosmology on basis of Einstein's equations, including: The stationary line element $d\sigma_{SUM}$ deduced from two postulates (2.1); the self-restoring validity of SRT within local inertial frames (2.2); Motion of free particles, galaxies, clusters in the gravitational field of the background universe (2.3); the energy-stress tensor and a negative gravitational pressure of matter statistically at rest (2.4); time independence of both Hubble constant and redshift (2.5); the stationary magnitude-redshift relation (2.6); 'dark' matter and the ultra-large scale distribution of universal objects as a function of z (2.7); mean radiation density from thermal sources in the universe (2.8); transformation of the SUM line element to the corresponding FLRW form (2.9); the intrinsic limitations of proper length and proper time explained using two alternative systems S' and S of 'integrated' or 'adapted' coordinates (2.10); the universal embedding of local gravitational fields (2.11); cosmic evolution in a stationary background universe (2.12); some remarks on the SUM concept, its origin and related earlier attempts (2.13).

II. The comparative part – In *SECTION* 3 there is given a heuristic approach to the Cosmological Concordance Model, in particular: modeling different homogenous densities of energy, matter and radiation by one scale factor (3.1); SUM 'boundary' conditions matching the CCM density parameter $\Omega_\Lambda$ (3.2). – *SECTION* 4 is dedicated to the Supernovae Ia data in view of the SUM, including: A first comparison with the [Riess et al. 2004/07] 'gold' sample as well as with the CCM and its 'parents' EdS and SST (4.1); straight SUM accordance with 'the world's supernova distance-redshift data' on scales z > 0.1 (4.2); full scale compatibility 2008 given a local Hubble contrast (4.3); additional adaptability from effects like faint dimming by dust (4.4). – In *SECTION* 5 there is consi-dered a chance of having already observed parts of a stationary universe, with regard to: 'Primordial' nucleosynthesis and the law of entropy restricted to evolutionary processes (5.1); large-scale structure, quasar distribution, and a mass-to-radius relation (5.2); the CMB as a Black-Body (BB) Stationary Microwave Background (SMB) of redshifted components (5.3); the BB-SMB as only a special part of universal radiation (5.4).

III. Discussion and conclusions – After in SECTION 6 some rivaling aspects are discussed, SECTION 7 will briefly summarize the results and show future scope between two thinkable approaches to relativistic cosmology with straight SUM instead of the SST as an arguable alternative to the CCM.

Since originally the paper has been finished before the *Planck*-2013 data release, there is added only a brief preliminary appendix on these results in view of the SUM.

## I. THE DEDUCTIVE PART

It is impossible to do cosmology without appropriate principles which – besides the indispensable compatibility to observational facts – should fulfill the criteria of simplicity, adequacy and clarity. In the absence of such criteria not even the decision between a heliocentric and a geocentric conception of our planetary system would be possible within GRT because of legitimate mutual coordinate transformations. The simple idea leading to the SUM as the stationary cosmological solution of Einstein's equations is that no universal horizons must limit physical reality, where in the interplay with gravitation, quantum mechanics may locally reveal its full creative potential.

## 2. A STATIONARY BACKGROUND COSMOLOGY ON BASIS OF EINSTEIN'S EQUATIONS

When Einstein developed his first relativistic cosmology, he tacitly took for granted an eternal universe according to what was later called the 'perfect cosmological principle' in the SST. This homogeneous and isotropic large scale universe should be determined by its average densities of energy and pressure. Unfortunately he was focused on a static solution solely, whereas the relativistic model of a background universe developed here, will prove stationary instead.

Since evolution affects our own cosmos from a joint beginning, it may be appropriate to distinguish cosmos from universe, stationary the last and including all that is, was, and will be. Correspondingly *our* cosmos may be the largest structure of conjoint local origin surrounding the solar system today. Considering such a difference between cosmos and universe and regarding horizon problems or coincidences unacceptable for the latter, one will find the solution for a stationary relativistic cosmology without unnecessary peculiarities.

It will prove an essential feature that GRT represents gravitation, whereas SRT represents quantum mechanics in that the behavior of natural atomic clocks and spectral rods, which are displaying SRT 'proper' time and SRT 'proper' length, is governed by quantum mechanics undoubtedly.

As usual, the underlying concept is based on a strong simplification assuming idealized spatial homogeneity. A more realistic model should take into account statistical inhomogeneities already from its foundation, as addressed e.g. by [Buchert 2000/01], [Wiltshire et al. 2007], [Coley 2010a/b], [Buchert & Ostermann 2012]. Nevertheless, starting from a deductive SUM approach, now Einstein's equations of (G)RT will be found ready to support the lively model of a stationary ultra-large scale background universe. Once accepted a negative gravitational pressure and some new properties of 'dark' matter, there seem to be only 'geometric' reasons that the unexpected features of this model have been ignored so far.

The SUM may help clarify history and shape of our cosmic environment – or even 'our' cosmos – by distinguishing the peculiar features of e.g. a more and more fine tuned CCM from the ultra-large scale background as described here.

Of course, any physical theory of the universe cannot be based as of mathematical certainty. The intention of the following sections is, instead, to formulate the basics of a necessarily incomplete, but improvable SUM just as concisely and precisely as possible. Therefore, though several of the relations below are well-known, they are derived explicitly to make this Section 2 a self-contained presentation.

### 2.1 The Line Element $d\sigma^*_{SUM}$ Deduced from Two Postulates

According to the intention of a deductive approach one may start from two simple postulates which are sufficient to determine the relativistic SUM line element for an ultra-large scale background universe:

*Postulate I* – The universe is statistically stationary, homogeneous, and isotropic on sufficiently large scales.

*Postulate II* – Except for deviations caused by local inhomogeneities the universal coordinate speed of light $c^* = c$ is constant.

Obviously, the first postulate is equivalent to what has been called the perfect cosmological principle in the SST framework, while in contrast the second postulate will prove to imply several unexpected features.

If the universe – all-embracing by name – looks the same at all points, in all directions, at all times, this must apply particularly to the redshift of starlight emitted from sources statistically at rest with respect to universal coordinates. According to both postulates above, these coordinates $x^{*i} \equiv (t^*, l^{*\alpha})$ are fixed by the following definition:

The universal spatial coordinates $l^{*\alpha}$ are understood to be those of a Euclidean space filled with a stationary, homogeneous, and isotropic ultra-large scale distribution of matter, momentum and energy, while the universal time $t^*$ – where $t^* = 0$ may stand for 'today' – is determined by the condition of the constant universal coordinate speed $c^*$ introduced above. Any asterisk '*' always means *universal*.

These universal coordinates are a special representations of what is usually called 'conformal' time $t^*$ and 'comoving' space $\vec{r}^*$. Evidently, they apply to a universal frame, which not necessarily has to be identified exactly with the rest frame of the CMB as will be discussed later. Making statistical use of the Doppler effect, one may always define a preferred frame ranging as far as the most distant objects ever observed. In principle, already with Hubble´s discovery one could have referred to maximal isotropy of the cosmic redshift or, before that time, to a mean peculiar velocity zero of stars, though over at least extra-cluster distances only.

In contrast to intervals of the universal coordinates $(t^*, l^{*\alpha})$ themselves, their approximate local realizations are $(dt_{SRT},$

d$l_{SRT}$). These intervals of proper time and length are directly measurable within sufficiently small regions, using spectral clocks and rods, for example, which are local with respect to universal space *and* time. Both intervals are defined necessarily together according to the line element of SRT in local inertial frames

$$d\sigma_{SRT}^2 = c^2 dt_{SRT}^2 - dl_{SRT}^2 . \qquad (13)$$

To avoid unnecessary assumptions, it is sufficient to understand 'proper time' as the display of atomic clocks, and 'proper length' as a number of spectral rods, both correspondingly affected by gravitational potential and motion.

The mere definition of universal time $t^*$ above already presupposes a spatially Euclidean universe, implying that every universal line element can be written as a conformal relation

$$d\sigma_{SUM}^{*2} = \zeta_{SUM}^{*2} \left\{ c^2 dt^{*2} - dl^{*2} \right\}, \qquad (14)$$

where $dl^{*2}$ may stand for $dx^{*2} + dy^{*2} + dz^{*2}$ or for equivalent forms. Evidently the general expression of the line element (14) is required by postulate II of a constant universal (coordinate) speed of light

$$c^* \equiv \frac{dl^*}{dt^*} = c , \qquad (15)$$

which results from $d\sigma_{SUM}^* = 0$ and is not given with any form other than that above. It is obviously covariant (form-invariant) under Lorentz transformations and only this form allows to maintain the modified definition of a length unit on basis of the two-way time of light [Ostermann 2002, 2008b]. Dealing with universal distances, such a definition is not only convenient but essential for a clear and straightforward mathematical treatment (s. [v. Laue 1961] or e.g. [Durrer 2008]).

In addition, (14) is the simplest of all possible extensions leading from special to general relativity theory, if the intention is to take into account a non-empty universe. In particular, with the stationary assignment

$$\zeta_{SUM}^* = e^{Ht^*} \qquad (16)$$

valid from here, the SUM line element is fixed uniquely now, where $H$ is another macroscopic constant in addition to $c$ and $G$ only. In contrast to any other 'conformal' line element, the difference is made by the time scalar (16), in that it excludes any 'horizon' of the background universe. Together with its covariant energy-stress tensor $T_{ik}^*$ and all universal consequences to be drawn, the SUM line element (14), (16) does not only turn out to be non-singular, but to fulfill postulate I of stationarity, because:

Due to the exponential form of the time scalar $e^{Ht^*}$, all relative temporal changes depend on differences $\Delta t^* = t^* - t_R^*$ solely. This allows to set any reference point of universal time $t_R^* = 0$ for arbitrary complexes of observation. No matter how far in the past or in the future, adapting appropriate units once in each epoch, no special reference point $t_R^*$ – but possibly a local direction – of the universal time scale is preferred.

One may always substitute the universal time according to $t^* \equiv t_R^* + \Delta t^*$. With the legitimate assignment $t_R^* = 0$, then each isolated occurrence of universal time $t^*$ has tacitly to be taken for an interval $\Delta t^*$ only. Therefore, with respect to universal time $t^*$ – but not with respect to local proper time – there actually exists a stationary line element of relativistic cosmology (it is the combination of both postulates above which excludes – among all others – also the SST; after a transformation of its coordinate time $t'$ to a corresponding conformal time necessary to fulfill postulate II, there would appear an inacceptable singularity in its line element).

Explicit aspects of the actual stationarity will be shown in the following sections step by step, where the stationary universal line element (14), (16) may finally be written as

$$d\sigma_{SUM}^* = e^{Ht^*} d\sigma_{SRT}^* . \qquad (17)$$

The expression $d\sigma_{SRT}^*$ in (17), however, is different from the usual line element $d\sigma_{SRT}$ (13) of special relativity in that the elements of local proper time and length ($dt_{SRT}$, $dl_{SRT}$) have to be replaced in (17) by the elements of universal coordinates ($dt^*$, $dl^*$). In contrast to the first ones, the latter are not directly displayed by atomic clocks or spectral rods, except approximately for times $|t^*| \ll T_H \equiv 1/H$. In particular, the line element (17) shows the obvious transition from SRT to SUM as a key to the new cosmological model.

If one had started alternatively without using the above postulates I, II, but placing (17) as the evidently simplest ansatz for a cosmological line element of GRT with a non vanishing homogeneous Einstein tensor, one might have included SRT as a temporary approximation at times $t^* \approx 0$ in the neighborhood of any arbitrarily chosen reference point of universal time set $t_R^* = 0$.

In addition to the features already stated here, relation (17) suggests a possible extension (Sect. 2.11), to cover local inhomogeneities of matter and energy, too.

*2.2 The Self-Restoring Validity of SRT within Local Inertial Frames*

Although on basis of the stationary line element (17) the Maxwell vacuum equations remain valid, natural atomic clocks do not tick intervals of universal time $dt^*$ but intervals $dt_{SRT}$ of local 'proper' time.

According to the basics of GRT, the line element (13) of SRT applies approximately within freely falling local inertial frames. However, a continuous validity of SRT cannot be kept without interruptions. Instead, the observed validity of SRT is self-restoring again and again. The reason is the *non-integrability* of proper length and time, which has been found mathematically by [Einstein 1912a/b/c] himself once trying to disprove [Abraham's 1912a/b/c/d] claim SRT to be valid in infinitesimal small regions of the gravitational field. Retrospectively this might have suggested his final understanding of the equivalence principle, here shown to imply the *approximate* validity in freely falling inertial frames.

It may be of importance for the actual understanding of Einstein's RT today, that the disproof of Abraham's assumed claim had happened before his breakthrough to the ultimate equations of GRT some years later [Einstein 1916], where he – after Grossmann's start-up assistance [Einstein & Grossmann 1913] – transferred the mathematical apparatus of Riemannian geometry on gravitation. The first one who successfully introduced non-Euclidean geometry into relativity theory has been [Kaluza 1910] (rediscovered by [Stachel 1989]) in his pioneering treatment of [Ehrenfest's 1909] paradox of the rigidly rotating disk, the latter intensively discussed at that time. The varied genesis of GRT may have been the reason that Einstein's insight into the non-integrability of proper length and time apparently passed into oblivion. On the other hand, how can this feature go together with Einstein's equivalence principle, if not as a – repeatedly interrupted – self-restoring validity of SRT concluded here?

Since obviously the processes within local inertial frames – freely falling like space labs with varying relative velocities – cannot stay compatible continuously, deviations from SRT behavior might actually increase with time. How does nature manage to restore it again and again? To give the impression of an uninterrupted macroscopic validity of SRT, it seems sufficient that SRT is strictly valid for each process connecting any two quantum leaps, i.e. between emission and absorption of photons underway in a Michelson interferometer for example. Any quantum leaps, however, may imply an appropriate adaption of involved proper quantities to restore SRT again and again. In the context of the well-known reduction of quantum mechanical wave functions, such normalization processes are not out of place.

Now, comparing the local line element (13) within a freely falling local inertial frame on the one hand with the universal line element (17) ≡ (14),(16) on the other hand, Einstein's equivalence principle applies in the form

$$c^2 \mathrm{d}t_{\mathrm{SRT}}^2 - \mathrm{d}l_{\mathrm{SRT}}^2 \stackrel{!}{\approx} \mathrm{e}^{2Ht^*} \left\{ c^2 \mathrm{d}t^{*2} - \mathrm{d}l^{*2} \right\}. \quad (18)$$

This immediately leads to fundamental relations between universal coordinates ($\mathrm{d}t^*$, $\mathrm{d}l^*$) and local proper coordinates ($\mathrm{d}t_{\mathrm{SRT}}$, $\mathrm{d}l_{\mathrm{SRT}}$). Measuring a sufficiently small constant interval of universal time $\mathrm{d}t^*$ using atomic clocks at rest, or measuring a sufficiently small constant interval of universal length $\mathrm{d}l^*$ using spectral rods at rest, these measurements will obviously result in increasing intervals of local proper time $\mathrm{d}t_{\mathrm{SRT}}$ and local proper length $\mathrm{d}l_{\mathrm{SRT}}$, displayed as

$$\mathrm{d}t_{\mathrm{SRT}} \approx \mathrm{e}^{Ht^*} \mathrm{d}t^* , \quad (19)$$

$$\mathrm{d}l_{\mathrm{SRT}} \approx \mathrm{e}^{Ht^*} \mathrm{d}l^* . \quad (20)$$

Because of the non-integrability stated above it is used the symbol '≈' and not an equals sign here. Both relations imply the effects of the gravitational potential on atomic clocks and spectral rods in case of the stationary line element (17), where (19) means a relative time dilation with respect to universal intervals. Due to the constant universal speed of light $c^* = c$, constant intervals of universal length or universal time are evidently present in the wavelengths $\delta l^*$ and the oscillation periods $\delta t^*$ of free radiation for the time between emission and absorption. There do not exist, however, any local standards of constant universal length and time. While according to (19), (20) constant universal intervals $\delta t^*$, $\delta l^*$ measured as $\mathrm{d}t_{\mathrm{SRT}}$, $\mathrm{d}l_{\mathrm{SRT}}$ temporarily increase with time, the latter if taken constant seem to decrease relative to universal coordinates.

On the other hand, any such constant local proper time intervals – like the oscillation periods of spectral lines at place of their origin – do *not* change with time, of course, when measured with spectral ticks of atomic clocks at rest. It is self-evident that this natural constancy also applies to any constant local proper lengths – like the distance of neighboring nodes of a standing light wave at the place of the source – when measured with spectral rods. Natural quantities together with their natural standards are respectively changed by the same stationary time scalar $\zeta^* = \mathrm{e}^{Ht^*}$. Therefore in the case of local physics, the displayed *numbers* – i.e. the quotients of natural quantities and units – will be independent of universal time. This can also be concluded from Einstein's equivalence principle directly, stating that the influence of the gravitational potential is not measurable within freely falling local inertial frames.

Regarding any line element of relativistic cosmology, the essential non-integrability of proper length and time is obvious from the fact that it is simply impossible to write it down using *both* coordinates ($t_{\mathrm{SRT}}, l_{\mathrm{SRT}}$) only.

It makes a decisive difference in the SUM approach against the conventional treatment that the term *local* is not only related to space but also to time. In particular, it will be shown mathematically below that any local inertial frame governed by quasi-SRT if large in time, can only be small in space, and if it is large in space it can only be small in time. Therefore all SRT concepts like proper distance or proper time are limited to regions sufficiently small at least in space or in time, what immediately restricts the concept of a 'cosmic proper time' – presupposed since the first beginnings of relativistic cosmology – to comparably small regions of universal space (s. Sect. 2.10).

*2.3 Motion of Free Particles, Galaxies, Clusters
in the Gravitational Field of the Background Universe*

It is necessary to verify the basic assumption that the stationary line element (17) is compatible with an average distribution of matter and energy at rest. Therefore, the relativistic equations of motion will here be solved for the SUM. The result confirms an ultra-large scale universe at rest, as well as it gives the motion of free particles against this background as deduced from

$$\delta \int \mathrm{d}\sigma_{\mathrm{SUM}}^* = 0, \quad (21)$$

what is called Einstein's 'geodesic' law. The equations of gravitational motion resulting from (21) base directly on Einstein's equivalence principle. In addition, as is well-known, the following derivation from the phenomenological kinetic energy-momentum tensor

$$\mathbf{K}_{\mathrm{N}\,i}^{*k} = \boldsymbol{\mu}_{\mathrm{N}}^* c^2 u_i^* u^{*k}, \quad (22)$$

where the individual index 'N' may refer to a corresponding number density $n$, applies to the motion of any particle in the gravitational field given by all others, too. Bold non-italic symbols like $\mathbf{K}_{\mathrm{N}}{}^{*k}_i \equiv \sqrt{g^*}\, K_{\mathrm{N}}{}^{*k}_i$ or $\boldsymbol{\mu}_{\mathrm{N}}^* \equiv \sqrt{g^*}\, \mu_{\mathrm{N}}^*$ always include the square root of the negative determinant of the fundamental tensor as a prefixed factor only, where in case of the SUM $\sqrt{g_{\mathrm{SUM}}^*} = \mathrm{e}^{4Ht^*}$. Since here is $E_i^{*k} = \kappa K_i^{*k}$, the contracted Bianchi identities $E^{*k}_{i\,;k} \equiv 0$ yield

$$\partial_k^* \mathbf{K}_{\mathrm{N}\,i}^{*k} = \tfrac{1}{2} \mathbf{K}_{\mathrm{N}}^{*kl} \partial_i^* g_{kl}^{*\,\mathrm{SUM}}, \quad (23)$$

where $\partial_i^*$ stands for $\partial/\partial x^{*i}$. This equation (23) obviously results in the explicit form

$$\frac{d u_i^*}{d \sigma_{SUM}^*} = \tfrac{1}{2} u^{*k} u^{*l} \partial_i^* g_{kl}^{SUM} \qquad (24)$$

though only if a conservation of rest mass according to the continuity equation

$$\partial_k^* \left( \mu_N^* c^2 u^{*k} \right) = 0 \qquad (25)$$

is fulfilled there. Except for collision processes, this applies to the motion of test particles in any external field.

Actually, the variation of (21) with respect to the stationary universal line element (17) yields as solutions of (24) the temporal component of the universal four-velocity $u^{*i}$

$$u^{*0} \equiv \frac{c\,dt^*}{d\sigma_{SUM}^*} = e^{-Ht^*} \sqrt{1 + u_{(0)}^{*2} e^{-2Ht^*}}, \qquad (26)$$

and the spatial components

$$u^{*\alpha} \equiv \frac{dx^{*\alpha}}{d\sigma_{SUM}^*} = u_{(0)}^{*\alpha}\, e^{-2Ht^*}, \qquad (27)$$

where $u_{(0)}^{*2} \equiv \sum [u_{(0)}^{*\alpha}]^2$ ($\alpha = 1,2,3$). Obviously the integration constants $u_{(0)}^{*\alpha}$ are the initial values of the spatial components at time $t^* = 0$. From this simple calculation the components of the ordinary spatial velocity referring to universal coordinates are $v^{*\alpha} \equiv dx^{*\alpha}/dt^*$. Such velocities

$$\frac{v^{*\alpha}}{c} \equiv \frac{u^{*\alpha}}{u^{*0}} = \frac{u_{(0)}^{*\alpha}\, e^{-Ht^*}}{\sqrt{1 + u_{(0)}^{*2} e^{-2Ht^*}}} \qquad (28)$$

regarded as deviations from the state of statistical rest decrease with time.

Here it may be pointed out that the $u^{*i} = u^{*i}(x^{*i})$ in (22) and (25) are rightly related to a medium like a fluid, whereas in (24), (27), (28), for example, $u^{*i}$ should actually be replaced by $U^{*i} = U^{*i}(t^*)$ related to particles. The transition occurs by spatial integration of the original factor $\mu_N^*$ on both sides, which in case of particles effectively applies as a $\delta$-function respectively.

Only for zero-rest-mass particles like photons where, because of $d\sigma_{SUM}^* \to 0$, relation (27) implies $u_{(0)}^{*\alpha} \to \infty$, a constant velocity results in the universal speed of light $|v^{*\alpha}| \to c$ directly. On the other hand, for all particles of non-vanishing rest-masses this apparently means a deceleration with respect to universal coordinates. Therefore – though concluded from the stationary line element (17) – even in intergalactic space a freely falling inertial frame would not keep on moving uniformly with respect to these coordinates. This again implies that there is no physical situation where SRT can be valid otherwise than locally and approximately only.

There raises the question how an object leaving any Schwarzschild region shall turn continuously to the universal motion as derived here (s. Sect.s 2.10.2, 2.11 for a corresponding modification of Galileo's law of inertia).

The relativistic equations of motion (28) support the idea of galaxies statistically at rest in universal Euclidean space. This even applies to long-term averages of peculiar motions like that of galaxies bound in clusters, for example. According to (28), the special solution describing this situation is

$$\overline{v}^{*\alpha} = 0 \;, \qquad (29)$$

where – here as an exception – a bar means averaging over time. This solution (29) is actually that of matter statistically at rest, since in the SUM there is no need for the otherwise established concept to interpret the same feature as an 'expansion' presupposing a 'comoving' universal coordinate frame. The results (26), (29) show one non-vanishing component of the mean four-velocity $\overline{u}^{*i} = (\overline{u}^{*0}, 0, 0, 0)$, which is

$$\overline{u}^{*0} = e^{-Ht^*} = \frac{1}{\overline{u}_0^*} \qquad (30)$$

implying a universal accelerating time rate of atomic clocks at rest. This may have been also approximately concluded from (19) in the form $dt^*/d\sigma_{SRT}^* \approx e^{-Ht^*}$.

Evaluating (27), (28) completely, the universal four-velocity $u^{*i}$ may be written in a form analogous to that of SRT at last, namely

$$\left( u^{*0}, u^{*\alpha} \right) \equiv \frac{\left( 1, \dfrac{v^{*\alpha}}{c} \right)}{\sqrt{1 - \dfrac{v^{*2}}{c^2}}} e^{-Ht^*}, \qquad (31)$$

where in $v^{*2} \equiv \sum [v^{*\alpha}]^2$ the summation has to be carried out for $\alpha = 1, 2, 3$ again. Relation (31) is formally different from the SRT assignment by multiplication of the reciprocal time scalar $e^{-Ht^*}$ only, while according to (28) $v^{*\alpha}$ is not constant here. The result (31) proves the consistency of the relations above, since it may be alternatively derived using the definitions of 4-velocity $u^{*i} \equiv dx^{*i}/d\sigma^*$ and that of ordinary velocity $v^{*\alpha} \equiv dx^{*\alpha}/dt^*$ directly.

Now, given the stationary line element (17), relation (25) yields in case of free particles at rest

$$\mu_N^* = \mu_N^{*\,const}\, e^{-3Ht^*}, \qquad (32)$$

where evidently

$$\mu_N^{*\,const} = \frac{dm_N}{dV^*}. \qquad (33)$$

Therefore the rest mass $\delta m_N$ of such a 'particle' is constant, whether taking it from the universal volume $\delta V^*$ or from the local proper volume $\delta V = \delta V^* e^{3Ht^*}$ according to

$$\delta m_N = \mu_N^{*\,const} \delta V^* = \mu_N^* \delta V. \qquad (34)$$

Here 'particles' mean locally bound systems including idealized galaxies or clusters, for example. The result of constant mean rest masses is in accordance with the stationarity of the universal matter-energy distribution (though with regard to the energy exchange by radiation or collision processes individual universal objects do not obey a rest mass conservation law, there may be a statistical equilibrium, s. also 2.8).

Since given the statistically averaged number density of 'particles' as presupposed independent of time with respect to universal coordinates, now together with the constant rest masses just derived, also the SUM matter density may be regarded statistically independent of time.

### 2.4 The Energy-Stress Tensor and a Negative Gravitational Pressure of Matter Statistically at Rest

Trying to describe the ultra-large scale universe, it is usually asked which line element might follow from an assumed cosmological energy-stress tensor or from other observational facts. Here – the other way round – discussing the relativistic principles of cosmology proposed above, it is the question which energy(-momentum)-stress tensor follows from the stationary line element (17) instead.

Given the corresponding fundamental tensor $g_{ik}^* = \mathrm{e}^{2Ht^*}\eta_{ik}$, the universal energy-stress tensor $\bar{T}_{ik}^*$ actually proves constant in case of the SUM as deduced from Einstein's original equations (1), and may be written here in the form

$$\bar{E}_{ik}^* = \tfrac{2}{3}\kappa\varepsilon_c \begin{pmatrix} 1 & 0 & 0 & 0 \\ 0 & 0 & 0 & 0 \\ 0 & 0 & 0 & 0 \\ 0 & 0 & 0 & 0 \end{pmatrix} + \tfrac{1}{3}\kappa\varepsilon_c\,\eta_{ik} = \kappa\bar{T}_{ik}^*, \quad (35)$$

where $\varepsilon_c\,\eta_{ik} \equiv \bar{\rho}^* c^2 g_{ik}^*$ with $\bar{\rho}^* = \rho_c \mathrm{e}^{-2Ht^*}$ and as a rule, if not otherwise stated, a bar means averaging over space. In the framework of the SUM, the critical energy density $\varepsilon_c \equiv 3H^2/(\kappa c^2)$ is a real constant where $\kappa \equiv 8\pi G/c^4$. The original covariant form of the Einstein tensor $\bar{E}_{ik}^*$ in (35) above and thus the corresponding stationary energy-stress tensor $\bar{T}_{ik}^*$, too, are obviously independent of time, what also applies to their contravariant tensor densities $\bar{\mathbf{E}}^{*ik}$ and $\bar{\mathbf{T}}^{*ik}$ after all. In this context it may be mentioned that Einstein's 'geodesic' law of motion does not only result as usual from the mixed form $T_{i;k}^k = 0$ but from the contracted Bianchi identities $T^{ik}_{\;;k} = 0$, too, where the last would include the constant $\bar{\mathbf{T}}^{*ik}$ explicitly. Furthermore, in contrast to $T_0^{\,0}$, only $T_{00}$ seems necessarily positive as stated in [Landau & Lifschitz 1992].

Remarkably, in the time-independent covariant equations (35) the Hubble constant appears via $\varepsilon_c$ only in the quadratic form $H^2$, what means that given the same mean universal energy density, solutions of (35) would be conceivable with a different sign of $H$ in various regions.

To apply Einstein's equations to macroscopic gravitation according to the conventional treatment, one may formally define another scalar $\bar{\mu}_0^* \equiv \bar{\mu}_N^* \mathrm{e}^{Ht^*}$ in addition to the density given by (32). Then the commonly used mixed form $\bar{T}_i^{*k} = \varepsilon_c \mathrm{e}^{-2Ht^*}\cdot\mathrm{diag}(1,\,^1/_3,\,^1/_3,\,^1/_3)$ of (35) looks like the well-known purely phenomenological energy-stress tensor

$$\bar{P}_i^{*k} \equiv \bar{\mu}_0^* c^2 \bar{u}_i^* \bar{u}^{*k} - \bar{p}^*\delta_i^k\,. \quad (36)$$

Note that inserting $\bar{p}^* = 0$ into (36), the tensor $\bar{P}_{i\,(\bar{p}^*=0)}^{*k}$ is not the same as $\bar{K}_{Ni}^{*k}$ in (22) of the last section, because the first one is that of an idealized 'perfect-fluid', whereas the second is that of a universal 'particle' distribution in its mutual gravitational field. That the latter is the appropriate representation immediately reflecting stationarity has been shown above.

It may remain some small range for modifications in the assignment of Einstein's phenomenological energy-momentum-stress tensor of a perfect fluid, whose provisional nature has been already addressed in Section 1.1. Traditionally, a generalization $P_i^k$ without asterisks and bars is used to replace $T_i^k$ in the mixed form of Einstein's equations directly, what means 'locally' with respect to universal scales of space and time. This replacement, however, was actually bound to the condition $p \geq 0$ corresponding to all laboratory experience with ordinary matter only. On the other hand, even astrophysical experience can never firmly include the totality of a background universe as a whole. Therefore the conventional ansatz $T_i^k = P_i^k$ may not necessarily apply to the SUM without any modification.

Mistaking $\bar{\mu}_0^*$ of (36) for $\bar{\mu}_N^*$ in (22) would seemingly contradict the stationarity of the matter-energy distribution stated in the previous section. The concept of rest mass, however, is tied not at all to uniquely defined quantities, just as the rest mass of an H-atom is different from that of proton plus electron, for example. In addition, Section 3 will show that other models make effectively also use of a corresponding modification. Now, formally accepting the traditional assignment

$$\bar{T}_i^{*k} = \bar{P}_i^{*k} \quad (37)$$

instead of $\bar{T}_i^{*k} = \mathrm{e}^{\nu Ht^*}\bar{P}_i^{*k}$ (with $\nu$ here an appropriate number) or correspondingly $\bar{T}_i^{*k} = \bar{P}_i^{*k} + \bar{W}_i^{*k}$, for example, it would follow

$$\bar{\mu}_0^* c^2 = \tfrac{2}{3}\varepsilon_c\,\mathrm{e}^{-2Ht^*}, \quad (38)$$

$$\bar{p}^* = -\tfrac{1}{3}\varepsilon_c\,\mathrm{e}^{-2Ht^*}, \quad (39)$$

where it has to be kept in mind that also $\bar{\mu}_0^*{}_{(t^*=0)}$ and $\bar{p}^*{}_{(t^*=0)}$ are effectively representing unchanged values for arbitrary reference points of universal time, since $t^*$ means $t_R^* + \Delta t^*$ with $t_R^* = 0$ again and again.

In any case, however, relation (39) means the existence of a *negative* universal gravitational pressure which is evidently required by Einstein's equations for a stationary universe. In contrast to the ordinary positive pressure $p$ of fluids or gases, a negative universal gravitational pressure is not unacceptable, quite the contrary; in the textbooks of relativity theory this has not been considered a physical option for a long time. To state it explicitly, however, a stationary universal gravitational pressure $\bar{p}_N^*$ must be negative because:

Let a subvolume of a large hall filled with ordinary dust be separated by a box. Since the situation in the box will stay the same after all matter outside the box is removed, this implies a positive pressure of the dust because the walls of the box are exerting force *inwards* to bar the dust from diffusion. Now in contrast, consider a separate subvolume of a stationary universe including a plenty of galaxy clusters without peculiar velocities. Then there must be a negative pressure equivalent to hypothetical walls which in this case had to pull *outwards*, to prevent the homogenous ultra-large scale distribution of galaxy clusters inside from massing together due to their mutual attraction, after those outside had been removed.

In this context, there also appears some relationship between the negative gravitational pressure and a local decrease of entropy since – the other way round – the well-established increasing entropy of ordinary gas is clearly associated to its positive pressure causing all well-known diffusion.

With respect to (36), (38), the phenomenological density of matter without the contribution of the negative gravitational pressure may be only $\bar{\mu}_0^* = 2/3\,\bar{P}_0^{*0}$ according to the assignment (38) above. Though even taken this matter density rather than the full tensor component $\bar{P}_0^{*0}$, however, it would be by far too much regarding the small amount of universal matter observed directly or indirectly today. Therefore in view of the SUM, there should be in addition an almost homogeneous part of presently 'dark' matter in the universe.

With the stationary universal energy-stress tensor $\bar{T}_{ik}^*$ on hand, it is possible to verify once more the equilibrium of the universal matter-energy distribution derived from Einstein's 'geodesic' equations in the previous section. The contracted Bianchi identities $\bar{T}_{i;k}^{*k} \equiv 0$ imply

$$\partial_k^* \bar{P}_i^{*k} - \tfrac{1}{2} \bar{P}^{*kl}\,\partial_i^* \bar{g}_{kl}^* \;=\; 0\,. \tag{40}$$

Actually, taking the generalization mentioned above, in case of an extended object with non-vanishing variable pressure $p$ the 'geodesic' equations of motion corresponding to (24) cannot apply to each of its elements independently, except for a special kind of 'free fall' where

$$u_i \partial_k (\mu_0\,u^k) = \sqrt{g}\,\partial_i p\,. \tag{41}$$

Thus, a conclusion from $\bar{\mu}_0^*$ instead of $\bar{\mu}_N^*$ on rest masses of 'particles' is impossible since (41) shows that there is no continuity equation of matter valid here. Therefore, though a galaxy or cluster may be regarded as 'particles' in the universal gravitational field, this does not apply to arbitrary parts of the ultra-large scale matter-energy distribution described by the stationary tensor $\bar{T}_{ik}^*$.

In spite of these complications, the only non-vanishing relation of (41) in case of the SUM is the one indexed $i = 0$. Inserting the corresponding universal quantities $\bar{\mu}_0^*$ for $\mu_0$ and so on, however, this relation is fulfilled taking into account (30) and (36). For mass-energy with $\bar{u}^{*\alpha} = 0$ statistically at rest, this yields according to the results of the previous section

$$\frac{d\,\bar{u}_i^*}{d\sigma_{\text{SUM}}^*} \;=\; \left(\frac{H}{c},0,0,0\right), \tag{42}$$

what means it will stay at rest. The same result would follow from the alternative ansatz

$$\bar{T}_i^{*k} \;=\; e^{Ht^*}\,\bar{P}_{Ni}^{*k}\,, \tag{43}$$

where instead of (36) it is

$$\bar{P}_{Ni}^{*k} \;\equiv\; \bar{\mu}_N^*\,c^2\,\bar{u}_i^*\,\bar{u}^{*k} - \bar{p}_N^*\,\delta_i^k\,, \tag{44}$$

and instead of (38), (39), but according to (32), now is set

$$\bar{\mu}_N^*\,c^2 \;=\; \tfrac{2}{3}\varepsilon_c\,e^{-3Ht^*}, \tag{45}$$

$$\bar{p}_N^* \;=\; -\tfrac{1}{3}\varepsilon_c\,e^{-3Ht^*}. \tag{46}$$

The new assignment (43) may be plausible because taking into account (45), (46), the contracted Bianchi identities $\bar{T}_{i;k}^{*k} \equiv 0$ now read

$$\bar{P}_{Ni;k}^{*k} \;=\; -\frac{H}{c}\,\bar{P}_{Ni}^{*0} \tag{47}$$

instead of (40), where the non-vanishing right-hand side obviously would mean a sufficiently 'small' modification. For universal particles statistically at rest, this again would lead to (42). As a result from $\bar{T}_{i;k}^{*k} \equiv 0$, this is largely independent of isolating a factor $e^{\nu Ht^*}$ within $\bar{T}_i^{*k}$ as mentioned above.

Thus, the feature of an average star velocity $\bar{u}_\alpha^* = 0$ is confirmed by (42) as well as by the absence of any energy flux in (35), (37) or (43). It may be emphasized that – in contrast to the problematic assignment of the phenomenological 'perfect fluid' energy-stress tensor to the SUM's Einstein tensor – the solutions (30) or (42) are in case $\bar{v}^{*\alpha} = 0$ already unambiguously determined by a direct evaluation of $d\sigma_{\text{SUM}}^*$.

Besides the conservation of universal mass-energy stated in the previous section, the complete conservation laws of GRT are in general $\partial_k V_i^k = 0$, strictly valid for the mixed bi-tensor density $V_i^k = T_i^k + t_i^k$, which would allow for an exchange with the energy of gravitational waves, too. That $t_i^k$ is more than a 'pseudo'-tensor density, but with regard to the universal frame a true bi-tensor density of the gravitational field, has principally been explained in Section 1.1 above. In case of the SUM, here the total matter-energy bi-tensor density including that of the gravitational field is

$$\bar{V}_i^{*k} \;\equiv\; \bar{T}_i^{*k} + \bar{t}_i^{*k} \;=\; \tfrac{4}{3}\varepsilon_c\,e^{2Ht^*}\begin{pmatrix}0 & 0 & 0 & 0\\ 0 & 1 & 0 & 0\\ 0 & 0 & 1 & 0\\ 0 & 0 & 0 & 1\end{pmatrix}. \tag{48}$$

In particular, the obvious result $\bar{V}_0^{*0} = 0$ holds for Einstein's original definition of $t_i^k$, s. [Einstein 1916], as well as for some alternatives, see e.g. [Landau & Lifschitz 1992] or [Weinberg 1972]. At first glance it may look strange that the total energy density of matter and gravitational field should be zero. In the system S' of integrated coordinates (Sect. 2.10.1), however, there will be found $\bar{V}'^k_i = \bar{T}'^k_i$ as another result, the latter with non-zero total energy and even fulfilling the ordinary conservation laws there.

Independent of questions caused by the traditional assignment (37), with the presupposed constant number density of universal objects, the rest mass conservation stated in the previous section does not only apply to microscopic particles but also to gravitationally bound systems up to galaxies or even clusters. Therefore – regarding those structures statistically at rest – this means a conservation of universal mass-energy, too, thus corresponding to the evidently stationary covariant energy-stress tensor (35) or its contravariant density immediately. The conventional perfect-fluid interpretation based on the time-dependent mixed tensor $T_i^k$, however, might account together with the bi-tensor $t_i^k$ of the gravitational field for 'local' processes of emergence and disappearance instead.

*2.5 Time Independence of Both
Hubble Constant and Redshift*

The following deduction will show that in spite of the time-dependent scalar $\zeta^*_{SUM}$ (16) the line element (17) proves stationary again. In particular, the universal redshift of galaxies as the fundamental observational fact of cosmology is found independent of time. This feature then also applies to all other quantities which are functions of $z$ like the apparent magnitude of SNe Ia used as standard candles, for example, or naturally to the Hubble constant $H$ in the SUM framework itself.

Starting from the assumption that – according to the solution (29) in Section 2.3 – galaxies are statistically at rest with respect to universal coordinates, now the redshift

$$z \equiv \frac{\lambda_A}{\lambda_E} - 1 \qquad (49)$$

will be calculated in complete analogy to the well-known gravitational redshift in local fields, where the indices 'E', 'A' mean emission or absorption respectively.

As usual, consider the crest of a light wave emitted at universal time $t_E^*$ anywhere at a distance $l^*$ in the Euclidean ('comoving') space and arriving at universal time $t_A^*$. Then the following crest, emitted at the same place as before but at time $t_E^* + \delta t^*$, will arrive at $t_A^* + \delta t^*$ because of the constant universal speed $c^* = c$ of light. This means that the interval $\delta t^*$ – which is nothing but the oscillation period $\tau_0^*$ of propagating starlight with respect to universal time $t^*$ – has been transported and kept unchanged over an intergalactic distance $l^* = c\Delta t^*$ where $\Delta t^* \equiv t_A^* - t_E^*$.

On the other hand, a proper time interval $\tau_0$ of a natural atomic clock at rest is related to the corresponding interval $\delta t^* = \tau_0^*$ of universal time according to (19). Hence at the time $t_E^*$ of emission and at the time $t_A^*$ of arrival, the corresponding proper time intervals are

$$\tau_{A/E} = \tau_0^* e^{H t_{A/E}^*} \qquad (50)$$

respectively. With regard to relation $\lambda = c\tau$ for wavelength and period of light, it follows immediately that the corresponding intervals of proper length and time will be different in a proportion

$$\frac{\lambda_A}{\lambda_E} = \frac{\tau_A}{\tau_E} = e^{H\Delta t^*}, \qquad (51)$$

where

$$\Delta t^* = l^*/c \qquad (52)$$

is just the positive transit time of extragalactic light. Obviously, the result (51) does not depend on the single absolute values $t_E^*$ or $t_A^*$ of universal time, but only on their positive difference $\Delta t^*$ and the constant $H$. This is one more detailed example fulfilling the postulate of stationarity because after having inserted $t_A^* = t_R^*$ and $t_E^* = t_R^* - \Delta t^*$ into (50), the physical result (51), (52) prove the non-occurrence of the arbitrary reference time $t_R^*$ directly..

So far, though, $\tau_E$ in (51) is only the proper time interval at the time $t_E^*$ of emission whereas $\tau_A$ is a proper time interval at the time $t_A^*$ of absorption. But the actual question is to compare the oscillation period $\tau_A$ with that oscillation period $\tau_0$ of new spectral radiation of the same type both emitted at place and time of absorption. It is obvious, however, that with respect to local proper time – in contrast to $\tau_0^*$ – the oscillation period of one particular spectral line is $\tau_0 = \tau_E = constant$ at place and time of its origin. This is a direct consequence of Einstein's equivalence principle. Using natural atomic clocks, the same statement would be a mere tautology, because the design of those clocks is just based on this constancy. Since measuring means comparing, the common constant factor $e^{H t_R^*}$ which would appear in (50) cancels out because displayed on clocks is the *quotient* of measured natural quantities and corresponding local natural units only, which are always changed at the same rate

Now, inserting the 'infinitesimal' wavelengths $\lambda_{A/E} = c\tau_{A/E}$ according to (51) into (49) respectively, the redshift parameter $z$ is found completely independent of time for starlight emitted from sources at rest,

$$z = e^{Hl^*/c} - 1 \quad \Leftrightarrow \quad l^* = \frac{c}{H}\ln(1+z), \qquad (53)$$

where $l^* = c\Delta t^*$ is the covered universal distance, after all. Therefore, to get a simple explanation for the redshift of galaxies it is sufficient to make the difference between proper intervals ($\delta t_{SRT}^*$, $\delta l_{SRT}$), and universal intervals ($\delta t^*$, $\delta l^*$) of time and length according to (19), (20). Not only the redshift, but also the local time dilation is clearly confirmed by the SNe-Ia measurements quoted in Section 4 below.

Now both relations (53) – since applying to sources statistically at rest – actually prove the physical relevance of universal spatial coordinates. This means, in addition to intervals of local 'proper' length $dl_{SRT}$, the quantity $l^*$ is a physical distance since it is an immediate universal measurand determined by time-independent values of $z$. Thus, the SUM makes the difference to all other flat space models of GRT. The striking proof of stationarity according to (53), of course, is clearly what was aimed at by introducing the exponential ansatz for the universal time scalar (16).

On the other hand, it is usually concluded from (20) that fixed values of $l^*$ should mean increasing proper distances, what indeed is suggesting the universal coordinates as 'comoving' ones. As already mentioned above and will be explicitly shown in Section 2.10, however, any proper distances are inappropriate to cover the universe. In view of the SUM, the concept of literally 'comoving' coordinates therefore seems rather misleading. For example, the stationarity of (53) stands also in clear contrast to the SST, where – to avoid any confusions again – the redshift parameters would be of the form $z = z_0 (l^{*\alpha}) e^{Ht'}$, thus depending on time in the 'expanding universe' presupposed there.

Concerning an unexpected problem of relativistic cosmology, it may be stated already that there is a subtle but far-reaching difference between a time-dependent conventional Hubble parameter and the significant Hubble constant occurring in (53). This will be cleared up in a general FLRW context, s. Section 2.9 below.

Now, from the quantum mechanical energy-frequency relation for photons – but also deducible from classical electrodynamics in GRT – and with

$$\nu_E \equiv \nu_A (1+z) \qquad (54)$$

according to (49), the extended form (53) of Hubble's linear approximation shows that the redshift also applies to photon energies as

$$\delta\varepsilon_A = \delta\varepsilon_E \, e^{-Hl^*/c} . \qquad (55)$$

Re-substituting $l^*$ by $c\Delta t^*$ here, the cosmic redshift apparently requires the energy of free photons to decrease relative to local absorbers with universal time. Such a time-dependent energy loss of free photons might look like a violation of an overall energy conservation, but – given a stationary universe – with respect to ultra-large scales it is not. In this case, with statistically constant values of $l^*$ relation (55) may be understood a stationary energy loss affecting the whole of free photons respectively. Its mathematical form is exactly that of the familiar laws of ordinary attenuation what, amazingly, would include the hypothetical absorption once assumed by [Olbers 1823] in his proposal to solve the famous paradox at the beginning of modern cosmology. The main objection made against Olbers' absorption hypothesis will be taken up in Section 2.8, and questioned essentially or even disproved in Section 5 at last.

Furthermore, relation (55) may be also regarded as completion of both relations (19), (20) in that it affects mass as the third basic quantity of physics. In this context, on the one hand it has to be taken into account that statements about homogeneously distributed matter at rest are not applicable straightforwardly to the energy of propagating photons. But on the other hand, the energy differences of atoms at rest before and after emission, naturally agree with the energy of the corresponding photons at place and time of their origin. In any case, however, a relation corresponding to (55) does neither apply to the rest mass of particles constituting cosmic rays nor to that of galaxies, for example, which all are conserved according to (32).

Altogether, with respect to universal coordinates – now measurable by their constant redshift parameters according to (53) – and except for peculiar motions or any processes of reformation, galaxies as well as other universal objects may statistically stay where they are.

*2.6 The Stationary Magnitude-Redshift Relation*

Given a universal object (U) of the absolute radiation power $L_U^*$ at a constant distance $r^*$ with respect to universal coordinates, the SUM implies as apparent luminosity

$$I_U^* = \frac{L_U^*}{4\pi r^{*2}} \, e^{-(2+\kappa)\frac{r^*}{R_H}}, \qquad (56)$$

with $R_H \equiv c/H$, which is the bolometric intensity of the radiation observed per square unit and locally measured per unit of proper time. Here from the redshift (53), (55), a first factor $e^{-r^*/R_H} = 1/(1+z)$ results as usual by application of the quantum mechanical energy-frequency relation of photons, and a second factor $e^{-r^*/R_H}$ from the relative dilation (50) of the local proper time. Furthermore, taking into account possible effects of attenuation like extinction, absorption, scattering, or obscuring, there is a corresponding coefficient $\kappa$ in (56) which is set constant here, though applying to spectral distributions it may be taken a function $\kappa(\nu)$ of frequency if necessary. Obviously $\kappa/R_H$ corresponds to the reciprocal of a mean free path of the respective radiation. Inserting

$$r^* = R_H \ln(1+z) \qquad (57)$$

taken from (53) leads to

$$I_U^*(z) = \frac{L_U^*}{4\pi R_H^2} \left[ (1+z)^{1+\frac{\kappa}{2}} \ln(1+z) \right]^{-2} . \qquad (58)$$

This relation is neglecting any 'local' cosmic evolution and does not yet take into account thinkable effects of inhomogeneities or any systematic peculiar flow of our cosmic environment. To compare the result (58) obtained here with the SNe-Ia apparent magnitude-redshift data in Section 4 directly, it has to be converted to the distance modulus

$$m - M = 5 \log\left(\frac{d_L^*}{\text{Mpc}}\right) + 25 , \qquad (59)$$

where $m$ is the apparent magnitude, $M$ represents an appropriate value for the absolute standard brightness of e.g. SNe Ia, and $d_L$ is the *luminosity distance* defined by

$$d_L^* \equiv \sqrt{\frac{L_U^*}{4\pi I_U^*}} = r^* (1+z) \, e^{\frac{\kappa}{2R_H} r^*} , \qquad (60)$$

which then may be written as a pure function of redshift

$$d_L^*(z) = R_H (1+z)^{1+\frac{\kappa}{2}} \ln(1+z) . \qquad (61)$$

Inserting this into (59) yields

$$m - M = 5 \log\left[(1+z)^{1+\frac{\kappa}{2}} \ln(1+z)\right] + 25 + 5\log\left(\frac{R_H}{\text{Mpc}}\right) \qquad (62)$$

Since for sources at rest in universal ('comoving') coordinates the redshift parameters $z$ are independent of time, so are the magnitudes and all other quantities, too, which are functions of $z$. It is relation (62) for the distance modulus which will be shown in Section 4.2 to fit the SNe-Ia magnitude-redshift observations on universal scales. That this accordance applies straightforwardly in the high redshift range $z > 0.1$, is just reflecting the intention that (17) should describe the universe on ultra-large scales where it is justified to assume the averaged densities to be homogenous and isotropic. More details and possible effects due to a local Hubble contrast $\delta H/H$ or to dimming by a small amount of intergalactic 'gray dust' will be explicitly addressed in Section 4, too.

In particular the distance modulus is of fundamental interest for each cosmological model in question, since it establishes a nearly unobjectionable relation between the directly measurable values of the apparent magnitudes $m$ and the redshift parameters $z$. It may be also remarkable in this context, that the SNe-Ia data did not prove any significant cosmic evo-lution for a long time, thus indicating a self-restoring validity of local physics again.

### 2.7 'Dark' Matter and the Ultra-Large Scale Distribution of Universal Objects as a Function of z

The difference between the estimated universal matter density and its the theoretical value, as stated in Section 2.4, is essentially reduced taking into account some 'dark' matter which is indirectly observed inside galaxies and clusters or by gravitational lensing so far.

Now in view of the SUM, a still remaining deficit might be explained by an additional contribution of a sufficiently homogeneous background $\bar{\mu}^*_{\text{background}}$ of that 'dark matter', too, where the homogeneous part (hDM) may be associated to what is called 'dark energy' today. All that 'dark' matter would not necessarily consist of only one fraction of particles, but there may be various components also including unseen macroscopic objects with an effectively transparent distribution.

As is well-known, the universe seems irregularly structured by filaments, superclusters, voids, and walls, interfused with corresponding densifications of 'dark' matter and an Inter-Galactic Medium (IGM). According to the tentative approach considered in parts of Section 2.4 – though in contrast to the conventional view – the main contribution to universal matter might exist in form of a 'dark' background. Except for field galaxies, most of other ones seem gravitationally bound to dark-matter halos of clusters with an Intra-Cluster Medium (ICM), where hot gas is emitting X-ray radiation. Several types of galaxies seem dominated by dark-matter and otherwise composed of stars and various amounts of an Inter-Stellar Medium (ISM) primarily containing cosmic rays, gas, or dust. While stars are the sources of stellar radiation, dust clouds seem the main source of (far-)infrared radiation. According to such a tentative straight SUM approach, it has to be taken into consideration that 'dark' matter may be the main source of a universal microwave radiation, where – in contrast to the mm-range of the non-Planckian Cosmic Infrared Background (CIB) – what is called 'CMB' would be only one part of it. A theoretical distribution of universal objects U may be roughly estimated here as a function of $z$.

Considering an idealized uniform number density $n^*$ of homogeneously distributed objects like stars, galaxies, quasars or clusters, for example, the number of them included within a spherical shell between $r^*$ and $r^* + dr^*$ is

$$dN_U^* = n_U^* dV^* = 4\pi n_U^* r^{*2} dr^* \qquad (63)$$

with

$$n_U^* = \frac{\Omega_U \rho_c}{M_U}, \qquad (64)$$

where as usual $\Omega_U$ is the parameter of a mean matter density given by $\mu_U^* \equiv \Omega_U \rho_c$, and $M_U$ the mass of a typical object.

Inserting (64) as well as $r^*$ and $dr^*$ taken from (57) into (63) yields

$$\frac{dN_U^*}{dz} = 4\pi n_U^* R_H^3 \frac{\ln^2(1+z)}{(1+z)} \qquad (65)$$

not yet taking into account any effects of possible absorption, selection, or local evolution. The total number of respective objects is $N_U = \infty$ as easily verified by integration. This natural result corresponds directly to the concept of the SUM, where the underlying stationary line element (17) does not imply any horizons of the universe as a whole.

The idealized distribution (65) shows a flat peak at $z_{\text{peak}} = e^2 - 1 \approx 6.4$ while it is approximating zero in the limit $z \to \infty$. The value $z_{\text{peak}}$, though, seems clearly above the observed maximum at $z_{\text{observed}} \approx 2$ (s. Sect. 5.2) of Quasi Stellar Objects (QSOs/quasars). However, the steep decrease of the QSO distribution in the interval $2 < z < 4$ to almost zero as shown in [Schneider et al. 2010], for example, does not necessarily mean a steep decrease in the actual number density, since there is implied a selection bias due to a magnitude limit of about 20.2 mag. In particular Section 5.2 will come back to this subject.

Summarizing the various aspects, the universal 'dark' matter distribution may be similar to a fluid medium of high viscosity filling all the space, though with local overdensities in form of bulges, halos or clusters gathering stars and galaxies, while in huge seeming voids between them the density is low but yet high enough to make the dominant fraction of universal matter and energy. Besides the well-known forms of stellar and interstellar or intergalactic matter there might be, in addition, two main sorts of that 'dark' matter, one of them consisting of non-baryonic particles like possibly e.g. thermalized neutrinos, the other one consisting of unknown baryonic objects cold and small enough to be 'invisible' for telescopes, both making up a universal non-lensing background together with its local inhomogeneities.

### 2.8 Mean Radiation Density from Thermal Sources in the Universe

In a stationary universe there is necessarily a stationary distribution of temperature. Its mean value should be largely determined by that of the homogeneously distributed part, while local DM inhomogeneities may be of only approximately the same temperature.

The SUM order of magnitude for mean bolometric flux densities $F_U^*$ coming from remote universal objects like galaxies, halos, or clusters may be roughly estimated now. It is appropriate to start from

$$dF_U^* = I_U^* \cos\vartheta \, dN_U^*, \qquad (66)$$

where $dN_U^* = n_U^* dV^*$ has to be written in spherical coordinates leading to

$$dF_U^* = I_U^* n_U^* r^{*2} dr^* \cdot \int_0^{\pi/2} \cos\vartheta \sin\vartheta \, d\vartheta \int_{-\pi}^{\pi} d\varphi \qquad (67)$$

as an integration over the hemisphere what yields

$$dF_U^* = \pi I_U^* n_U^* r^{*2} dr^*. \qquad (68)$$

After insertion of $I_U^*(z)$ from (58) and $r^{*2} dr^*$ according to (57) again, this relation reads

$$dF_U^* = \frac{1}{4} R_H n_U^* L_U^* \frac{dz}{(1+z)^{3+\kappa}} \tag{69}$$

with $L_U^*$ the mean absolute radiation power of a typical universal object U, while their average number density $n_U^*$ is given by (64) with $M_U$ the mass of a typical member. Integration of (69) leads as an average flux density coming from the sources with universal redshifts between $z_1 < z < z_2$:

$$F_U^*(z_1, z_2) = \frac{1}{4} \frac{R_H \rho_c}{(2+\kappa)} \Omega_U \frac{L_U^*}{M_U} (1+z)^{-(2+\kappa)} \Big|_{z_2}^{z_1} . \tag{70}$$

Correspondingly, the total flux density from all point-like U-objects in the infinite universe is

$$F_U^* = \frac{1}{4} \frac{R_H \rho_c}{(2+\kappa)} \Omega_U \frac{L_U^*}{M_U} . \tag{71}$$

In spite of an infinite number of objects U, the intensity (71) is obviously finite and, again, independent of time. According to Stefan-Boltzmann's law, the radiance $B_U^{*SB}$ of any Black-Body (BB) radiation at an absolute temperature $\Theta_U$ is

$$B_U^{*SB} = \frac{2\pi^4}{15} \frac{k^4 \Theta_U^4}{c^2 h^3} . \tag{72}$$

If there was a representative special sort U of objects, a comparison of $B_U^{*SB}$ with the mean radiance $B_U^* \equiv F_U^*/\pi$ as taken from (71) shows that an equivalent black body radiation would be of an effective temperature

$$\Theta_U^{\text{effective}} = \frac{1}{k} \sqrt[4]{\frac{15 h^3 c^2}{16\pi^5} R_H \rho_c \frac{L_\odot}{M_\odot} \frac{\Omega_U}{X_U \left(1 + \frac{\kappa_U}{2}\right)}} , \tag{73}$$

where $X_U$ means the corresponding mass-to-light ratio according to $M_U/L_U^* = X_U \cdot M_\odot/L_\odot$ in units of the sun. Thus, Olbers' paradox is solved in the SUM framework even for $\kappa = 0$, what means without taking into account ordinary extinction yet. It is unnecessary to assume any 'big bang' here.

In addition to relation (71) which approximately may also apply to superclusters, filaments, or walls if these are sufficiently far away, now corresponding to (56) the local intensity $I_U$ (here without an asterisk) at the surface of an idealized spherical gravitationally bound universal object may be set

$$I_U \equiv \frac{L_U}{4\pi R_U^2} \tag{74}$$

with $R_U$ an effective radius. In case of $L_U = L_U^*$, a comparison of (71) and (74) yields a mass-to-radius relation, which in general reads

$$M_U = \frac{\pi R_H \rho_c}{(2+\kappa)} \frac{\Omega_U}{(F_U^*/I_U)} R_U^2 . \tag{75}$$

Remarkably, already an unpretentious tentative approach based on the results derived here, will show some unexpected relationship with observations in Section 5.2. There will also be taken into consideration that $\kappa$ may not only depend on ordinary extinction but on obscuring by at least partially opaque spheres. Therefore the attenuation coefficient may be split up into

$$\kappa = \kappa_U + \kappa_{\overline{U}}, \tag{76}$$

where the last summand would correspond to an additional extinction by non-U objects or other matter distributions. Now for idealized spherical universal objects of constant absorptivity $\alpha_U$ it would apply

$$\frac{\kappa_U}{R_H} = \pi \alpha_U \rho_c \Omega_U \frac{R_U^2}{M_U}, \tag{77}$$

where $\alpha_U$ may be associated to an effective cross section $\alpha_U \cdot \pi R_U^2$. This leads from (75) to

$$M_U = \frac{\pi}{(2+\kappa_{\overline{U}})} \left[ \frac{1}{(F_U^*/I_U)} - \alpha_U \right] R_H \rho_c \Omega_U R_U^2 , \tag{78}$$

Obviously the expression in square brackets of (78) has to be positive what means $F_U^* < I_U$ for $\alpha_U = 1$. In general an absorptivity or opacity $\alpha_U(\nu) = 0$ means completely transparent, and $\alpha_U(\nu) = 1$ completely opaque structures.

Given that a distant spherical source has been emitting some grey-body radiation of spectral distribution $\beta_U \rho_{BB}$ with $\beta_U$ a constant emissivity – where in this case it is $\beta_U = \alpha_U$ corresponding to the absorptivity in Kirchhoff's law – then coming back to (69) it is possible to extend this relation for the observed spectrum. At first, analogously to (74) with $L_U$ replaced by $L_U^*$ and $I_U$ by $\beta_U \cdot \pi B_U^{*SB}$ it is

$$L_U^* = \beta_U \cdot \pi B_U^{*SB} \cdot 4\pi R_U^2, \tag{79}$$

while the bolometric radiance of non-polarized thermal radiation (72) is the integral value of

$$B_U^{*SB} = \frac{c}{4\pi} \int_0^\infty \rho_{\nu_U, \Theta_U} d\nu_E . \tag{80}$$

In this context $\rho$ means the Planck spectrum where $\nu_U$, $\Theta_U$ are frequency and temperature of the corresponding radiation at place and time of emission. Inserting (79) with (80), (64), and (77) into (69) where $\kappa$ is reduced to $\kappa_U$ yields

$$\Delta F_U^* = \pi \kappa_U \cdot \left( \frac{c}{4\pi} \int_0^\infty \rho_{\nu_U, \Theta_U} d\nu_U \right) \cdot \frac{\Delta z}{(1+z)^{3+\kappa_U}} . \tag{81}$$

For the spectral density of a redshifted BB radiation it is

$$\rho_{\nu,\Theta_U/(1+z)} \, d\nu = \frac{1}{(1+z)^4} \rho_{\nu_U,\Theta_U} \, d\nu_U, \quad (82)$$

since the emitted frequency has to be replaced by $\nu_U = \nu(1+z)$ according to (54). Taking this into account in connection with

$$B^*_{U\nu} \equiv \frac{1}{\pi} \frac{dF^*_U}{d\nu} \quad (83)$$

according to the identity $F_U^*/\pi \equiv B_U^*$ again, relation (81) may finally be written as

$$B^*_{U\nu} = \kappa_U \cdot \frac{2h}{c^2} \cdot \nu^3 \int_0^\infty \frac{(1+z)^{1-\kappa_U} \, dz}{e^{\frac{h\nu(1+z)}{k\Theta_U}} - 1}, \quad (84)$$

what is no longer a grey-body spectrum, of course. According these results, considering the observation of distant thermal sources in the universe, the influence of redshift has to be taken into account.

Before coming back to relation (84) in Section 5.4 again, now a natural question is whether there may be some radiation of non-resolvable sources, too. From the huge amount of 'invisible' matter stated in Sections 2.4, 2.7, might come a temperature radiation making most of the CMB. Keeping in mind that unknown forms of homogeneously distributed 'dark' matter should make up most of the critical-density needed for the SUM line element deduced in Section 2.1, such a radiation should actually exist and it should be of BB type. Apparently a plausible reason for such a feature is:

Consider an empty unheated space capsule in form of a hollow sphere which is at a sufficiently large distance from localizable radiation sources. As a consequence of all the radiation being absorbed and emitted, the walls of the capsule may reach a constant temperature as shown in relation (73) above. If a fictitious observer inside this capsule drills a fictitious hole into the wall, this observer will be in the situation of a physicist primarily measuring the cavity radiation of the stationary universe *enclosed outside* the sphere.

With regard to (73), taken a tentative mean value $X_U \approx 10$ for the mass-to-light ratio of the universal energy-matter distribution with respect to its critical density, the effective temperature of the corresponding radiation would result near the temperature of an equivalent black body of 3 K roughly. This seems appropriate order of magnitude straight off, though the problem remains how nature might accomplish the microwave thermalization in detail. It would be one more strange coincidence, however, which seems hardly a matter of pure chance. In any case, a black test-object under the only influence of the stationary universal radiation apparently would cool down or heat up to about such a temperature.

Therefore it seems a suggestive attempt to question the CMB possibly radiated from distant unresolvable spherical sources in radiation equilibrium with itself. From (71), (79), and (77), it follows that the mean radiance $B_U^* \equiv F_U^*/\pi$ is

$$B^*_U = \frac{\beta_U}{\alpha_U} \frac{\kappa_U}{(2+\kappa)} B^{*\,SB}_U. \quad (85)$$

Replacing $\kappa$ according to (76), however, shows that given constant values $\beta_U = \alpha_U$ of 'grey' sources there cannot be a stationary equilibrium for pure redshifted thermal radiation. An exception would be the unrealistic case $\kappa = \kappa_U \to \infty$ of an almost completely microwave-opaque medium, which is definitely excluded, however, by quasar observations in the mm-range, see e.g. [Haas et al. 2006]. Thus, in the SUM framework, there should be some additional contribution to the emissivity $\beta_U$ of universal CMB sources.

It may be remarked in addition, that without redshift there would lack the summand "2" in the parentheses of (85). Then even for only one single sort of thermal sources like stars, with $\beta_U = \alpha_U$ and $\kappa = \kappa_U$, there would be that unsettling bright night sky, resulting from (85) in $B^*_{stars} = B^*_{sun}{}^{SB}$ as already concluded in Olbers' paradox once.

Now, taking into consideration that, given a frequency-dependent emissivity $\beta_{SMB}(\nu_E)$ for a stationary microwave radiation at a constant mean temperature $\Theta_{DM}$, the locally emitted radiation itself has not necessarily to be of simple grey-body type, the following composition

$$\rho^*_{SMB\,\nu} = \frac{8\pi h}{c^3} \nu^3 \int_0^\infty \beta_{SMB}[\nu(1+z)] \frac{(1+z)^{1-\kappa}}{e^{\frac{h\nu(1+z)}{k\Theta_{DM}}} - 1} \, dz, \quad (86)$$

where

$$\beta_{SMB}(\nu_E) = \frac{\frac{h\nu_E}{k\Theta_{DM}}}{1 - e^{-\frac{h\nu_E}{k\Theta_{DM}}}} \quad (87)$$

would lead to the expected result. Here, according to (54), $\nu_E = \nu(1+z)$ is the frequency of the SMB radiation at place and time of its origin. In fact, it is easily verified that in case of $\kappa = 2$ an integration of (86) yields exactly Planck's law

$$\rho^*_{SMB\,\nu} = \frac{8\pi h}{c^3} \frac{\nu^3}{e^{\frac{h\nu}{k\Theta_{DM}}} - 1}. \quad (88)$$

The ansatz in (86) where the coefficient $\kappa$ is still dealt with as a constant though $\beta_{SMB}(\nu_E)$ is set a function of frequency, takes preliminarily into account that emission is a local effect where even microwave synchrotron radiation may be involved, while absorption happens all the way from distant sources to the observer. Therefore this ansatz may be one among others leading to approximately the same result. Here, however, it is primarily to show that the SUM is well compatible to the existence of some SMB radiation of BB character.

Comparing the local radiance $dB_{SMB}^{*\,local}$ in a shell of thickness $dr^*$ to the local attenuation $dA_{SMB}^{*\,local}$, the first is found from (86), (87) after re-substituting $z$ according to (53) and then setting $r^* = 0$, $\nu = \nu_E$ as

$$\mathrm{d}B_{\mathrm{SMB}}^{*\,\mathrm{local}} = \frac{\mathrm{d}r^*}{R_H} \int_0^\infty \frac{\frac{h\nu_E}{k\Theta_{\mathrm{DM}}}}{1-e^{-\frac{h\nu_E}{k\Theta_{\mathrm{DM}}}}} \frac{2h}{c^2} \frac{\nu_E^3}{e^{\frac{h\nu_E}{k\Theta_{\mathrm{DM}}}}-1} \mathrm{d}\nu_E \quad (89a)$$

resulting in

$$\mathrm{d}B_{\mathrm{SMB}}^{*\,\mathrm{local}} \,\, '=' \,\, 4 B_{\mathrm{SMB}}^{*\mathrm{SB}} \frac{\mathrm{d}r^*}{R_H}. \quad (89b)$$

Remarkably, the factor "4" on the right-hand side of (89b) seems to result exactly, though only found by numerical integration ('='). On the other hand, the effective total attenuation of the SMB radiation (88) is due to local absorption *plus* local redshift

$$\mathrm{d}A_{\mathrm{SMB}}^{*\,\mathrm{local}} = (2+\kappa) \frac{\mathrm{d}r^*}{R_H} \int_0^\infty \frac{2h}{c^2} \frac{\nu_E^3}{e^{\frac{h\nu_E}{k\Theta_{\mathrm{DM}}}}-1} \mathrm{d}\nu_E \quad (90a)$$

resulting in

$$\mathrm{d}A_{\mathrm{SMB}}^{*\,\mathrm{local}} = (2+\kappa) B_{\mathrm{SMB}}^{*\mathrm{SB}} \frac{\mathrm{d}r^*}{R_H}. \quad (90b)$$

Now, unexpectedly in these details, if given the solution above $\kappa=2$ again, there would be an energetic equilibrium for emission and total local attenuation in the same shell

$$\mathrm{d}B_{\mathrm{SMB}}^{*\,\mathrm{local}} = \mathrm{d}A_{\mathrm{SMB}}^{*\,\mathrm{local}}. \quad (91)$$

This, however, seems to imply a strange consequence. The factor $(2+\kappa)$ may be regarded something like an effective 'extinction coefficient' $\kappa_{\mathrm{effective}}$, where according to (56) its first summand "2" clearly originates from redshift. As stated before, one part is caused by local time dilation and the other part by the energy loss of photons.

Furthermore as a generalization, this result (91) would even suggest the possibility of a tentative answer to the question, where the energy of any redshifted photons might be lost before they are absorbed anywhere in the universe. In view of the SUM – here presupposed naturally – there must be a statistical energy re-cycling from radiation to stars to keep them shining, though not forever the same.

As seen in (71), for example, according to the SUM the flux density coming from sources all over the universe is finite and independent of time. Therefore even in case of an energy loss without traceable re-cycling governed by a continuity equation, there would be a stationary overall statistical distribution of radiation, though without detailed information about how the radiation losses turn back to the sources. On the other hand, probably most of the starlight seems already absorbed by ordinary matter building new stars or returning to SMOs in galaxies, for example.

After all, because of Heisenberg's uncertainty principle and the Bose-Einstein statistics any attempt would fail to treat all photons of the background radiation as mere particles on traceable ways from emission to absorption across the universe. Furthermore, in a lively – locally non-isothermal though otherwise stationary – universe, what is called CMB has not to be the only background radiation. Section 5.4 will come back to this issue. Various additional contributions – as for example the CIB overlapping the CMB in parts – will be also addressed there.

*2.9 Transformation of the SUM Line Element to the Corresponding FLRW Form*

To directly compare the stationary SUM line element with today's CCM in Section 3 below, it is convenient to rewrite (17) in a traditional FLRW form which – given spatial flatness and keeping $l^{*\alpha}$ the universal ('comoving') coordinates – may be written here as

$$\mathrm{d}\sigma_{\mathrm{FLRW}}'^2 = c^2 \mathrm{d}t'^2 - a^2 \mathrm{d}l^{*2}, \quad (92)$$

where $a \equiv a(t')$ is the general scale factor. Obviously $t'$ is the FLRW coordinate time which will be referred to as the *integrated* coordinate time, since it is given by direct integration of (19) after having replaced $\mathrm{d}t_{\mathrm{SRT}}$ by $\mathrm{d}t'$. and the sign '$\approx$' by '='. This replacement is necessary because the local intervals of proper length $\mathrm{d}l_{\mathrm{SRT}}$ and proper time $\mathrm{d}t_{\mathrm{SRT}}$ are not integrable without changing their character. The integrable FLRW time $t'$, though, cannot be understood as a valid 'cosmic proper time', otherwise the expression $a^2\mathrm{d}l^{*2}$ of (92) had to be identical to $\mathrm{d}l^2_{\mathrm{SRT}}$. If in the valid relation

$$\mathrm{d}l_{\mathrm{SRT}} \approx a\,\mathrm{d}l^*, \quad (93)$$

however, an equal-sign '=' was used instead of the approximate-sign '$\approx$', the whole relation (92) would be nothing but the line element of SRT itself – whose Riemann, Ricci, Einstein tensors and therefore the entire universal mass energy density would vanish to zero. There are intrinsic limitations of proper length and time as consequences of this necessary distinction.

Now a determination of the stationary scale factor $a_{\mathrm{SUM}}$ can be done by simply transforming the universal time $t^*$ to the integrated quasi-proper time $t'$ or $T' \equiv T_H + t'$ where $T_H \equiv 1/H$ without thereby changing any physical results, of course. Using the relation

$$t^* \equiv \frac{\ln(HT')}{H} \quad (94)$$

found from (19) by direct integration as mentioned above, a corresponding coordinate transformation of (17) yields the stationary FLRW-form

$$\mathrm{d}\sigma_{\mathrm{SUM}}'^2 = c^2 \mathrm{d}T'^2 - a_{\mathrm{SUM}}^2 \left( \mathrm{d}r^{*2} + r^{*2} \mathrm{d}\Sigma^{*2} \right), \quad (95)$$

where $r^*$ is the radial distance and $\mathrm{d}\Sigma^*$ the element of a Euclidean spherical surface in universal ('comoving') coordinates. Then the SUM scale factor

$$a_{\mathrm{SUM}} \equiv HT' \equiv 1 + Ht' \quad (96)$$

equals the stationary time scalar $\zeta^*_{\mathrm{SUM}}$ (16) as is obvious from (94). In contrast, the SST scale factor $a_{\mathrm{SST}} = e^{Ht'}$ would result in a horizon problem – corresponding to a mathemati-cally small, but physically essential difference – which in view of the SUM is regarded an unacceptable feature.

The apparent singularity of (95) at $T' = 0$ cannot disprove the *universal* SUM stationarity found in the previous sections, because: According to the covariance of GRT, the alternative FLRW representation (95) must yield the same physical results as the original stationary line element (17) of the ultra-large scale background universe. It is easily verified, for example, that from (95) the exact Hubble relation (53) holds in its time-independent form, too. In fact, taking into account the FLRW coordinate velocity of light $c' = c/a_{SUM}$, the covered radial distance $l^* \equiv l_A{}^* - l_E{}^*$ between the time of emission $t'_E$ and the time of absorption $t'_A$ is

$$l^* = \int_{t'_E}^{t'_A} c' \, dt' = \frac{c}{H} \ln \frac{1+Ht'_A}{1+Ht'_E}. \qquad (97)$$

From this result, calculating the universal redshift in mathematical analogy to the original derivation done by [Lemaître 1927/31], or e.g. [Weinberg 1972], the parameter $z$ is found the same again as in relation (53) of Section 2.5 above, keeping the full stationarity of the magnitude versus redshift relation, the distribution of galaxies, and the mean radiation density as well. Thus it may be emphasized here, that this stationarity is a coordinate-free statement resulting from the universal line element (17) as well as from (95) now, or even from (104) below, for example.

To argue along the traditional lines of relativistic cosmology, the stationary 'deceleration' parameter defined as $q(t') \equiv -a\ddot{a}/\dot{a}^2$ is found $q_{SUM} \equiv 0$ as it must be (some more remarks in 2.13).

In contrast, however, since the *conventional* Hubble parameter $H_c(t') \equiv \dot{a}/a$ is yielding the time-dependent value $1/T'$ in case of the SUM, it might be confusing to have found the stationary redshift (53) actually *independent* of time. Therefore, it seems necessary to show explicitly, that in general the *significant* FLRW Hubble parameter is $H_s(t') \equiv \dot{a}$, what – given the stationary scale factor $a_{SUM} \equiv HT' \equiv 1+Ht'$ – actually means a true Hubble constant $H_{s\text{-}SUM} = H$ again.

In view of far-reaching consequences this is shown in more detail now. With regard to the FLRW-form (92), the definition (49) of redshift $z \equiv \lambda_A/\lambda_E - 1$ may be transferred to

$$z \equiv \frac{a(t'_A)}{a(t'_E)} - 1 \equiv \frac{\Delta a_{AE}}{a(t'_E)} \approx \frac{\dot{a}}{a} \Delta t', \qquad (98)$$

where a dot means differentiation with respect to $t'$. Since light propagates according to $d\bar{\sigma}_{FLRW} = 0$ and a local element of proper length is assumed to be $\Delta l' \approx a \Delta l^*$, it is

$$\Delta t' \approx \frac{a \Delta l^*}{c} \qquad \text{or} \qquad \Delta t' \approx \frac{\Delta l'}{c}. \qquad (99)$$

Inserted both into (98) it follows at first Hubble's law in its *significant* form

$$cz \approx \dot{a} \Delta l^* \equiv H_s \Delta l^*, \qquad (100)$$

as well as approximately the same law in its *conventional* form

$$cz \approx \frac{\dot{a}}{a} \Delta l' \equiv H_c \Delta l', \qquad (101)$$

where according to (99) the expression $\Delta l' \approx c \Delta t'$ is commonly regarded the 'proper' distance to the light source.

Even in view of traditional cosmology, however, the conventional assignment of the Hubble parameter $H_c$ on the right hand side of (101) is misleading, because with respect to 'comoving' coordinates it is not the proper distance $\Delta l'$ which is presupposed to be independent of time, but the universal ('comoving') distance $\Delta l^*$ instead, thus clearly confirming relation (100). Independently of the respective scale factor $a(t')$ this means the significant assignment $H_s \equiv \dot{a}$ after all.

### 2.10 The Intrinsic Limitations of Proper Length and Proper Time

Though the SUM stands for a stationary ultra-large scale background universe, the FLRW form (95), (96) of its universal line element (17) does no longer look stationary at all. In spite of the time-independent redshift verified in the last section, it seems that at the negative Hubble time $t_H' = -T_H$ (i.e. $T' = 0$) all proper lengths $dl_{SRT}$ had been zero, all proper densities infinite, and therefore the whole universe a mere singularity.

Today at $t^* = t' = 0$, the time scalar $\zeta^*$ as well as the FLRW scale factor $a$ are usually fixed to the value 1 by the choice of appropriate units. This choice is a need according to the approximate validity of SRT in our freely falling local region of space and time.

But consequently, SRT requires this scale factor to be adapted to that value 1 again and again, whereas pure GRT would obviously require different values at different times. In the SUM framework this implies an eternal struggle of self-restoring local SRT against strict universal GRT. Here might appear in outlines an interplay of evolutionary and revolutionary processes instead of any ongoing expansion of universal 'space' as concluded from an overstrained concept of proper length. It has been shown in Section 2.2 that without the self-restoring aspect, SRT could not effectively stay valid since any infinitesimal Lorentz transformations are not integrable in gravitational fields.

Within the SUM framework the proper-distance relation $dl_{SRT} \approx (HT')dl^*$ taken from (93), (96) can only apply on scales which are local with respect to space and time. Accordingly the FLRW coordinate time $t'$, $T' \equiv T_H + t'$ cannot be – or at least, has not necessarily to be – understood as one uniform proper time valid all over the universe.

In analogy, the limited local validity of SRT at different regions of the universe – with its own approximate proper time each – corresponds to the fact that although a perfect spherical surface is seen approximately flat at each point, the respective contact planes are not the same. In spite of the strong concepts of SRT locally valid of course – as may even also be seen in constant periods and radii of the planetary orbits both measured by means of atomic clocks – it is not justified to apply these concepts to universal distances or universal periods of time straightforwardly.

It will be explicitly shown here, how on basis of Einstein's equations the idea of an *infinite* stationary universe turns out to imply clear indication that individual cosmic structures are necessarily of *finite* dimensions in space and time. This con-clusion arises from the interplay of local SRT (macroscopically representing quantum mechanics) and universal GRT (representing gravitation). Such a scenario may suggest a universal '*tohu w'a-bohu*' (with all due respect) where the origin of our cosmos once might have taken place.

### 2.10.1 The Local System S' of Integrated Coordinates

To demonstrate the intrinsic limitations of proper length and proper time in more detail, one may define an alternative set of coordinates by transformation. The resulting quasi-proper approximation will prove that no real proper time interval can be suitable to hold at and beyond proper distances $r_{SRT} \to R_H$. To this end the transformation obtained by direct integration of (19) and – though only with respect to $t^* =$ *constant* – of (20) is followed by replacing the variables ($t_{SRT}$, $r_{SRT}$) by ($t'$, $r'$) and the signs '≈' by '='.

The fundamental feature in this context is that according to SRT, the proper time interval $dt_{SRT}$ is always locally defined *together* with $dr_{SRT}$ according to relation (13) of Section 2.1. That the mathematical integral $t'$ of $dt_{SRT}$ cannot be exactly an overall universal proper time, is evident from the fact that otherwise a suitable transformation of (17) had to result in $c^2 t'^2 - r'^2 \equiv \sigma^2_{SRT}$, what is impossible because the Riemann tensor of (17) is non-vanishing, in fundamental contrast to that of SRT.

*Definition* – The *integrated* coordinates ($r'$, $T'$) – where $T' \equiv T_H + t'$ with the time $t' = 0$ meaning today – are related to the universal coordinates implicitly by

$$t^* \equiv \frac{\ln(HT')}{H} \Leftrightarrow dT' \equiv dt^* e^{Ht^*}, \qquad (102a)$$

$$r^* \equiv \frac{r'}{HT'} \Leftrightarrow dr' \equiv dr^* e^{Ht^*} + Hr^* dt^* e^{Ht^*}. \qquad (102b)$$

Considering both relations on the right hand sides of (102a), (102b), the second one shows the non integrability of (20) directly. On the other hand, using the relations on the left – where the variable $T' \equiv T_H + t'$ is obviously identical to the FLRW coordinate time above – the stationary line element (17) is transformed exactly into the quasi-proper line element

$$d\sigma'^2 = \left[1 - \left(\frac{r'}{cT'}\right)^2\right] c^2 dT'^2 + 2\left(\frac{r'}{cT'}\right) c dT' dr' - dl'^2 \quad (103)$$

with $dl'^2 \equiv dr'^2 + r'^2 (\sin^2\vartheta d\varphi^2 + d\vartheta^2)$. In contrast to the stationary line element (17), the quasi-proper line element (103) shows a local character now. Furthermore, while in the FLRW form (95) the temporal component $g_{00} = 1$ appears normalized to that of SRT, in the quasi-proper-form (103) it is the spatial component $g_{rr} = -1$ instead. Here the reason for the term 'quasi-proper' coordinates is obvious, since the stationary line element in the form (103) as an approximation to that of SRT today ($T' \approx T_H$) can be valid only in local cosmic regions limited by $r' < cT'$.

Tough at first glance this region may look like an expanding area of SRT applicability, now a replacement of ($r'/cT'$) by ($r^*/R_H$) according to (102b) with $R_H \equiv c/H$, yields the same quasi-proper line element (103) in its significant form

$$d\sigma'^2 = \left[1 - \left(\frac{r^*}{R_H}\right)^2\right] c^2 dT'^2 + 2\left(\frac{r^*}{R_H}\right) c dT' dr' - dl'^2 . (104)$$

Here the limitations of applicability refer to the universal ('comoving') coordinate $r^*$ associated to mean fixed positions of galaxies. They culminate in the simple condition:

$$r^* \overset{!}{<} R_H . \qquad (105a)$$

The line element $d\sigma'$ of (104) can coincide approximately with that of SRT only in the 'neighborhood' of arbitrarily choosable coordinate origins. This may be indication for a stationary universe seeded with local cosmic areas limited to extensions $r^* \ll R_H$, where a transfer of SRT concepts may coherently apply. In particular, relation (105a) is the reason why the integrated time $T'$ as a quasi-Minkowskian coordinate approximation to a local proper-time integral $t_{SRT}$ is not at all suitable to hold at or beyond universal distances $r^* \to c/H$.

Considering the FLRW form (95), (96) concerning the limit $T' \to 0$ in this context, now relation (105a) evidently indicates that the 'big bang' concept

$$0 \leq T' \leq T_{H_{(0)}} \qquad (105b)$$

of today's cosmology would only affect local cosmic regions of at most a radius $r^* \approx R_H$ or even distinctly smaller. Instead of one singular universal origin, local regions of gravitational re-creation may be spread all over the universe.

Inserting $dT' = 0$ into (104) would result in the spatial line element of SRT exactly. Then, however, given a real homogeneous distribution of matter for some standstill instant of time all over the universe, only the smallest process – necessarily implying $dT' \neq 0$ – would be sufficient to break the symmetry leading to cosmic regions of radii $r^* < R_H$ where 'local bangs' might apply. This raises the question how large those areas as for example our actual evolutionary cosmos would be. Not only $r^* \approx R_H$ equivalent to $z \approx e - 1 \approx 1.7$ seems possible as a maximum value for coherent structures, but even values down to $r^* \ll R_H$ may not yet be excluded definitely. From (103) compared with (13) it is

$$\left. \begin{array}{l} dt' \approx dt_{SRT} \\ dr' \approx dr_{SRT} \end{array} \right\} \quad (r^* \ll R_H), \qquad (106)$$

where because of (102b) the condition $r^* \ll R_H$ corresponds to $|r'/cT'| \ll 1$. Consequently the interval $dt_{SRT}$ of proper time – though mathematically integrable if taken isolated – is not integrable over universal spatial distances uniformly.

With his development of GRT, [Einstein 1916] explicitly had drawn the conclusions from Ehrenfest's paradox that in particular it is impossible to apply SRT without any coordinate time in addition to local proper time. Thus, one can refer to the integrable FLRW coordinate $t'$ only as a quasi-proper time, which is displayed temporarily and approximately by atomic clocks within limited regions of universal space. Correspondingly it remains indispensable as well, to distinguish local proper time $t_{SRT} \ll T_H$ from universal time $t^*$.

Therefore the conventional conclusion from (95), (96) to a singularity of the entire universe cannot be regarded as an un-disputable paradigm. Instead it turns out to be inherent in the principles of relativity theory itself that – as already assumed by Einstein – the mathematical singularity at $T' = 0$, seems to indicate rather a limitation of local applicability than an origin of space and time.

According to (94), (102a), the point $T' = 0$ actually means infinite past $t^* = -\infty$ with respect to universal time, what the other way round means that there cannot exist any individual macroscopic structures older than $T_H \equiv 1/H$ with respect to their proper time. On the other hand this may not yet exclude for example a cluster to get older than $T_H$, though it would not be the same after this time since in particular galaxies and stars would have arisen and gone by.

While $T_H$ has not necessarily to be understood the age of the universe as a whole, now what is called expansion of space might be understood a universal condensation of material structures arising again and again.

Calculated from (103) and taking into account $g' = 1$ and $t'^k_i = 0$, the complete energy-momentum bi-tensor density of matter and gravitational field $\mathbf{V}'^k_i$ – corresponding to (48) above – coincides with $E'^k_i$ here. In the system S' of integrated coordinates it is

$$\overline{\mathbf{V}}'^k_i = \frac{\varepsilon_c}{(HT')^2} \begin{pmatrix} 1 & \frac{2}{3}\frac{x'}{cT'} & \frac{2}{3}\frac{y'}{cT'} & \frac{2}{3}\frac{z'}{cT'} \\ 0 & \frac{1}{3} & 0 & 0 \\ 0 & 0 & \frac{1}{3} & 0 \\ 0 & 0 & 0 & \frac{1}{3} \end{pmatrix}. \quad (107)$$

In contrast to the situation described in Section 2.4 and to $\mathbf{V}^{*k}_i$ found in (48) with respect to universal coordinates, here is a non-vanishing energy density $\mathbf{V}'^0_0$ of matter and gravitational field with respect to integrated coordinates, though locally limited by (105a) to $r^* < R_H$. Considering an 'expanding universe' one might argue, that actually S' should be the appropriate frame instead of S*, since – taking into account the density $\mu' = 2/3\, \rho'$ – it appears a flux of mass-energy density with just the right velocity $Hx'^\alpha$ away from the origin. Moreover this flux would just match the mass-energy loss in the interior of a sphere around the origin, too.

This view, however, is unsustainable for the universe as a whole because of the arbitrarily chosen but respectively preferred origin of the system S' in the line element (103). The direction of an energy flux through a test surface would depend on which side of this surface the coordinate origin was chosen. It might cause little difficulties, however, to understand (107) as only one individual representation of an infinite number of locally expanding cosmic regions instead.

### 2.10.2 The Local System S of Adapted Coordinates

Though $t'$, $T'$ is the simplest quasi-proper coordinate time, it is not yet the best one approximating $t_{SRT}$ in local spacetime regions of our present epoch given by $Ht^* \ll 1$. On basis of the line element (103) with respect to integrated coordinates, the Maxwell equations are not even valid in first order $Hr'/c$. This becomes obvious considering the coordinate speed of light $c'_\pm \approx Hr' \pm c$ there. Hence, to avoid such first-order errors, it may be appropriate to refine the transformation formulae (102a), (102b) to *adapted* coordinates $x^i$ now:

*Definition* – The adapted coordinates $(r, t)$ – where $T \equiv T_H + t$ with $t = 0$ for today – are implicitly given by

$$t^* \equiv \frac{\ln(1+Ht)}{H} - \frac{1}{2}\frac{Hr^2}{c^2} \Leftrightarrow dt^* \equiv \frac{dt}{1+Ht} - \frac{Hr\,dr}{c^2}, \quad (108a)$$

$$r^* \equiv \frac{r}{1+Ht} \Leftrightarrow dr^* \equiv \frac{dr}{1+Ht} - \frac{Hr\,dt}{(1+Ht)^2}. \quad (108b)$$

As an improvement in comparison with the integrated coordinates of (102a), (102b), the adapted coordinates according to (108a), (108b) are suitable to transform the stationary line element (17) up to second order $O^2(Ht, Hr/c)$ into that of SRT here ($r \ll R_H$) and today ($T \approx T_H$). The result may be written in the form

$$d\overline{\sigma}^2 = C\{C_{00}c^2 dT^2 - C_{10}c\,dT\,dr - C_{11}dr^2 - r^2 d\Sigma^2\}, \quad (109)$$

where

$$\begin{aligned}
C &= e^{-\frac{H^2 r^2}{c^2}} &&= e^{-\left(HT\frac{r^*}{R_H}\right)^2} \\
C_{00} &= 1 - \frac{H^2 r^2}{c^2(1+Ht)^2} &&= 1 - \left(\frac{r^*}{R_H}\right)^2 \\
C_{10} &= \frac{2H^2(2+Ht)tr}{c(1+Ht)} &&= -2(1-H^2T^2)\frac{r^*}{R_H} \\
C_{11} &= 1 - \frac{H^2 r^2 (1+Ht)^2}{c^2} &&= 1 - \left(H^2 T^2 \frac{r^*}{R_H}\right)^2
\end{aligned} \quad (110)$$

and $d\Sigma^2 \equiv \sin^2\vartheta\, d\varphi^2 + d\vartheta^2$ as before. In close neighborhood to the coordinate origin ($t=0$, $r=0$), this adapted line element coincides approximately with the integrated form (103), since obviously $t \approx t'$, while $x^\alpha \equiv x'^\alpha$. To show the spatial limitations of this best quasi-proper representation with respect to the universal distances $r^* (\equiv r/HT)$, the assignments in (110) are also given in corresponding expressions there. It is important to see from $C_{00}$ that also the adapted time $T$ as the best quasi-proper coordinate approximation to a local proper time integral $t_{SRT}$ is not suitable to hold at and beyond universal distances $r^* \to R_H$ again. In contrast to strict SRT, however, the line element (110) keeps the non-vanishing Einstein tensor of a universal background of matter and energy.

Here it is possible, briefly to come back to the apparent violation of Galileo's law of inertial motion with respect to universal coordinates as stated in Section 2.3. To this end the universal 4-velocity $u^{*i}$ (31) is transformed to the system S of adapted coordinates yielding $u^i$ as well as the ordinary spatial velocity $v^\alpha \equiv u^\alpha/u^0 \equiv dx^\alpha/dt$, too, what finally results in

$$v = \frac{v^* + Hr^*}{1 + \frac{Hr^* v^*}{c^2} e^{2Ht^*}} \quad (111)$$

for any radial motion along the x-axis (or x*-axis) of appropriately chosen coordinate systems. Now taking the component $v^{*1} \equiv v^*$ from (28) as well as calculating $r^* = \int v^* dt^*$ starting from the origin, a series expansion of relation (111) to second order $O^2(Ht)$ yields

$$v \approx v^*_{(0)} \left\{ 1 - 2H^2 t^2 \frac{v^{*2}_{(0)}}{c^2} \right\}, \qquad (112)$$

where is $v^*_{(0)} \equiv v^*(t^* = 0) = v(t = 0) \equiv v_{(0)}$ of course. In comparison, a direct expansion of (28) would have led to $v^{*\alpha} \approx v^{*\alpha}_{(0)}(1 - Ht^*)$ meaning an approximately constant deceleration $-Hv^{*\alpha}_{(0)}$ with respect to *universal* coordinates. Nevertheless, relation (112) ensures the local validity of Galileo's law of inertial motion in frames of *adapted* coordinates. Therefore, with respect to an atomic clock at rest passed by any test particle, the law of inertia may be kept approximately though only piecewise. This seems another aspect of the self-restoring validity of local SRT again, which has generally been stated in Section 2.2 above.

Similar to the considerations in the context of (107) at the end of 2.10.1, also in the system S of adapted coordinates the corresponding Einstein tensor would rather meet the view of recent cosmology than (35) of Section 2.4. Here, up to corrections of fourth order $O^4(Ht)$, the adapted energy-momentum bi-tensor density $\overline{V}_i^k$ is of a similar form as (107) in the system S' of integrated coordinates before [Ostermann 2003, 2008b]. It is found a non-vanishing total energy density $\overline{V}_0^0$ of matter and gravitation as before. The question, which of the various forms may be the appropriate one, might be answered with: all of them. They are suggesting many locally expanding cosmic complexes of matter again which may be well compatible with an ultra-large scale stationary universe.

*2.11 The Universal Embedding of Local Gravitational Fields*

Imagine an ideal spacecraft started with an adequate acceleration from an ideal single star in extragalactic space, now left to itself with all engines switched off. If the motion of the spacecraft – neglecting all friction losses as well as all other conceivable disturbances – would stay determined by the Schwarzschild solution of GRT, it should reach uniform velocity once the distance from the star will be sufficiently large. But uniform motion, though only in principle here, would apparently contradict the universal deceleration of free particles derived in (28), (31). According to the results of the previous section, however, a natural explanation for this dilemma seems the conclusion that the Schwarzschild solution corresponds to locally adapted coordinates rather then universal ones. Not only this consideration, but primarily astronomical experience within the solar system making use of atomic clocks, are supporting this view.

The SUM line element (17) is ready by structure for a natural extension, trying to include inhomogeneities of the energy-matter distribution by embedding local gravitational fields into the stationary background universe. An ansatz of the form $d\sigma = \zeta d\sigma_{(SRT)}$ had been discussed by [Einstein & Fokker 1914] following an early relativistic theory of gravitation [Nordström 1912, 1913a/b], before an extension to GRT was applied by [Weyl 1918/19] in another context.

Now, correspondingly, taking into account the argumentation above, a Local Embedded (LE) line element may be

$$d\sigma^*_{LE} = \zeta^* d\sigma^*_{GRT}, \qquad (113)$$

where again $\zeta^* = \zeta^*_{SUM}$, and in comparison to (17) above $d\sigma^*_{SRT}$ has to be replaced straightforwardly by the corresponding line element of local GRT, which has to be transferred according to

$$d\sigma^*_{GRT} \equiv d\sigma_{GRT}(t := t^*, r := r^*, [GM] := [GM]^*). \qquad (114)$$

As usual, $d\sigma^2_{GRT} \equiv g_{ik} dx^i dx^k$ may represent a solution of Einstein's vacuum field equations $R_{ik} = 0$ for the space outside ordinary matter, neglecting the background universe. In (114) it seems necessary not only to replace the effectively adapted coordinates $(t, r)$ of local GRT by the universal ones $(t^*, r^*)$ – analogously to (17) – but in addition also the natural *constant G* formally by

$$(GM)^* \equiv (GM) e^{-Ht^*}, \qquad (115)$$

since any Newtonian potential $c^2\Phi$ which appears in the conventional fundamental tensor $g_{ik}^{GRT}$ has to be replaced by $c^2\Phi^*$ according to

$$c^2\Phi \equiv -\frac{GM}{r} \approx -\frac{GM\,e^{-Ht^*}}{r^*} \equiv c^2\Phi^*. \qquad (116)$$

The reason is that given a circular orbit, for example, the variable $r$ in the expression $c^2\Phi = -GM/r$ is actually representing a constant quasi-proper distance with respect to the time scalar $e^{Ht^*}$, and therefore it has to be approximately substituted by $r \approx r^* e^{Ht^*}$ to get only universal coordinates in (114) at last. The other way round, the mathematical consistency of the third replacement in (114) may be verified by transforming the adapted coordinates of the conventional line element of local GRT to universal ones with respect to the inverse transformation relations calculated from (108a), (108b). From (113), the covariant equations of motion take the form

$$\frac{du_i^*}{d\sigma^*} - \frac{1}{2} e^{2Ht^*} u^{*k} u^{*l} \partial_i^* g_{kl}^{*GRT} = \left( \frac{H}{c}, 0, 0, 0 \right). \qquad (117)$$

Once calculated these covariant components $u_i^*$, the contravariant components result from $u^{*k} \equiv g^{*ik} u_i^*$ and finally the actual velocity is found $v^{*\alpha} \equiv dx^{*\alpha}/dt^* \equiv u^{*\alpha}/u^{*0}$. Obviously, this relation (117) corresponds to the special case described by (42).

By inverse transformations according to (108a), (108b) also the universal motion of planets, moons, satellites may be found from the conventional Schwarzschild solution already. With respect to universal coordinates, the orbital radii and periods of planets should temporarily decrease in time. Given circular planetary orbits in contrast to linear motion, the velocity is approximately constant there with respect to both the adapted as well as the universal coordinates.

The application of the embedded line element may also be verified, showing that essentially the same correlations as derived in the context of (112) follow approximately if – here the other way round – the track $X^\alpha(t)$ of a test body is given according to conventional GRT. Then this track is understood a sufficient approximation within the local system S of adapted coordinates. Therefore the corresponding track $X^{*\alpha}(t^*)$ is found with respect to universal coordinates according to (20) by re-substitution of $t$ taken from the first identity in (108a)

$$X^{*\alpha}(t^*) \approx X^{\alpha}(t)\,\mathrm{e}^{-Ht} \quad . \tag{118}$$

It may be worthwhile to consider the special case of uniform motion in an appropriately chosen system S of adapted coordinates once more, where the one-dimensional linear motion is simply described by $X = vt$. Now according to (118) derived twice with respect to $t^*$, the result is a universal deceleration of approximately

$$\dot{v}^*_{\mathrm{SUM}} \approx -Hv\left(1 - \frac{v^2}{c^2}\right)\mathrm{e}^{-Ht^*} \tag{119}$$

except for errors of second order $O^2(Ht^*)$ or higher. Actually, this result holds in addition to possible local accelerations, too, since temporarily any motion may be understood as approximately of constant velocity $v$. Furthermore, relation (119) is perfectly compatible to (28) of Section 2.3 again. In particular, universal deceleration $\dot{v}^*_{\mathrm{SUM}} = 0$ for both, photons as well as bodies at rest. – If necessary, the concept of local embedding according to (113) may be even extended replacing $\zeta^* = \zeta^*_{\mathrm{SUM}}$ by

$$\zeta^* = \mathrm{e}^{H(t^*,\vec{r}^*)\,t^*} \quad , \tag{120}$$

where now $H$ may depend on space and time. According to the stationary homogeneous line element given in (17) the only constraint might be

$$H^2 \equiv \overline{H^2(t^*,\vec{r}^*)} \overset{!}{\equiv} const.\big|_{H\Delta t^*,\,H|\Delta \vec{r}^*|/c\,>\,X}\,, \tag{121}$$

if averaged over sufficiently large universal scales of space and time, what should mean $0.1 < X \leq 1$. In this case the equations (53) and (62) for universal redshift and the distance modulus would need at least a substitution of $H$ by

$$H_{\mathrm{E}} \equiv H\left(t^*_{\mathrm{A}} - \frac{|\vec{r}^*_{\mathrm{E}} - \vec{r}^*_{\mathrm{A}}|}{c},\, \vec{r}^*_{\mathrm{E}}\right), \tag{122}$$

where again $(t^*_{\mathrm{A}},\,\vec{r}^*_{\mathrm{A}})$ would mean time and place of absorption, and 'E' indicates emission or an event. This generalization might allow to take partially into account local cosmic inhomogeneities of the Hubble 'constant', though only neglecting even large peculiarities possibly caused in such scenarios.

An objection in principle against the embedded line element could be that apparently two views seem mixed here which are possibly incompatible within the same approach: a local one (planetary system) with the universal one (homogeneous distribution of matter and energy). In fact, the energy-momentum-stress tensor derived from (113) does no longer fulfill Einstein's condition $E_{ik} = 0$ outside the sources of local gravitational fields, but is approximating the universal distribution according to (35), of course. To illustrate this difficulty, one may consider to enlarge the extent of the embedded gravitational field till it reaches at first that of our whole galaxy. Then, if still enlarged further, finally the mean universal matter density seems represented *twice* in the embedded energy-stress tensor, what cannot be reasonable. The phenomenological energy-momentum-stress tensor, however, does not try to represent the microscopic reality but only its corresponding average densities on macroscopic scales.

In addition, another question arises, too, why embedding should not be carried out into the mean matter density of the Milky Way, for example.

These objections against the embedding of local gravitational fields would be weakened essentially, if there is a large amount of 'dark' mass-energy distributed much more homogeneously than 'visible' baryonic matter. It seems a challenging idea to think about the time scalar of the embedded line element representing the homogeneous background of 'dark' matter and energy in this context, whereas the statistical account of local Newtonian potentials might stand for directly visible or indirectly observable matter only.

On the other hand, given the local embedded line element (113), there might occur peculiar deviations from conventionally expected stellar motion as in particular at the peripheral regions of galaxies, for example, where acceleration sinks down to and below order $Hc$. Therefore in only this view, the embedded line element could not yet be firmly excluded as another possible explanation for various well-known astrophysical phenomena instead of dark matter.

*2.12 Cosmic Evolution in a Stationary Background Universe*

One SUM consequence – leading from a questionable singular 'big bang'-origin of the whole universe to a concept of 'local bangs' – is worth to be considered explicitly again:

The redshift parameter (53) calculated from SUM depends only on the 'coordinate' distance of the radiation source. This distance is uniquely related to the coordinates which in the context of an expansion are called 'comoving'. Thus, even those who argue in favor of an expansion of the universe, do accept that here galaxies or quasi-stellar objects are statistically at rest. Now because of (53), however, by measuring the values of redshift also these universal coordinates are measured unambiguously except for peculiar motions. Therefore, in contrast to the interpretation accepted so far, they have immediate physical relevance. This conclusion requires the enlightenment of a misleading conception, though. In Section 2.9 it has been shown, that the conventional parameter $H_c$ is not the same as the significant redshift constant $H$.

Since the concept of universal proper time and length were shown to break down at $t' \to -T_H$ and $r' \to R_H \equiv cT_H$, these values seem to represent upper limits for macroscopic joined structures. In this view, given a stationary universe, the singularity of its FLRW-form (95), (96) 'only' means that there cannot exist clocks, stars, invariable galaxies or clusters older than $T_H$ with respect to any associated local proper time.

Besides the – now physically relevant – intergalactic 'coordinate' distances, there are also the 'proper' extensions of local objects and standards, which because of a reciprocal dependency on time do come to agree with the universal values temporarily again and again. Thus there is a mathematically derivable conflict between local and universal dimensions. According to SRT on the one hand and GRT on the other hand, both of these scales should retain their dimensions, but this is evidently impossible over long periods of time. Hence, from this perspective the origin of local *physical evolution* becomes recognizable here, based on an eternal struggle of individual cosmic structures against the overall universal distribution.

Considering the natural fact that obviously everything comes into existence and passes away, there is no longer a physical reason for a beginning of space and time. That which was previously interpreted as the age of universe, now becomes – once again by mathematical deduction in Section 2.10 – the maximum life time. Therefore, even in a stationary universe there must have been a beginning of our own evolutionary cosmos, obviously billions of years ago.

This cosmological concept as suggested by a new combination of Einstein's equations of SRT and GRT, seems attractive because: particularly within physics, neither a beginning of the universe as a whole nor any eternal macroscopic joined structures would be plausible options.

The upper limits of proper length and proper time may not be transgressed. These may correspond to lower limits represented by the Schwarzschild radius or even the Planck length and time on the short end of the scales. Just behind the limits of macroscopic applicability there might take place unknown processes leading to a statistical stationary equilibrium of all cosmic events.

Because of the limited lifetime of stars, individual galaxies, or clusters, the stationary nature of the universe, which is concluded here, suggests that new structures will emerge from time to time – if not even every once in a while. In reverse symmetry to the commonly known conflict of living organisms against decline and decay, there would be an interplay between large-scale entropic balance and local gravitational re-creation – though at first sight this seems fundamentally to contradict the principle of entropy. Taking a closer look, however, in Section 5.1 there will be shown that this cannot be excluded on grounds of physical experience.

Local re-creation of that matter and energy which is described by $T_0^{\,0}$ as directly related to the Einstein tensor $E_0^{\,0}$ might possibly occur in 'local-bang' events from the amount of absorbed gravitational radiation described by $t_0^{\,0}$, too, while the sum of both parts stays zero as shown in Section 2.4.

Particularly in view of the universal bi-metric interpretation of GR summarized in Section 2.2, the possibility of 'local bangs' according to the SUM seems clearly favored instead of the one 'big bang' which might have been concluded by overstraining the concept of 'proper' quantities of SRT in the framework of GRT.

In view of the fact that apparently only to a small part the average universal energy density can be associated with ordinary stellar matter, it is no surprise that some of the abovementioned aspects are not yet understood or not even directly observed so far. The concept outlined here is only to show that a stationary universe is not unthinkable, but instead actually suggested by Einstein's equations, after all.

Therefore, in contrast to most of recent attempts, the major Part I of this paper was focused on what (G)RT might be able to tell without additional hypotheses about the universe. The energy-stress tensor of Einstein's equations deals with purely phenomenological densities only. In view of the spatially homogeneous SUM line element (17), this is not necessarily implying any information about its composition at all. On the other hand, however, just the same feature allows the equivalence principle to apply to any freely falling local inertial frames. To gain more deducible information about the otherwise unspecified universal densities of energy and pressure, it will be necessary to combine gravitation with quantum mechanics in more detail, including particle physics as well as electrodynamics or the weak and strong interaction, too.

In view of straight SUM, the observational indications so far to a joint common evolution on ultra-large scales z > 1.7 corresponding to $r^* > R_H$ – as consequently concluded within the CCM framework – might partially be artefacts of selection effects whether these are caused by various frequency-dependent absorption, local peculiar evolution in observer's cosmic neighborhood, or other unknown limitations of (statistical) observability. Altogether, the existence of essential obstacles for an unambiguous identification cannot be firmly excluded given an infinite universe.

*2.13 Some Remarks on the SUM Concept, its Origin and Related Earlier Attempts*

Independent of an actual applicability of the SUM, this model may be formally classified in the context of various approaches to relativistic cosmology. When in 1917 Einstein developed his first approach, he tacitly took for granted an eternal universe according to what in the SST framework has been called the perfect cosmological principle later. This homogeneous and isotropic large scale universe should be completely determined by its average densities of energy and pressure. Unfortunately he was focused on a static solution solely. Then, with [Friedman(n) 1922/24] relativistic cosmology turned to temporal evolving solutions of Einstein's equations, which were supported by [Hubble's 1929] law – actually discovered by [Lemaître 1927] before – after Slipher's early discovery of the redshift of galaxies. With the 1917 cosmological constant discarded, a pressureless flat-space model (EdS) was proposed in [Einstein & de Sitter 1932]. Partially in contrast to previous approaches, [Bondi & Gold 1948] as well as in particular [Hoyle 1948/49] tried to reconcile [Lemaître's 1931a/b/c] 'expanding' universe with the concept of a 'steady state', which model soon after turned out to conflict with observational facts. What might have been misunderstood with Einstein's equations?

The SUM has been developed [Ostermann 2003, 2004, 2008b, 2012a/b], looking for a *stationary* line element of GRT in infinite Euclidean space and in infinite universal time, both mathematical concepts without any physical properties. The stationarity of this model – primarily evident from its unchanging redshift parameters – is supported by the self-restoring validity of SRT within 'local' inertial frames shown above.

The feature, that this stationarity does not mean a 'steady state' but a lively process, finds its expression in that the term 'stationary' does not mean 'static'. In contrast to the latter, the first may be understood to describe an eternal process where – necessarily in an ongoing interplay with quantum mechanics resulting in local gravitational re-creation – each evolutionary cosmos may take a limited life time.

Correspondingly, the SUM line element (17) is not static of course. That in spite of its obvious dependence on time, it is rightly qualified as stationary, however, follows from calculating its most important characteristic features. In particular, it has been shown in the previous sections, that …

a) … the redshift parameters $z = e^{Hl^*/c} - 1$ are independent of time for galaxies statistically at rest in universal coordinates;

b) … all cosmic observables which are pure functions of $z$, are also independent of time;

c) ... all universal distances $l^*$ – statistically measurable by stationary values of $z$ – simply stay unchanged;

d) ... the magnitude-redshift relations for 'standard candles' like type Ia SNe are independent of time;

e) ... because of the exponential form of the time scalar $e^{Ht^*}$ in (17), all relative temporal changes depend solely on differences $\Delta t^* = t_A^* - t_E^*$, what allows to set any reference point of universal time $t_R^* = 0$ for coherent complexes of observation;

f) ... the stationary SUM line element (17) implying a constant universal speed of light corresponds to the simplest of all FLRW-forms (95), (96) without cosmological constant;

g) ... the stationary 'deceleration' parameter is $q(t') \equiv 0$;

h) ... both the covariant EMS tensor $T_{ik}$ as well as its contravariant density $\mathbf{T}^{ik}$ are constant, what – taken together with galaxies statistically staying where they are – coincides with a conservation of universal mass-energy as shown above;

i) ... in addition to $c$ and $G$ – completed by the microscopic constants $e$ and $h$ – the law of universal redshift includes $H_{\text{s-SUM}} \equiv H$ as a significant Hubble *constant*, which seems determined by the claim that the Schwarzschild radius $2GM_H/c^2$ of the 'Hubble mass' $\rho_c \cdot 4/3\pi R_H^3$ should equal the 'Hubble radius' $R_H \equiv c/H$ most naturally. The other way round, this claim may be understood as a determination of the gravitational constant from $H$ and $\rho_c$, the latter as density necessary for a flat-space background universe.

Regarding earlier attempts, as for example various versions of what has been called 'Steady-state Theory' (SST) or a 'Coasting Cosmology' [Kolb 1989] there are essential differences, some of which may be summarized here.

At first sight, it appeared that any new stationary approach must fail since the SST had turned out to conflict with observational facts, see e.g. [Weinberg 1972]. In spite of its reasonable intention – which according to earlier concepts has been concentrated to a 'perfect cosmological principle' – that theory is not stationary at all. For example, its redshift parameters as the fundamental cosmological observables are $z = e^{H\Delta t'} - 1$ with $\Delta t'$ the light time, which due to the time-dependent SST coordinate velocity $c'$, however, are not constant at all. Correspondingly a simple calculation yields $z = H/c \cdot r^* e^{Ht'}$ with $t'$ the 'cosmic' time of a respective redshift measurement. Furthermore, this reveals an immanent contradiction concerning the non-SUM framework to the presupposition of galaxies at rest with respect to constant universal ('comoving') coordinates $r^*$, since the first expression for $z$ above depends only on an interval $\Delta t'$, while the second one would depend only on a special point $t'$ of time.

In contrast to the SUM, the SST kept on claiming proper distances together with proper time as valid measures for arbitrary intervals of universal space and time, what has been disproved in Section 2.10 above. This claim is already clear from the titles *"The Steady-State Theory of the Expanding Universe"* [Bondi & Gold 1948] and *"A New Model for the Expanding Universe"* [Hoyle 1948/49] of their papers.

Concerning the CMB there has been discussed an origin from stellar radiation thermalized by iron whiskers. It would be impossible, however, to keep a Planck spectrum of redshifted pure BB radiation coming from cosmic distances except for the inapplicable case of an almost complete universal opacity in the corresponding frequency range.

There are other contradictions to the SUM as for example regarding a SST horizon for light signals in the future. According to that model this horizon would equal the Hubble length $c/H$, while according to the SUM there is no horizon concerning the universe as a whole. If in the table of Section 6 below, the CCM had been compared with the SST instead of the SUM, the picture would have looked completely different in many points.

Quite naturally, however, there are not only differences but according to clearly related aims also several things in common of the SUM and the SST including the later Quasi-Steady-State Cosmology (QSSC) respectively. In particular, the SST approach has lead to exceptional achievements like understanding the synthesis of heavy elements in stars [Hoyle, Burbidge, & Narlikar 2000]. In one of both SST papers quoted above, Hoyle introduced a universal scalar field into the framework of GRT, thus anticipating the concept of cosmic 'inflation'. Some features of the QSSC may possibly prove applicable though with essential modifications in the SUM. For example, 'creation centers' of the QSSC might correspond to 're-creation centers' there.

Instead of the C-field of the SST, however, in the SUM framework there might be rather an equivalent in the energy contribution of gravitational radiation given by $t_i^k$ converted to that ordinary energy-stress-momentum which is described by the tensor $T_i^k$ or the phenomenological perfect-fluid tensor $p_i^k$ respectively.

Since early times of relativistic cosmology, the FLRW form is dominating corresponding line elements. There, unfortunately, its time coordinate $t'$ is taken as 'cosmic proper time' where the word cosmic would mean universal, while the SUM concept of proper time – as shown in Section 2.10 – doesn't apply without limitations.

A more general FLRW form than (95) including spatial curvature, once has been named 'coasting' cosmology [Kolb 1989], before a closely related concept has been discussed several years later in [Melia & Shevchuk 2012, and references therein], both fundamentally different from the SUM.

If that line element is specialized to flat space, it takes the FLRW-form of the SUM. In view of an assumed coasting expansion of the entire universe fixed to the old concept of universal 'proper' length and 'proper' time, however, most stationary features of the SUM shown in the previous sections remained unrevealed. The central consequence of redshift parameters independent of time, for example, has not been stated there.

Taken together, how could the clear stationarity of the SUM line element (17) happen to escape its discovery even in those times, when the SST attempt was developed and then widely discussed? Three main reasons may be:

α) The coordinate time $t'$ of the FLRW form has been misunderstood as a universal proper time whereas according to the SUM this concept of proper time applies 'locally' only as stated above.

β) A negative gravitational 'dark' pressure $p^* = -1/3 \cdot \varepsilon_c^*$ of one third the critical energy density has not been considered a physical option, what only happened after the breakthrough of the SNe-Ia observations. Now it is shown to be a plausible feature in the SUM framework (s. Sect. 2.4). In view of Kolb's 'coasting cosmology' – mathematically but scarcely physically related – this negative pressure was interpreted as a property of a 'K-matter' instead.

γ) The stationarity of SUM's statistically unchanging redshifts (53) might have been concealed by the rather mislead-ing conventional Hubble parameter $H_{c\text{-SUM}}(T') \equiv (da/dT')/a$ which equals $1/T'$ in case of the SUM. Unfortunately this parameter is indicating a dependence on time where actually no such dependence exists. Instead, its significant Hubble constant is $H_{s\text{-SUM}} \equiv H = da/dT'$, as shown in Section 2.9. This conclusion is evident from the unquestioned presupposition that the universal sources of radiation are statistically at rest with respect to the universal ('comoving') distances $l^*$, but definitely not with respect to corresponding proper distances $l'$. Therefore it is a wide spread mistake assuming the conventional Hubble parameter $H_c$ to be the basic observable of redshift. Though this conventional Hubble parameter would be a constant of SST, for example, its redshift is not.

According to the SUM, the 'deceleration' parameter $q \equiv -a \, (d^2a/dT'^2)/(da/dT')^2$ is naturally $q_{\text{SUM}} \equiv 0$ which value has been interpreted according to that 'coasting' expansion, though without the postulate of spatial flatness equivalent to a constant universal speed of light $c^* = c$. In addition, there is also missing the universal line element (17) corresponding to a stationary embedding of SRT, or other essential features of the approach presented here.

The first development of the SUM was given in an e-print [Ostermann 2003], while both the 'Coasting Cosmology' [Kolb 1989] or related later attempts, as well as the earlier 'chaotic inflation' approach of [Linde 1983], [Linde, Linde, & Mezhlumian 1994], [Mezhlumian 1993/94], [Linde & Mezhlumian 1993] has been unknown to author at that time. The reason may be, that both an expanding space as well as a concept of many totally separated 'parallel universes' basically contradict essential presuppositions of the SUM. A fundamental line element of GRT like (17) to describe *one* coherent background universe is missing there. Nevertheless, it cannot be firmly excluded that an attempt to 'embed' the evolutionary CCM cosmos into the stationary universe will bring different approaches together in the end.

The fact, that the stationary solution SUM was found only after the SNe-Ia data had been published but without knowledge of them, may be why it remained nearly unnoticed so far. Otherwise these data might have been regarded to confirm the straight SUM prediction on universal scales, whereas an appropriate amount of 'dark energy' corresponding to a cosmological constant has been established in the meantime instead. On the other hand, without the invaluable SNe-Ia measurements, the SUM concept would have not been developed to an arguable level after all.

## II. THE COMPARATIVE PART

At least in parts, the well-known pillars of relativistic cosmology referred to in the introduction might also support the new model as will be shown now.

In contrast to quite a few attempts to explain the SNe-Ia data afterwards, the SUM as a deductive model is based on two simple postulates only. Therefore its magnitude-redshift relation is rather a 'prediction', though it has been derived some years after the observational breakthrough achieved on the one hand by the *Supernova Cosmology Project* (SCP) from [Perlmutter et al. 1999] to [Kowalski et al. 2008] with references therein, as well as on the other hand by the *High-z Supernova Search Team* (HZT) from [Riess et al. 1998] to [Riess et al. 2004/07] and references therein.

These SNe-Ia data – in addition to those of the Hubble Space Telescope (HST) Key Project [Freedman et al. 2001], the Type Ia Supernova HST Calibration Program (STS) of [Sandage et al. 2006], and the Sloan Digital Sky Survey (SDSS) of e.g. [Abazajian et al. 2009], [Kessler et al. 2009], [Schneider et al. 2010] and references therein are commonly seen in the context of the Cosmic Background Explorer (COBE) results [Mather et al. 1990] and those of the Wilkinson Microwave Anisotropy Probe (WMAP) as from [Bennett et al. 2003], too. In particular in case of the latter it seems nearly impossible to work out high precision cosmology without fundamental priors including essentially unknown physics. The CMB anisotropies are only determined after subtraction of some 'unsuitable' microwave radiation as small part of the CIB, s. [Kashlinsky 2005], [Ade et al. 2011]. – At first, however, it may be appropriate to recall briefly some essentials of the CCM for comparison in the new context.

## 3. A HEURISTIC APPROACH TO THE COSMOLOGICAL CONCORDANCE MODEL

In the hypothetical case of a strictly isotropic and homogeneous distribution of matter and energy without local inhomogeneities, the stationarity of a background universe according to the SUM would be mathematically perfect, because without macroscopic local structures, there would be no possibility to detect any change in such a universe. Of course the physical reality looks quite different.

What astronomers really see, is an inhomogeneous space seeded with stars, galaxies, clusters, filaments and voids apparently showing a *phenomenological* pressure $p_M \approx 0$. Since cosmology has to take into account the actual composition of matter and energy, it is reasonable to try and outline a more realistic picture of 'our' evolutionary cosmos, though such an approach should not impair stationarity on sufficiently large scales of space and time.

The negative gravitational pressure (39) with its pressure-parameter $w_M = -1/3$ found in Section 2.4 seems completely 'dark' in that it is not directly detectable so far. This difficulty would apply to a homogenous background density, too. This negative gravitational pressure $\bar{p}^* \equiv \bar{p}^*_{\text{background}}$ might correspond to some aspect of 'dark energy'. In the CCM framework it is related to a 'cosmological constant'. The analogy to the negative pressure $\bar{p}^*$ is quite obvious from (36).

It is necessary to provide some CCM relations to compare the SUM with current cosmology in the following sections.

### 3.1 Modeling Different Homogenous Densities of Energy, Matter, and Radiation by One Scale Factor

In view of the CCM framework, the reciprocal proportionality $\Theta_{\text{radiation}} \sim 1/a$ between temperature and FLRW scale factor – which is needed here to have kept the CMB a black body radiation over cosmic times $T' \approx T_{H_0}$ – would imply $\varepsilon_{\text{radiation}} \sim 1/a^4$. This means that the CCM scale factor should yield an energy-stress tensor including at least one part for radiation and another part for matter. In addition, with regard to a flat-space scenario, the curvature parameter had to vanish in the Friedmann solutions. Taking into account the observed densities of matter and radiation, however, this condition seems impossible to fulfill, unless one is ready to accept Einstein's extended equations, here in the form

$$E_i^{k\,(\Lambda)} \equiv E_i^k - \Lambda\delta_i^k = \kappa T_i^{k\,(\Lambda)}. \quad (123)$$

Since Einstein has found it an option to subtract an *ad-hoc* cosmological term $\Lambda g_{ik}$ from the left hand side of his equations, there is the chance to account for the negative gravitational pressure $\bar{p}^*$ (39) heuristically. Based on a flat-space FLRW form according to (92) – with all asterisks and bars suppressed in this section – the original Einstein tensor takes the familiar form

$$E_i^k \equiv \begin{pmatrix} \frac{3\dot{a}^2}{c^2 a^2} & 0 & 0 & 0 \\ 0 & \frac{(\dot{a}^2 + 2a\ddot{a})}{c^2 a^2} & 0 & 0 \\ 0 & 0 & \frac{(\dot{a}^2 + 2a\ddot{a})}{c^2 a^2} & 0 \\ 0 & 0 & 0 & \frac{(\dot{a}^2 + 2a\ddot{a})}{c^2 a^2} \end{pmatrix}, \quad (124)$$

while the energy-stress tensor is substituted by

$$T_i^{k\,(\Lambda)} = P_i^k, \quad (125)$$

where $P_i^k$ is given by (36) without asterisks and bars again, what for $u^\alpha = 0$ means as usual matter at rest with respect to 'comoving' universal coordinates. This implies $P_0^{\,0} \equiv \varepsilon \equiv \mu_0 c^2 - p$ in the following, what according to (123) means

$$\frac{3\dot{a}^2}{c^2 a^2} = \kappa\varepsilon + \Lambda \quad (126)$$

with $\Lambda$ the cosmological constant and $\varepsilon$ the time-dependent energy density of matter with radiation. The integration of (126) yields

$$a = C e^{\frac{c}{\sqrt{3}}\int\sqrt{\kappa\varepsilon + \Lambda}\,dt'}, \quad (127)$$

and inserted into (124) it follows

$$\bar{E}_i^k \equiv \begin{pmatrix} \kappa\varepsilon + \Lambda & 0 & 0 & 0 \\ 0 & X & 0 & 0 \\ 0 & 0 & X & 0 \\ 0 & 0 & 0 & X \end{pmatrix}. \quad (128)$$

with here $X$ temporarily

$$X \equiv \kappa\varepsilon + \Lambda + \frac{\kappa\dot{\varepsilon}}{c\sqrt{3}\sqrt{\kappa\varepsilon + \Lambda}}.$$

Taking into account that the phenomenological density observed seems to be composed of pressure-free matter together with radiation it is set

$$\varepsilon = \varepsilon_c \left(\frac{\Omega_M}{a^3} + \frac{\Omega_R}{a^4}\right), \quad (129)$$

where obviously the parameters $\Omega_M$ of matter and $\Omega_R$ of radiation mean respective fractions with regard to the critical energy density $\varepsilon_c \equiv \rho_c c^2$ today. Inserting (129) into relation (127) yields

$$\dot{a} = a H_0 \sqrt{\frac{\Omega_M}{a^3} + \frac{\Omega_R}{a^4} + \Omega_\Lambda} \quad (130)$$

with $\Omega_\Lambda = \rho_\Lambda/\rho_c$ defined analogously to the parameters $\Omega_M$ and $\Omega_R$ above, though with only a formal density $\rho_\Lambda = \Lambda/\kappa c^2$ of 'dark energy' here. With regard to (126) as a consequence of (123), the relations $\rho_c = \rho_M + \rho_R + \rho_\Lambda$ or $\Omega_M + \Omega_R + \Omega_\Lambda = 1$ mean a spatially Euclidean FLRW-form used in Einstein's extended equations, where obviously $\Lambda$ may compensate the effects of some negative pressure like $\bar{p}^*$ of (36) temporarily. In view of the SUM, it seems that a corresponding amount of pressure and energy is included into the cosmological constant $\Lambda$ to keep it outside the assume pressure $p_M \approx 0$. Thus, the remarkable feature that the local density $\mu_M$ is different from the component $E_0^{\,0}/\kappa$ resulting here, where the first is an additive part of the latter, may be another indication to the overall ambiguity of the 'rest mass' concept mentioned in Section 2.4.

To compare the CCM approach with the SNe-Ia measurements in the following Section 4 directly, the CCM distance modulus $m_{CCM} - M$ is derived making use of (130) now. Instead of the corresponding SUM expression (56), here the apparent luminosity given $\kappa = 0$ may be written in the form

$$I = \frac{L}{4\pi r_U^2} \cdot \frac{1}{(1+z)^2} \quad (131)$$

to calculate the radial coordinate distance $r_U(z)$ as a function of redshift, at first. From the identity of (98) in its normalized form $z = 1/a - 1$, with $a(t'_A = 0) = 1$ and $a(t'_E = t') = a$, it follows by differentiation with respect to $t'$

$$dt' = -\frac{a^2}{\dot{a}} dz. \quad (132)$$

Inserting this into

$$dr = \frac{c}{a} dt' \quad (133)$$

and making use of (130) ass well as of the normalized form $a = 1/(1+z)$ again, this yields

$$r_U = R_{H_0} \int_0^z \frac{dz'}{\sqrt{\Omega_M (1+z')^3 + \Omega_R (1+z')^4 + \Omega_\Lambda}}. \quad (134)$$

Since here – in analogy to (60) – the luminosity distance is

$$d_L \equiv \sqrt{\frac{L}{4\pi I}} = r_U (1+z), \quad (135)$$

now according to the general expression (59) the CCM distance modulus results in

$$m - M = 5 \log\left[(1+z)\frac{r_U}{R_{H_0}}\right] + 25 + 5 \log\left(\frac{c/H_0}{\text{Mpc}}\right) \quad (136)$$

with $r_U$ given by (134). This expression for the distance as a function of redshift given a spatially Euclidean model, though only a special case of the general one reported in [Perlmutter & Schmidt 2003] with references therein, for example, is completely sufficient here. The corresponding modulus (136) has been compared – among others – with the SNe-Ia measurements directly (s. Sect. 4).

It seems to be widely forgotten today, that the cosmological constant $\Lambda$ may be understood in two different ways. In view of the CCM it is interpreted to represent some enigmatic 'dark energy' contributing to a tensor $T^\Lambda_{ik}$, whereas the other way round, according to Einstein, such a constant may just mean only a (problematic) completion to an extended differential tensor $E^\Lambda_{ik} \equiv R_{ik} - (\tfrac{1}{2}R + \Lambda)g_{ik}$ which equals the real densities $\kappa T_{ik}$ on the right hand side of his equations without the $\Lambda$-term there. Since, however, physical results cannot depend on how essentially the same equations are read differently, now asking the nature of CCM's 'dark energy' might be a useless question.

### 3.2 SUM 'Boundary' Conditions Match the CCM Density Parameter $\Omega_\Lambda$

Given spatial flatness, assuming isotropy and homogeneity as well, all cosmological conclusions due to conventional GRT have to be drawn from the FLRW scale factor $a(t')$. Therefore the scale factor $a_\mathrm{SUM}$ may be compared with that of the concordance model $a_\mathrm{CCM}$ directly, before going on with the magnitude-redshift relations in the next section.

Radiation will be neglected here, what because of its density parameter $\Omega_R \ll 1$, in fact some orders of magnitude smaller than $\Omega_M$ or $\Omega_\Lambda$, proves legitimate in this context. Therefore according to a phenomenological pressure of matter $p_M \approx 0$ today and setting $\Omega_R = 0$, Einstein's extended equations yield the effective CCM scale factor for a spatially Euclidean model from (127), (129) by direct integration

$$a_\mathrm{CCM}(t') = \left[\left(\frac{1}{\Omega_\Lambda} - 1\right)\sinh^2 X\right]^{1/3} \qquad (137)$$

with here $X$ temporarily

$$X = \tfrac{1}{2}\ln\left(\frac{1-\sqrt{\Omega_\Lambda}}{1+\sqrt{\Omega_\Lambda}}\right) - \tfrac{3}{2}\sqrt{\Omega_\Lambda}\,H_0 t'.$$

Both relations (137) taken together are a valid approximation for the following. It is this reasonably constrained CCM scale factor, where in particular the magnitude-redshift relations obeying $\Omega_M + \Omega_\Lambda = 1$ may be readily calculated from. Even taking the CMB radiation density yet into account, this does not result in visible changes of the solid blue CCM-line in Figure 1, the latter already discussed in [Ostermann 2004].

There the actual CCM scale factor $a_\mathrm{CCM}(t')$, is compared with two other thinkable assignments corresponding to toy values $\Omega_\Lambda = 0.400$ and $0.900$, as well as with the SST and EdS scale factors, too. In contrast to the broken blue lines representing those alternative values of $\Omega_\Lambda$, this solid blue line shows the CCM scale factor according to its density parameters reported in [Bennett et al. 2003], for example. Of all thinkable assignments to $\Omega_\Lambda$ – in view of the CCM otherwise purely coincidental – it is evidently only one to match the SUM scale-factor zero just at $t' = -1/H$. According to (94), however, this special point of FLRW time does mean nothing but infinite past with respect to universal time $t^* = -\infty$.

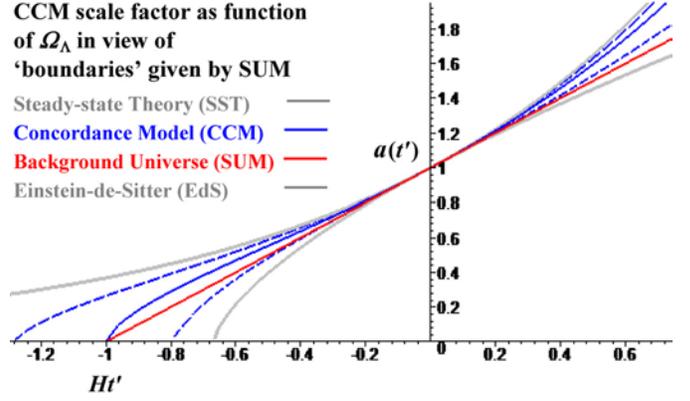

FIG. 1. – Top-down on the left: $(\Omega_M, w_M, \Omega_\Lambda) = (0, 0, 1)$, $(0.1, 0, 0.9)$, $(0.27, 0, 0.73)$, $(1, -\tfrac{1}{3}, 0)$, $(0.6, 0, 0.4)$, $(1, 0, 0)$, i.e.: Steady-state Theory $a_\mathrm{SST}(t') = e^{Ht'}$ [upper grey solid line, this model discussed as a possible option in the past], a first alternative to $a_\mathrm{CCM}(t')$ with higher value of $\Omega_\Lambda$ [blue broken line], today's concordance model $a_\mathrm{CCM}(t')$ [blue solid line, see (137)], stationary ultra-large scale universe $a_\mathrm{SUM}(t') = HT' = 1 + Ht'$ [red straight line, s. (95), (96)], a second alternative to $a_\mathrm{CCM}(t')$ with lower value of $\Omega_\Lambda$ [lower blue broken line], Einstein-de-Sitter model $a_\mathrm{EdS}(t') = (1 + \tfrac{3}{2} Ht')^{2/3}$ [lower grey solid line, favored before the SNe-Ia observational breakthrough]. In contrast to (all) other values (blue broken lines), the CCM best-fit parameter $\Omega_\Lambda = 0.737$ (blue solid line) seems determined by the condition that it should meet the SUM scale factor (red straight line) at its 'boundaries', i.e. at its beginning $Ht' = -1$ and at $Ht' = 0$ today.

Therefore, in view of the embedding background universe, it would be natural to claim the singularity of $a_\mathrm{CCM}(t')$ to be the same as the 'local' pseudo-singularity in the FLRW-form of the stationary solution (95), (96), i.e. $a_\mathrm{CCM}(t' = -T_H) \stackrel{!}{=} 0$ (s. Figure 1). This condition, equivalent to $a_\mathrm{CCM}(t^* = -\infty) = 0$, verifies again: with respect to universal coordinates, even in the CCM there is no universal singularity at all. From this self-explanatory claim, which simply corresponds to $T_0 \stackrel{!}{=} 1/H_0$ today, the numerical solution of (137) is

$$\Omega_\Lambda = 0.737, \quad \Omega_M = 0.263, \qquad (138)$$

thus almost perfectly matching the highly consistent CCM density parameters for 'dark energy' ($\Omega_\Lambda = 0.73 \pm 0.04$ reported in [Bennett et al. 2003]) and matter $\Omega_M = 1 - \Omega_\Lambda$, presupposed a spatially flat model according to the SUM. This aspect is pointed out again in e.g. [Melia & Shevchuk 2012].

In addition, that reason why it is *natural* to claim the singularity of $a_\mathrm{CCM}$ to be the same as that of $a_\mathrm{SUM}$, is illustrated by the following consideration, which may be legitimate in the context of the CCM: If a 'big bang' cosmos had started at Planck time for example (i.e. approximately at $Ht' = -1$) as a local 'quantum fluctuation' in a stationary universe described by (95), (96), then at the beginning of that process both scale factors $a_\mathrm{SUM}$ and $a_\mathrm{CCM}$ should have been the same, i.e. approximately zero. This, however, could only have matched if $\Omega_\Lambda$ had the 'right' value 0.737 today.

Therefore, obviously, the CCM scale factor coincides at its boundaries with that of a stationary ultra-large scale background universe as described by the SUM and originally deduced in (17) from pure (G)RT without any additional hypotheses.

Taken isolated, however, the CCM raises the well-known questions as among others: Why is the matter density $\rho_M$ of same order as that of 'dark energy' $\rho_\Lambda$ just in our time? Is there any physical reason why the 'age of the universe' $T_0$ should be extremely near the Hubble time $T_{H0}$ except from that one stated here? Apparently the CCM might be something like an ΛCDM approximation to the SUM.

Based on (137), the distance modulus $m_{CCM} - M$ results from (136) replacing $r_U$ by

$$r_{CCM} = R_{H_0} \int_0^z \frac{dz'}{\sqrt{(1-\Omega_\Lambda)(1+z')^3 + \Omega_\Lambda}}. \quad (139)$$

This corresponds to (134) taken $\Omega_R = 0$, of course, while for the stationary solution the FLRW scale factor $a_{SUM}$ and the distance modulus $m_{SUM} - M$ are given by (96) and (62) in Sections 2.9 and 2.6 above.

Accordingly for comparison, also the magnitude-redshift relations for the EdS model and the SST are readily calculated from their scale parameters

$$a_{EdS} = (1 + \tfrac{3}{2} H_0 t')^{2/3}, \quad (140)$$

$$a_{SST} = e^{H_{(0)} t'}, \quad (141)$$

yielding the distance moduli from relation (136) again replacing $r_U$ respectively by

$$r_{EdS} = 2\left(1 - \frac{1}{\sqrt{1+z}}\right) R_{H_0}, \quad (142)$$

$$r_{SST} = z R_{H_{(0)}}. \quad (143)$$

which values result as usual by integration (for a procedure see e.g. [Weinberg 1972] again).

Even though together with several SUM features otherwise described by a hypothetical inflation scenario, the heuristic deduction of the decisive density parameter $\Omega_\Lambda$ given in this section does not yet prove the CCM cosmos completely to fit into an embedding SUM background, it obviously indicates a close relationship, after all.

## 4. THE SUPERNOVAE IA DATA IN VIEW OF THE SUM

With the deductive approach taken in the main Section 2 above, Einstein's equations of (G)RT were found ready to outline the picture of a stationary background universe, uniquely determined on sufficiently large scales by some fundamental assumptions as few and as simple as possible. It is consequently indicated to test that SUM, looking for 'local' deviations from the universal requirements of stationarity, homogeneity, and isotropy now.

The SNe-Ia magnitude-redshift data gained by the HZT in [Riess et al. 2004/07] with references therein, and by the SCP in [Kowalski et al. 2008] with references therein, are probably those cosmological data whose evaluation is least contaminated by theory. They might represent the most valuable cosmological breakthrough of the last decade, because their confrontation with competing theories requires no preconception of unproven hypotheses about the universe. As a consequence, all relativistic models dominant before 1998 have been upset.

Instead, the completely unexpected results led to the CCM which is now fitting those SNe-Ia data numerically well. Today they are commonly understood to provide 'evidence' for a universal acceleration driven by 'dark energy'. Facing the SNe-Ia dilemma of theoretical cosmology at that time, this may have been a consequent conclusion within the development of pure FLRW-cosmology.

On the other hand, the SUM tries to describes the universal background on ultra-large scales, possibly embedding an evolutionary CCM cosmos therein. Taking into consideration its magnitude-redshift prediction, however, there appears a puzzling indication from the SNe-Ia data, in that a straight SUM turns out to represent those data in the high redshift range $z > 0.10$ without any ad-hoc hypotheses right away. In the following sections this model – highly adaptable if necessary – is represented in figures by red lines. Despite of their simple origin they mark a challenging alternative to the blue lines of the CCM there.

Only in the low redshift range $0.01 < z \leq 0.10$ the straight SUM luminosity predictions differ from those of today's CCM significantly. In view of the stationary model this local deviation may indicate e.g. a small Hubble contrast due to an inhomogeneity of the average density, or a corresponding peculiar flow in our cosmic neighborhood. There could also be a faint dimming of light due to 'gray dust', and perhaps a combination of some effects like the ones mentioned here.

Not only the existence of large inhomogeneities up to the Sloan Great Wall has been observed, but [Jha, Riess, & Kirshner 2007] reported an effective Hubble contrast of just the right order to possibly explain the local magnitude-redshift deviations. These appear only on basis of the new stationary solution SUM, however, which has been left unnoticed there.

Remarkably, such a local Hubble contrast – as for example according to $H_{universal} = 65$ km/s/Mpc with $H_{local} = 71$ km/s/Mpc – would just reflect the well-known uncertainty of about 10% from different values of the Hubble constant [Freedman et al. 2001], [Sandage et al. 2006] known today.

The data used in this Section 4 are primarily taken from the [Riess et al. 2004/07] SNe-Ia compilation (s. table in supplementary material) where for the first time – besides their own measurements – Riess and his coauthors have consistently reassessed the data by refitting the included light-curves with a single method. In addition for comparison, some figures below show the residuals obtained from SCP's Union Supernova Compilation [Kowalski et al. 2008] as an adequate confirmation (data in a table of supplementary material).

### 4.1 A First Comparison with the Riess et al. 'Gold' Sample, the CCM and its 'Parents' EdS and SST

From the 255 datasets of the full compilation including 47 +38 measurements from [Astier et al. 2006], the 'gold' sample of [Riess et al. 2004/07] consists of datasets for 187 ground-discovered plus 30 HST-discovered SNe-Ia. Like in the original papers, all datapoints of ground-discovered SNe Ia are plotted as black diamonds, whereas the data points of HST-discovered SNe-Ia are shown as red filled circles. The figures also include 11 HST-'silver' data for illustration again. The residuals of the SCP data [Kowalski et al. 2008] shown in the d), e) panels below will be handled analogously with color magenta instead of red there.

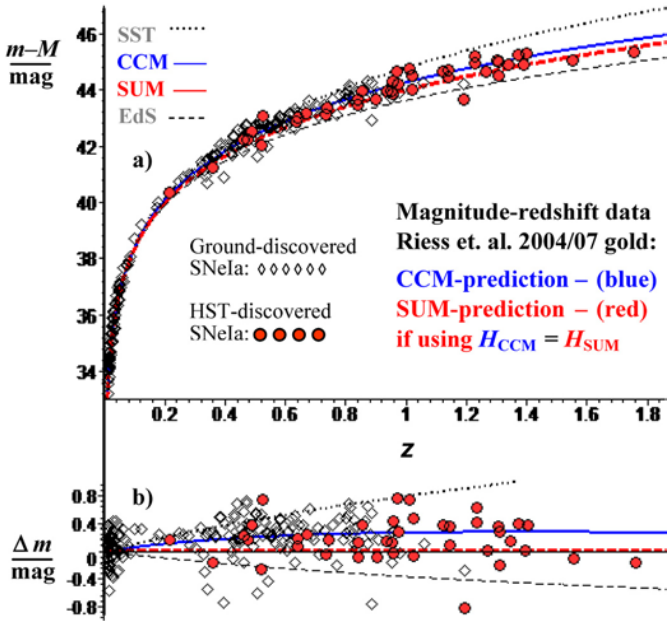

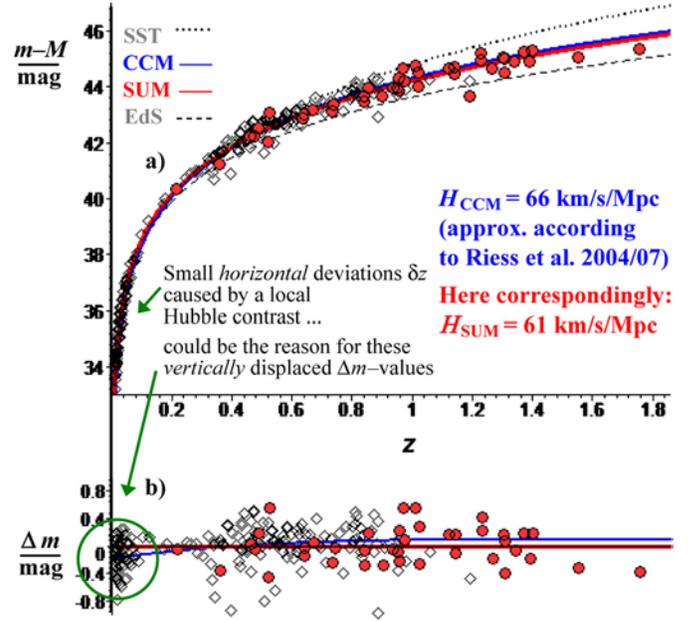

FIG. 2. – *Top panel (a):* The SNe-Ia data taken from [Riess 2004/07] are compared to the distance moduli $m-M$ of various models. Temporarily using the same parameter $H_0 = H$ for all models at first, the SUM magnitude-redshift prediction is naively compared (red broken line) to the CCM-prediction (blue line) which stands for the best fit representing a flat space model with $\Omega_\Lambda = 0.73$. In addition to the CCM there are also shown its 'parents' SST, EdS (grey broken lines above and beneath) and 11 HST-'silver' data for illustration. The red, blue and grey lines represent the predictions as derived from the scale factors $a_{SUM}$, $a_{CCM}$, $a_{SST}$, and $a_{EdS}$ as shown in Figure 1 and given above. According to the High-z Supernova Search Team papers of [Riess et al. 2004/07], the ground-discovered SNe Ia of their 'gold' sample are plotted as black diamonds whereas the HST discovered SNe Ia are represented by red filled circles. *Bottom panel (b):* The magnitude-redshift residuals and the CCM prediction are shown both with respect to the SUM prediction (neglecting any local peculiarities or dimming by gray dust). Since the blue CCM-line is a best-fit of the data and their $\Delta m$-residuals, the lower panel seemingly shows an unacceptable deviation from the red SUM-line here. This may be why such a model has not been taken seriously so far.

FIG. 3. – *Top panel (a):* A vertical shift of $\Delta m = 0.2$ mag is sufficient to remove all visible differences between the red SUM-line and the blue CCM-line here. This vertical shift means nothing but a reduction of about 8.3 % in the Hubble constant (if for example $H_{CCM} = 71$ km/s/Mpc then $H_{SUM} = 65$ km/s/Mpc). But now there are some hidden differences which come to light by plotting the new residuals. *Bottom panel (b):* Though this panel still shows significant deviations between the CCM- and the SUM-residuals, the picture has changed essentially, because now the remaining problem is only a local one concerning the low redshift-range $z \leq 0.10$, whereas CCM and SUM both describe the observed universal SNe-Ia-range $0.10 < z < 1.8$ comparably well (the SUM fits even slightly better than the CCM here).

The magnitude-redshift predictions (139) of the CCM with its best-fitted parameter $\Omega_\Lambda = 0.73$ for flat space are plotted blue, as well as those of the SUM (62) with $\kappa = 0$ as are plotted red. In the lower panels of some figures, showing residuals, the same colors are used for straight broken lines of least quadratic deviations. In the best case these lines should prove congruent with the z-axis respectively.

Figure 2.a shows the SUM prediction together with that of the CCM, which claims this diagram to prove an accelerated expansion of the universe.

In addition, two flat space models once prominent in the history of relativistic cosmology – the Steady-state Theory (SST) completely determined by a cosmological constant, and the Einstein-de-Sitter (EdS) model without such a term – are represented by grey broken lines. Obviously both the SST (upper grey line) as well as the EdS model (lower grey line) are disproved by the data. Nevertheless, they are still important as 'parents' of the CCM today, since the actual SNe-Ia data seemed to require a 'strange recipe'. Mixing about 3/4 of the old SST to about 1/4 of the EdS cosmology led to today's CCM whose blue solid line in the middle is certainly a very close 'best fit' to the measured values which evidently lie between the two mentioned before.

Assuming the same Hubble constant for both models over the full redshift range, e.g. $H_0 = H = 65$ km/s/Mpc, will prove an inappropriate approach. Nevertheless the red broken lines in Figure 2 show, that even straight off the SUM-prediction seems much less incompatible to the data than those of EdS or SST, though there are not yet considered any possible effects of our peculiar cosmic environment or of a weak dimming by 'gray dust', for example.

Figure 2.b shows the $\Delta m$-residuals of the SNe-Ia data themselves as well as those of the CCM, SST and EdS relative to the SUM prediction. It shows a global deviation of this SUM-line from the data. That may be the reason, why such a model has not been taken seriously so far. The upper panel of Figure 2, however, strongly suggests a small vertical shift to the blue CCM-line immediately.

Still neglecting all 'local' cosmic peculiarities, but based on two alternative Hubble constants $H_{SUM} = 61$ km/s/Mpc and $H_{CCM} = 66$ km/s/Mpc, however, the top panel (a) of Figure 3 shows the SUM prediction more suitably now. Since the absolute magnitude $M$ of the SNe-Ia 'standard candles' is not known precisely, the absolute values for the respective Hub-ble constants above are chosen to allow an almost straightforward comparison with corresponding figures of [Riess et al. 2004/07].

Though looking different, Figure 3 is physically equivalent to Figure 2. According to the new assignment of the universal Hubble constant, the SUM line is vertically shifted by $\Delta m = 0.2$ mag, what according to (62) with $\kappa = 0$ and in view of the actual range of measured values means a reduction of the CCM Hubble constant by about 5-6 km/s/Mpc. More realistic values than those above may be $H_{CCM} = 71$ km/s/Mpc and $H_{SUM} = 65$ km/s/Mpc, while only the relative difference $\Delta H/H \approx 9\%$ is relevant there.

In Figure 3.a, obviously the small vertical shift has been sufficient, to remove all visible differences between the blue and the red solid line. The predictions of both models seem to coincide almost completely now. Only when analyzed in detail – plotting the residuals with respect to the SUM – a difference appears in at the left of the vertical black broken line the lower panel 3.b.

In spite of these still significant deviations, however, the picture has changed essentially because the remaining problem is only a local one concerning the low redshift-range $z \leq 0.1$. In contrast, here the SUM covers the data in the universal range $z > 0.1$ comparably well, while the CCM still applies over the full range available so far. In view of the SUM, this strongly suggests that the deviations from its prediction may be caused by local effects in the range $0.01 < z \leq 0.1$.

*4.2 Straight SUM Accordance with 'The World's Supernova Distance-Redshift Data' on Scales $z > 0.1$*

Now taking into account the possible assignments for the Hubble constant as mentioned before, Figure 4 shows the consistency of the new stationary model with respect to the SNe-Ia data in the range $z > 0.1$ explicitly here. As stated by [Riess et al. 2005], employing SNe Ia to estimate $H_0$, the HST Key Project and the STS collaboration disagree by up to 20% in their attempts. The different values $H_{CCM} = 71$ km/s/Mpc and $H_{SUM} = 65$ km/s/Mpc used further in this section, are clearly within the range of observations.

Obviously the straight SUM accordance in Figure 4 applies to both, the data of the HZT reported in [Riess et al. 2004/07] as well as to those of the SCP reported in [Kowalski et al. 2008], the latter presented as world's supernova distance-redshift data once. The SUM fits seem even slightly better than those of the CCM, what can be seen from the lower panels 5.b – 5.e, where the red and blue straight broken lines are determined by the method of least quadratic deviations and in the best case should prove respectively zero.

Therefore, all relevant SNe-Ia data by 2008 obviously support straight SUM, which is directly applied to universal scales $z > 0.1$ here. In any case, however, the SUM is fitting these data not worse than the CCM. This is in accordance with the objective that the stationary line element (17) should describe the universe on sufficiently large scales.

*4.3 Full Scale Compatibility 2008 Given a Local Hubble Contrast*

Now it is a natural question whether the low range $0 < z < 0.1$ could be included into a SUM SNe-Ia agreement, too. In this section it will be shown how, instead of an accelerated expansion according to the CCM, a local Hubble contrast of about $-6.5\% \pm 1.8\%$ as reported by [Jha, Riess, & Kirshner 2007] might result in reasonable accordance with these low redshift data, too.

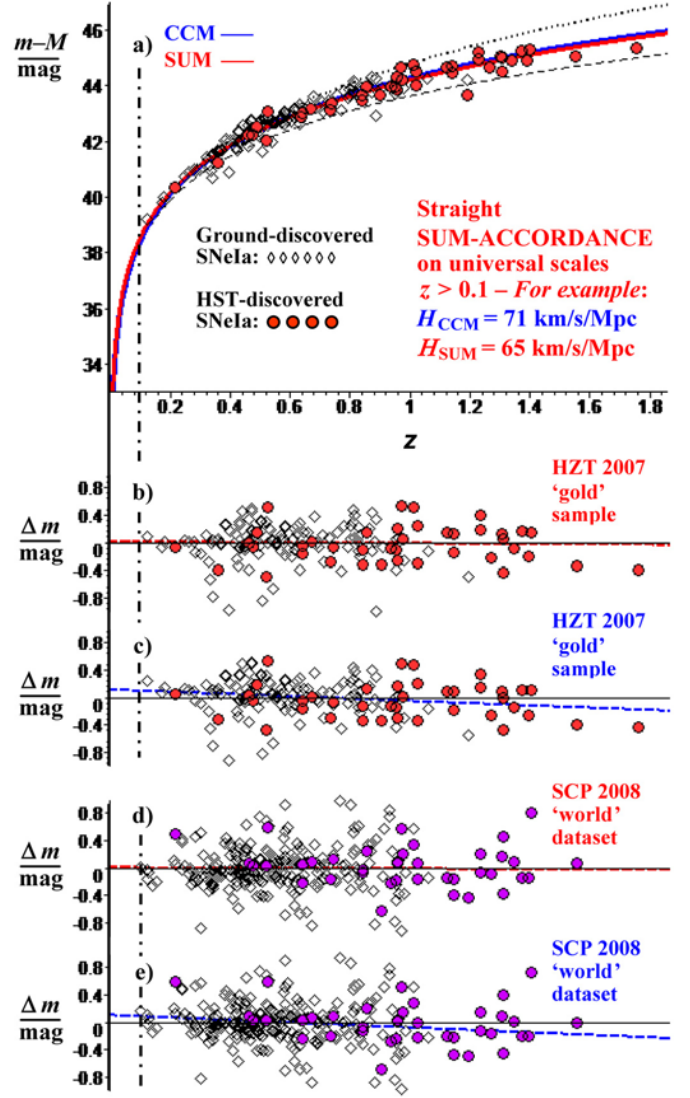

FIG. 4. – *Top panel (a)*: – Comparing the SUM magnitude-redshift prediction (62) for $\kappa = 0$ with the SNe-Ia data and the CCM-prediction, there is a straightforward SUM agreement on large scales $z \geq 0.1$ where the universe may be rightly regarded homogeneous and isotropic. The red SUM-line coincides almost completely with the blue CCM-line here. *Lower panels (b) – (e):* These Figures are of high importance, since here, in the high-redshift range $z > 0.10$, the pure model predictions are compared without any local corrections. Like the red broken lines, also the blue broken lines do not represent the predictions but the mean residuals in contrast to the respective z-axes, i. e. deviations from the data.

It seems not implausible that up to $z \approx 0.1$ the data may be affected by peculiar features of our 'local' cosmic environment. There are giant structures like the Great Wall or the Great Attractor, for example, whose dimensions extend up to several hundred Mpc. Therefore, only at about $z > 0.1$ the universe may be rightly regarded as homogeneous and isotropic. One might even wonder how the CCM can give the impression of nearly perfect homogeneity in closer vicinity.

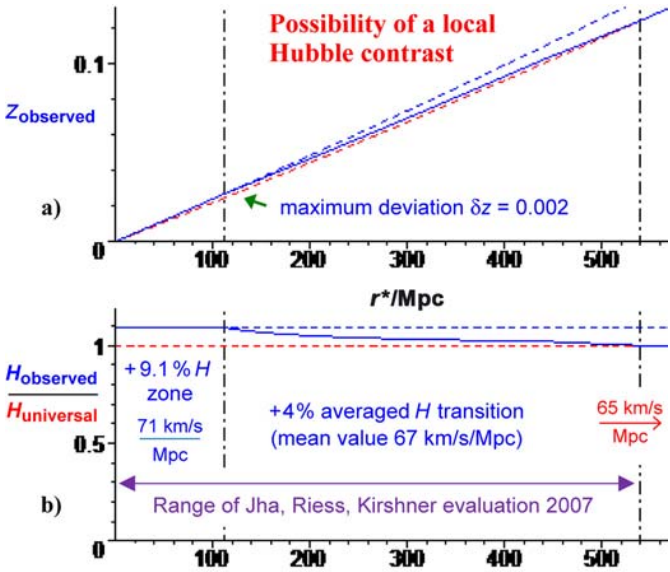

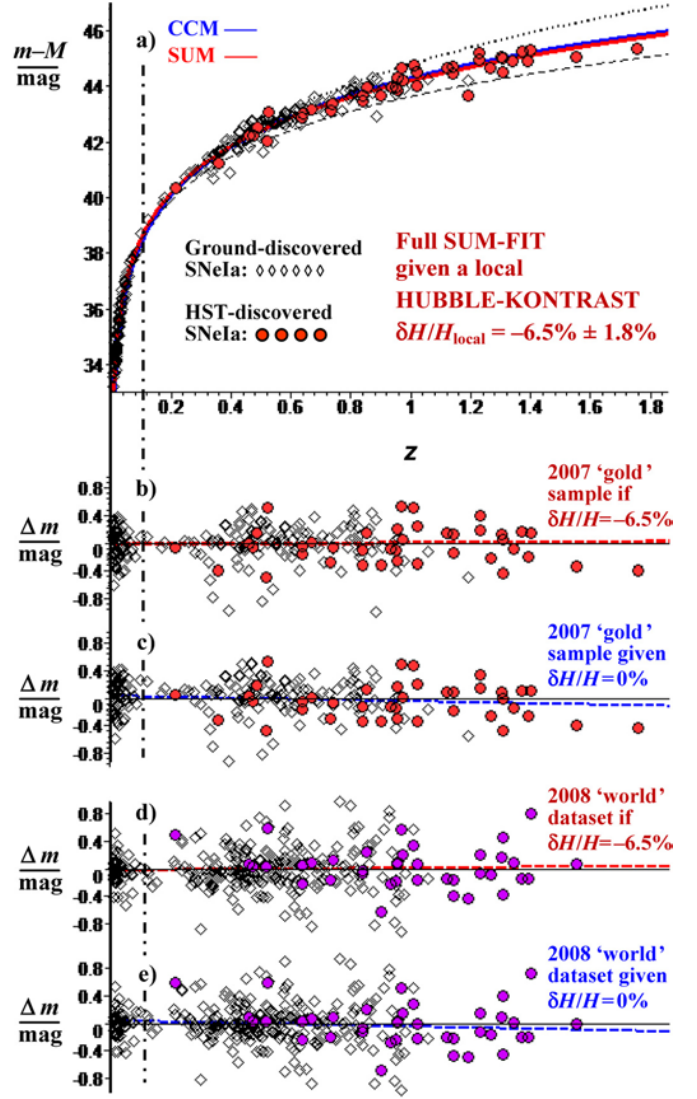

FIG. 5. – *Upper panel (a):* The blue solid line represents the real values $z_{observed}$ of the SNe-Ia measurements, the red broken line the SUM neglecting possible peculiar flows or local inhomogeneities. The maximum deviation $\delta z = 0.0022$ (= 660 km/s/c) within $z < 0.025$ corresponds to a Hubble contrast +9.1% at this point, though with respect to $H_{universal} = 65$ km/s/Mpc here. *Lower panel (b):* This corresponds to $H_{local} = 71$ km/s/Mpc within $r^* < 113$ Mpc, while the mean value in the transition zone up to $z \approx 0.13$ is about 67 km/s/Mpc. With about −5%, the difference 67 − 71 = −4 km/s/Mpc may correspond roughly to the lower limit of what [Jha, Riess, & Kirshner 2007] reported to be −6.5 % ± 1.8 %, here relative to $H_{local} = 71$ km/s/Mpc.

In the upper panel (a) of Figure 5 a maximum deviation $\delta z = 0.002$ would correspond to a local Hubble contrast of about +9%. The blue solid line on the top of the lower panel (b) might represent the real values of $H_0$ included in the SNe-Ia redshift measurements, the broken straight red line below the universal value $H$ according to the SUM. With $H_{universal} = 65$ km/s/Mpc, for example, this would mean $H_{local} = 71$ km/s/Mpc within $r^* < 113$ Mpc ($z < 0.025$), while the average value in the transition zone is about 67 km/s/Mpc. The difference ( 67 − 71 ) km/s/Mpc = −4 km/s/Mpc might correspond roughly to what [Jha, Riess, & Kirshner 2007] reported to be −6.5% ±1.8%. Please note that this result was given with respect to the larger local value, while here, in contrast, the quotients $H_{observed}/H_{universal}$ are taken with respect to the lower one.

Overcoming the apparent displacement of the low-redshift residuals seen in Figure 3, now Figure 6 shows that after taking into account the local Hubble contrast according to Figure 5 the SUM-residuals would result in reasonable agreement with the low redshift data, too.

The authors of [Jha, Riess, & Kirshner 2007] – as distinguished members of the HZT – stated in their paper that "regardless of whether the Hubble bubble is due to a real local void in the Universe or an artifact of SN Ia distances, the feature is present in the Hubble flow SN sample, and this has important implications for using SN Ia as tools for precision cosmology."

Concerning a full scale SUM compatibility according to Figure 6, the other way round, if such a Hubble contrast as shown in Figure 5 was real, then this might imply an addi-tional CCM problem. The blue straight broken lines of least quadratic deviations representing the respective residuals would show more deviation, because in the panels 5.c and 5.e such a feature – slightly impairing the plots – is not yet taken into account.

FIG. 6. – Taking into account a local Hubble contrast as shown in the low-redshift range $z \leq 0.10$ on the left of the vertical black broken line in Figure 5, there is a full scale SUM compatibility with not only the SNe-Ia data of the HZT, reported in [Riess et al. 2004/07], but also with those of the SCP's Union compilation, reported in [Kowalski et al. 2008], now. Obviously the corresponding correction of at most $\delta z \approx 0.002$ within $z_{corrected} < 0.025$ is sufficient to cause a reasonable accordance between the SUM and the data in the low redshift range, too. Here in the lower panels again, the red and blue broken straight lines are determined by the method of least quadratic deviations and should prove congruent with the z-axis respectively. In the very best case such a z-axis congruence should even apply piecewise, where the 0-level of the z-axis is bound to the best full-range fit respectively.

### 4.4 Additional Adaptability from Effects Like Faint Dimming by Dust

The SNe-Ia data of [Riess et al. 2004/07] and [Kowalski et al. 2008] have been used in the previous sections. The latter claimed to contain all relevant data of the first decade since the 1998 observational breakthrough. In fact, these are the data where that 'accelerated expansion' of the universe has been concluded from, whose discovery was awarded with the Nobel Prize 2011.

No more sophisticated statistical test or maximum likelihood analysis is needed to see that both models – the CCM as well as straight SUM – actually fit the data at least on universal scales $z > 0.1$ comparably well. Nevertheless besides this conclusion, here may be briefly mentioned some problematic corrections on the one hand, and an additional adaptability on the other hand.

Possibly an intergalactic extinction of starlight by gray dust over a mean free path of more than 3 - 4 times the Hubble radius $R_{H(o)} \equiv c/H_{(0)}$, might result in comparable agreement without any Hubble contrast, too. More probably, it would be also be sufficient to have a modest interplay of both effects, to provide considerable accordance. It cannot be excluded that there are even unknown features explaining small deviations of the SUM prediction to future data, if necessary.

The first SCP Union compilation [Kowalski et al. 2008] as used above has been updated to currently the "Union2.1" SN-Ia compilation [Suzuki et al. 2012]. In view of the SUM, a combination of a minimal local Hubble contrast with a tolerable amount of intergalactic (gray) dust might compensate some corrections small enough – otherwise ascribed to a "recently identified correlation between SN Ia luminosity and host galaxy mass" there – to have not changed the CCM picture essentially. Previously there seemed to be no relevant cosmic evolution affecting observed properties of the SNe Ia like the spectrum or the light curves, though in view of the CCM the high redshift SNe Ia had exploded in a stellar population much younger than the low redshift ones and with a smaller abundance of metals, for example.

In any case, given an extinction coefficient of only $\kappa \approx 1/7$ in (62), even a local Hubble contrast as considered by [Jha, Riess, & Kirshner 2007] without significant transition zone would model the updated data better now.

The stepwise correction of the SNe-Ia data in SCP 2010/11, resulting in an afterwards improvement of the CCM-fit – apparently necessary, see e.g. Fig.s 4.c/e – might in fact indicate the existence of an otherwise hardly detectable small amount of intergalactic dust whose detailed composition and properties remain unknown so far.

A corresponding SUM-adjustment – if necessary even to further SNe-Ia data improvements – seems always possible in two main steps. At first fit the SUM prediction by adapting a suitable extinction coefficient $\kappa$ in the universal range $z > z_{\text{transition}}$. Then model the local Hubble contrast according to the low-redshift data, where now, of course, is no longer any prior of a strict isotropy or homogeneity.

In contrast to the preliminary explanation of the SNe-Ia observations shown in the last Section 4.3, an adjustment due to dimming by dust has already been taken into consideration before. In particular, there has been discussed a 'replenishing gray dust' model by [Riess et al. 2004/07], who found a $\chi^2$ comparable to their CCM best-fit even in combination with the scarcely appropriate EdS-model there. It is clear already from Figure 2, however, that in case of the SUM such a $\chi^2$–test would have shown a much better result for an essentially smaller amount of dust. Furthermore, it seems remarkable in this context, that in view of the CCM a universal 'replenish-ing' of dust might be difficult to understand, whereas in case of the SUM it would be a matter of course. Furthermore, a corresponding dimming may be excluded for specific 'gray dust' models only.

Concerning the SCP 2010/11 Union 2/2.1 data, however, these have been changed afterwards by calibration fine tuning, for example, after the authors of Union 2.0 had stated, that there "could however still be unresolved NICMOS issues." Various corrections have folded the small CCM deviations shown in Fig.s 4.c/e up on the side of SUM now, though future improvements might disturb that subsequent perfect CCM accordance again. Furthermore, the data have been partially adjusted in view of a "fiducial cosmology". For the great advantage of the original data to be nearly free of model contamination this may mean a risk of being partially lost.

[Riess et al. 2004/07] have also discussed several alternatives to the "apparent acceleration of cosmic expansion (and dark energy)", too. Unlike other attempts, the explanation considered in this Section 4 does not need any ad-hoc hypotheses, but relies on the deductive one-parameter SUM in combination with an observed local Hubble contrast. Thus, there may actually be the chance even for a straight SUM universe without unnecessary coincidences, fundamental horizon problems or other peculiarities concerning an unfathomable entirety.

## 5. CHANCE OF HAVING ALREADY OBSERVED PARTS OF A STATIONARY UNIVERSE

While a mere background SUM would have to embed our cosmos according to a modified CCM, now in view of the straight SUM accordance to the SNe-Ia data at least on scales $z > 0.1$ there may be the chance for the new stationary cosmology, to apply to the observed parts of the universe. Some more concepts from the stationary line element (17) turn out to be at least not definitely incompatible with observational facts.

Though a straight SUM concept is suggesting itself, with regard to the wealth of cosmological observations this model cannot be fully developed at once, just as – step by step – today's CCM has grown with essential extensions and improvements from an in retrospect inapplicable first 'big-bang theory'. As a consequence, both the SUM and the CCM may temporarily be considered as legitimate alternatives in the following sections.

To check such a chance, these will try a preliminary tentative approach to corresponding observations, i.e. how nature might manage stationarity. As only a rough outline, this concept will not only be incomplete but may turn out even erroneous in parts as has been all models of the universe at least in first versions so far.

Now, within physics the universe has to be regarded a stationary actuality. It would be impossible to ascribe to its whole any peculiar properties locally found in our cosmic environment, just like no physicist would ever claim to have completely observed the entire universe.

To elaborate a straight SUM, however, there is a need to apply some unconventional concepts, too. One problem with a corresponding shift of hypotheses concerns the lack of 'dark energy' in the SUM. Thus the apparently missing mass-energy density is about three quarters of that total amount $\varepsilon_c =$ $\mu_c c^2 + p_c$ which is needed for spatial flatness. As already considered in Section 2.4, however, in view of straight SUM, this deficit might be explained by a sufficiently homogeneous DM background $\bar{\mu}^*_{\text{background}}$ which is associated to 'dark energy' in the CCM framework.

The next subsections will – in particular for some tentative detail assumptions – certainly need more specialized expertise. In spite of this caveat, however, it may be worth to go on with the preliminary straight SUM concept.

## 5.1 'Primordial' Nucleosynthesis and the Law of Entropy Restricted to Evolutionary Processes

Modern cosmology actually began with the paradox of [Chésaux 1744] and [Olbers 1823] who realized the problem to find out what it means for the night sky to be dark. It may be mentioned here that this problem actually would apply not only to light, but in the same way to the Newtonian gravitational force, too. Considering the SUM, the question is not, why the sky is dark at night, but the other way round, what follows from this fact.

Obviously, there has to be a statistically stationary attenuation of starlight – either by ordinary absorption, according to Olbers himself or by something else like the stationary energy loss of free photons considered in Section 2.5 – what over sufficiently large distances makes the universe effectively opaque in spite of an infinite number of stars. Against Olbers' approach to his paradox there has been – and occasionally still is – objected that any absorbing medium would heat up to stellar temperature. But such an objection is untenable, because stars do not radiate continuously. Just the other way round, there would be the question, how in a stationary universe new stars could arise again and again without any recycling of radiation and energy of their progenitors?

It is clear from the beginning, however, that in a stationary universe there must occur local violations of the second law of thermodynamics as this is commonly understood. Typical stars with finite lifetimes of about $10^{10}$ years due to their energy loss by radiation and particle emission, are formed newly over and over again, like all other 'temporary' structures, too. This might happen in the neighborhood of light and matter absorbing supermassive objects. In local processes of gravitational re-creation from previous energy and matter – these including parts of free and absorbed radiation – the entropy has to decrease locally, though remaining stationary with respect to an infinite universe.

The fact that – unlike electromagnetic force – gravitation always acts as an attractive force also supports the possibility of a local decrease of entropy. This is contrary to the increase of entropy by the processes of diffusion, in exactly the same way that the negative gravitational pressure present in the stationary model differs from the always positive, regular pressure exerted by gas.

The repeated re-creation of various elements – as well as the wealth of physically relevant knowledge and models – usually ascribed to a 'big bang' or the 'early phase of the universe' today, may be explained by or applied to such local processes. As a result, the temperature of an unknown interstellar 'dark' medium, cannot heat up on and on, but will adjust itself to a stationary value. As roughly estimated in Section 2.8, the mean intensity of starlight may be almost roughly same order as that of the CMB.

For all human experience, entropy as an indicator for statistical disorder or the probability of a natural system's state, seems in general increasing. The corresponding principle has proved to hold in every technical application without any exception. In comparison to local violations according to straight SUM, this principle does otherwise not apply to a CCM 'big bang' origin of the entire universe at all.

Therefore stationarity might demand space-time areas of decreasing entropy, temporarily delimited from empirical description so far. If not the alternative of a 'big-bang' creation out of nothing seemed even more problematic, it would be hardly justified to take such a consequence into consideration. But in this view there may be a statistical universal ultra-large scale balance of structure and chaos which, however, would not at all mean a monotonous thermal standstill in the future – but just the lively opposite of such a scenario.

Therefore a revised law of entropy may state that entropy always increases in evolutionary processes only. According to a concept of universal entropic compensation, under extreme conditions of local re-creation there may be reverse short-time processes where entropy decreases abruptly. In any case, this possibility cannot be ruled out, because even if it is true, no experiment would ever contradict a continuous increase within evolutionary structures. Just the contrary, all experiments in any laboratories all over the universe would respectively confirm that *there* are no exceptions to the law of increasing entropy.

Therefore, according to straight SUM, there should arise originative gravitational centers by sufficiently hot extreme densification processes in SMOs or quasars, for example, located at the centers of galaxies – partially ejecting jets in AGNi – which may work as sources for a statistically ongoing re-creation to a plasma of primordial entropy.

Correspondingly, such active centers – eluding direct insight – where may be a galactic flow of stars, dust, gas, and dark matter from close vicinity, are already releasing enormous amounts of energy continuously. As is well-known, there are Gamma-Ray Bursts (GRBs) which set free more energy in a few seconds than the sun over billions of years, thus millions of times brighter than the most luminous supernovae. Other hitherto unknown objects may run through various phases like quasars, black holes, or hypernovae.

The possibility seems not inconceivable that the prehistory of all matter entering such originative gravitational centers is erased, that the empirical law of entropy breaks down there, and that the light elements hydrogen and helium are partially re-created anew.

It is only evident, that matter before it allegedly disappears in 'black holes' would undergo phases of extreme density, pressure, and temperature similar to essential features of the assumed initial singularity called 'big bang'. Why should it not start a local 'expansion' instead of inevitably resigning from all creative interaction except for gravitation alone, which needs quantum mechanics to build up structures again? From this perspective, nothing should disappear for ever in 'black holes'.

With regard to theoretical physics, however, the assumed existence of literally 'black holes' remains speculation until Einstein's equations are solved for a quantized stress-energy-momentum tensor of matter. Unfortunately, Einstein's phenomenological approach breaks down at this point. Considering an inadequate relativistic description by the phenomeno-logical EMS tensor of matter there, an interplay with quantum mechanics might actually turn what is usually called 'black holes' into 'bright sources'.

Primordial nucleosynthesis – otherwise seen as a pillar of the CCM – may actually prove that there are 'local-bang' events at appropriate temperatures, pressures, and densities, for example, but may not prove that there was only one singular 'Big Bang', at only one time, in only one place.

Since the lifetimes of stars, galaxies and progenitor structures are finite, there would happen new formation of those structures, again a again – according to a heuristic principle of stationary cosmology: Given a stationary universe, all material components are determined by the requirement that they are gravitationally re-created according to the laws of quantum physics at the same rates as they have disappeared before in extreme gravitational centers, growing to cores of hot originative 'local-bang' events. These may compensate any corresponding deficits in the universal entropy balance, too.

Therefore the rates of material components in a stationary universe may be approximately those calculated from the big-bang model actually. In this view, a $^4$He abundance of about 25% – 30% is not necessarily evidence only for one singular 'big bang'. It might be sufficient that the core temperature reaches some $10^{11} - 10^{12}$ K in a gravitational collapse before an inevitable explosion may be caused by quantum laws.

According to stationarity, sufficiently large structures from clusters to filament-walls or void-cells should approximately stay in a statistical radiation equilibrium with the surrounding universe. This means, most energy radiated by stars should be absorbed by matter. To not limitlessly heat up, there may condense that 'warm' matter to primarily population I stars as well as SMOs may absorb all kind of matter and then eject it – continuously in parts – in form of 'primordial' protons and electrons, these building metal-poor population II/III stars, now all mainly burning hydrogen to helium for a long time, and ending in SNe explosions again, producing heavy elements and also building new populations of stars – until the next internal re-creation event in form of a 'local bang' might take place. In this view, there is a cycle of energy from nuclear burning in stars first to radiation, then after partial absorption by diffuse matter back into new stars.

### 5.2 Large-Scale Structure, Quasar Distribution, and a Mass-to-Radius Relation

At any point of universal time there should be extragalactic objects in any possible stadium of formation.

Still assuming the idealized uniform number density $n^*$ presupposed in Section 2.8, here may be compared some large scale distributions of universal objects, like e.g. galaxies, as predicted by the SST, the SUM, the EdS model, and the CCM, what means pressure-parameters $p^*/\varepsilon_c$ of $-1, -1/3, 0$, or $w_M \approx 0, w_\Lambda \approx -1$. These pressures correspond respectively to a cosmological constant, the stationary value, pressure-free matter, or two parameters of what is called the CCM's 'strange recipe'.

It is well-known, that the observed quasar distribution – as reported in the SDSS Data Release 7 by [Schneider et al. 2010], for example – shows a steep decrease to almost zero within the interval of about $2 < z < 4$, whose counterpart is not seen in relation (65) of Section 2.7. Therefore a comparison of this observed feature with the corresponding distributions of the SUM and the CCM – though both here neglecting quasar-specific evolutionary effects – may be illustrated roughly.

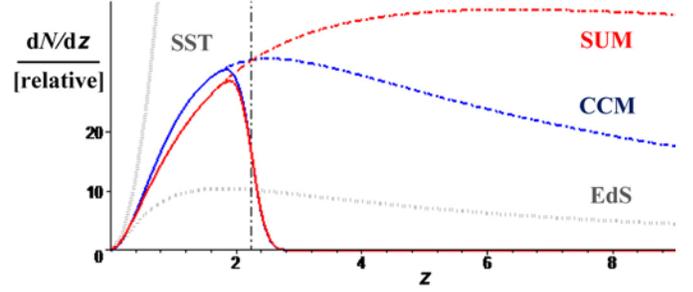

FIG. 7. – Theoretical distributions d$N$/d$z$ of universal objects assuming a homogeneous number density and neglecting effects of absorption or evolution. With regard to both solid lines, a smoothed Malmquist cut-factor is taken into account according to a magnitude limit of about 20.2 mag (corresponding to the vertical black broken line) as used in the SDSS Data Release 7 [Schneider et al. 2010] for example. It are shown the red SUM lines according to (65) together with the blue lines of the CCM-prediction (144) given $H_{CCM}$ and $H_{SUM}$ according to Section 4.2 and the best-fit CCM parameter $\Omega_\Lambda = 0.73$ as used there. The cut-factor above is ½{1−Φ[4($z$−$z_{\text{limit}}$)]} with Φ the Gaussian integral-function. For illustration are shown corresponding SST- and the EdS-predictions derived in [Ostermann 2003, 2008b] as grey dotted lines.

Analogously to the SUM prediction, a comparable quasi-CCM prediction d$N$/d$z$ is derived from (63) where $r^*$, d$r^*$ are replaced by $l$, d$l$ according to (134), what yields

$$\frac{dN_{CCM}}{dz} = 4\pi n^* R_H^3 \cdot X, \qquad (144)$$

with here $X$ temporarily

$$X = \frac{\left(\int_0^z \frac{dz'}{\sqrt{\Omega_M(1+z')^3 + \Omega_R(1+z')^4 + \Omega_\Lambda}}\right)^2}{\sqrt{\Omega_M(1+z)^3 + \Omega_R(1+z)^4 + \Omega_\Lambda}}.$$

Now, as can be seen from Figure 7, taking into account the magnitude limit of about 20.2 mag as used in the SDSS Data Release 7 quoted above, a corresponding smoothed Malmquist cut-factor – due to statistical scatter of absolute magnitudes, for example – may change both distributions from the broken lines to the solid red and blue line, which are showing a similar steep decrease at about $z_{\text{limit}} \approx 2.3$ now.

On the other hand, it may not be firmly excluded that the reason why quasars seem to occur only at redshifts $z > 0.05$ might partially find an explanation in a corresponding gravitational offset or in our coincidental local situation. The median redshift of $z \approx 1.5$ observed in Data Release 7 seems compatible to both the SUM or the CCM solid lines in Figure 7.

Apart from the idealized distributions of galaxies or quasars shown above, the extension and the formation of observed structures as for example the Sloan Great Wall, might enclose several CCM-riddles. According to straight SUM it bears no major difficulty, though in contrast to superclusters those structures may be rather described by $H(x^i)$ according to (120) in Section 2.11 than by local Newtonian potentials only.

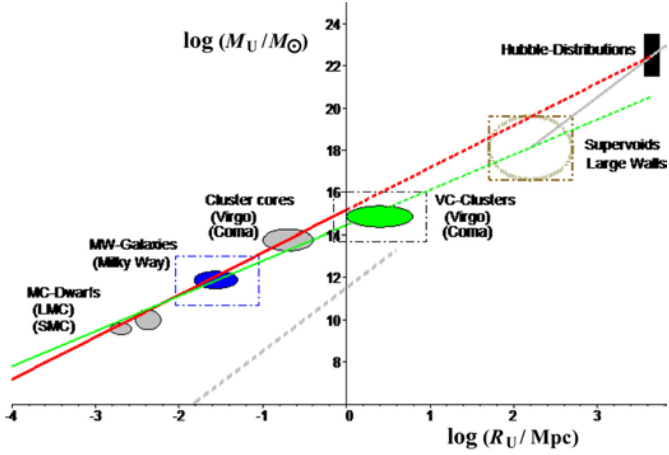

FIG. 8. – In this diagram of the SUM mass-to-radius relation (146) for gravitationally bound objects, the red broken line results from the constant assignment $Y = 16/9$ in (145). As an alternative, the green line is drawn for a tentative assumption $Y = 1/4 \cdot (R_U/2.5 \text{ Mpc})^{-1/3}$, while the grey line would apply to open cut-out parts of the homogeneous universe. The radius $R_U$ as well as the mass $M_U$ of universal objects, however, are not well-defined. In particular, the respective radii of galaxies and clusters may correspond only roughly in orders of magnitude to values between effective core radii of their dark-matter distribution and their visible sizes. Also the masses of these structures may be approximately known with large uncertainties. 'MW-Galaxies' refer to the Milky-Way type, 'V/C-Clusters' to Virgo or Coma type, for example. MC-Dwarfs are represented by the Large (LMC) or the Small Magellanic Cloud (SMC). In contrast to these universal objects, relation (146) does actually not apply to superclusters or 'cellular' voids surrounded by filaments or walls, since these structures are clearly no quasi-Newtonian gravitationally bound spheres as presupposed in Section 2.8 for the original form of the mass-to-radius relation (75). Therefore, this preliminary figure does not mean a claim, but a rather question.

In view of straight SUM, all radiation going out from an appropriate universal object should be part of an approximate energy equilibrium at its 'surface' with all matter and radiation coming in from all external sources. Coming back to (78) of Section 2.8, an order-of-magnitude diagram may be given in this context, though at most the DM halos of galaxies or clusters can be considered spherical objects, where in addition it is problematic to assign a clear corresponding radius $R_U$ respectively. With the dimensionless parameter

$$Y = \frac{\frac{1}{(F_U^*/I_U)} - \alpha_U}{1 + \frac{\kappa_{\overline{U}}}{2}} \Omega_U, \quad (145)$$

where $F_U^*/I_U$ means the fraction of universal radiation arriving from a single type U of objects at the 'surface' of one of them, the SUM mass-to-radius relation (78) now reads

$$M_U = Y \cdot \frac{\pi}{2} \rho_c R_H R_U^2. \quad (146)$$

This relation may be may illustrated for two simplifying tentative assumptions. In case $Y = constant$, $\kappa_{\overline{U}} \ll 1$, with $\alpha_U \leq 1$ relation (145) would obviously imply an approximate proportionality to the respective density parameter $\Omega_U$ for sufficiently small values of the fraction $F_U^*/I_U$. This seems one plausible option among others taken the special value $Y = 16/9$. The result is shown as red line in Figure 8 which would also apply to 'Hubble distributions' of density $\mu_c = 2/3 \rho_c$ and radius $R_H$ in this particular case.

According to such a provisional proportionality, Milky-Way type galaxies, for example, would receive on average significantly less radiation from the whole of their universal counterparts than they emit. Therefore their individual lifetimes should be limited, particularly as other relevant processes take place changing their constitution as well as their statistical distribution of course. In view of the SUM, this requires a stationary re-formation rate of those objects. Thus the red line shows, that single sorts of gravitationally bound objects like galaxies could not be each on its own at a radiation equilibrium of $I_U$ (emitted) and $F_U^*$ (received) if regarded as isolated objects respectively.

In contrast to the red line, Figure 8 shows as green line the mass-to-radius relation (146) taking into account an adjusted assignment $Y = 1/4 \cdot (R_{Cluster}/R_U)^{1/3}$ where a mean radius $R_{Cluster} = 2.5$ Mpc is chosen for illustration.

The grey line, however, would represent temporarily considered spheres of the critical mass density $\mu_c$ which only make sense on scales of at least voids and walls as parts of the homogenous universal distribution.

Altogether, Figure 8 is merely a preliminary schematic representation of another unexpected chance in the straight SUM framework. Accordingly, for example, whole clusters may not be regarded as something like those compact spheres presupposed in (74) rather in contrast to their cores, where only small fractions of associated galaxies may be within dense DM halos also containing X-ray emitting gas. Therefore, since $R_U$ in (74) is not well-defined, in particular $R_{Cluster}$ as an effective radius may lie between a mean core radius of order 0.2 Mpc and a mean visible radial cluster extension of order 5 Mpc. Representative values here taken approximately are $R_{core} \approx 0.1 - 0.4$ (0.2) Mpc and $M_{core} \approx 2 \cdot 10^{13} - 2 \cdot 10^{14}$ ($6 \cdot 10^{13}$) $M_\odot$ (the numbers in parentheses are taken as a provisional mean respectively).

Correspondingly, provisional values of masses and radii as assumed for other gravitationally bound objects in Figure 8 are $R_{VC} \approx 1 - 6$ (2.5) Mpc with $M_{VC} \approx 2 \cdot 10^{14} - 2 \cdot 10^{15}$ ($7 \cdot 10^{14}$) $M_\odot$ for Virgo or Coma type clusters, $R_{MW} \approx 15 - 50$ (27) kpc with $M_{MW} \approx 3 \cdot 10^{11} - 2 \cdot 10^{12}$ ($8 \cdot 10^{11}$) $M_\odot$ for the Milky Way, $R_{LMC} \approx 3 - 6$ (4) kpc with $M_{LMC} \approx 3 \cdot 10^9 - 3 \cdot 10^{10}$ ($1 \cdot 10^{10}$) $M_\odot$ for the Large, and $R_{SMC} \approx 1.5 - 2.7$ (2) kpc with $M_{SMC} \approx 2 \cdot 10^9 - 7 \cdot 10^9$ ($4 \cdot 10^9$) $M_\odot$ for the Small Magellanic Cloud, where all numbers may roughly apply in order of magnitude only.

Now considering the effective temperature (73) together with the mass-to-radius relation (75) again, there cannot be a single sort of SUM objects to stand in radiation equilibrium $F_U^* = I_U$ and to constitute the full matter density $\mu_c$ required for the SUM flat space solution at the same time. At most, an approximate energy equilibrium of radiation going out from one sort of universal object like clusters might at its effective 'surface' radius $R_U$ partially apply roughly with the total radiation coming in from all external sources.

Even independent of its SUM derivation relation (146) as illustrated here seems approximately to apply in order of magnitude, after all.

*5.3 The CMB as Black-Body SMB
of Redshifted Components*

With regard to a stationary universe, even in spite of various components at different temperatures there should exist an effective overall equilibrium of energy exchange. Here stars, gas, dust and dark-matter halos – the latter partially on their own or associated to galaxies and clusters – may be regarded as sources of various radiation in respective frequency ranges each. The universal radiation may partially appear also in form of non-thermal emission such as e.g. synchrotron, dipole, or other types of radiation from the various material components. Nevertheless, any prevailing homogeneous-isotropic background radiation other than of BB type with statistical fluctuations might rather need an explanation.

On the other hand, in contrast to a local cavity, it seems impossible to keep a redshifted Planck spectrum within a stationary infinite universe. Consequently, to observe such an expected universal BB background, there have necessarily to be emitted (also) non-thermal components.

As already considered above, because of the otherwise 'missing mass' there should be a large amount of 'dark' matter whether baryonic or not. According to arguments and results at the end of Section 2.8, the chance for an alternative explanation of the CMB may even look self-evident. The difficulty concerning a BB character of redshifted radiation from cosmic distances, has been solved for a special case $\kappa=2$ there.

This tentative straight SUM approach to the CMB as an alternative BB-SMB radiation shall be illustrated now. The basic idea is that such a microwave radiation might originate from 'dark'-matter possibly in combination with some microwave Synchrotron Radiation (SR), for example. Essential parts of DM may be distributed homogeneously, while a smaller inhomogeneous part seems gravitationally condensed to halos of galaxies or clusters up to DM-containing filaments at respectively similar temperatures of several K. With regard to these obvious inhomogeneities of matter and temperature, the large scale balance can be a statistical effect only.

In view of straight SUM, microwave radiation from (mostly) unresolvable sources at all distances – though redshifted on the way through intergalactic space – is obviously composing the almost perfect BB spectrum observed. One condition is that the absorption coefficient $\kappa$, limiting the mean free path, must not be too large to allow for the observations of distant quasars or other objects already made in the microwave or even in the CMB mm-range. Therefore extragalactic or extra-cluster space, where the dark matter density may be comparably small, should be of low microwave opacity, too. With the tentative assignment $\kappa=2$ leading from (86) to the SUM solution $\rho_{SMB\nu}$* (88), this condition might be sufficiently fulfilled.

Even given an almost perfect SUM Planck spectrum within each shell of the universal large scale distribution, only a small fraction $\beta \Delta r^*/R_H$ of the CMB-photons outcoming its surface may have been emitted from within, while the completing part comes from behind.

With regard to the local emission spectrum shown as red line in Figure 9, it may be taken into account that in addition to the thermal DM radiation assumed here, there is also some non-thermal universal microwave SR, as observed not only in the Spectral Energy Distribution (SED) of quasars and radio galaxies but also in the WMAP haze of our Milky Way, for example, see e.g. [Ade et al. 2012] and references therein.

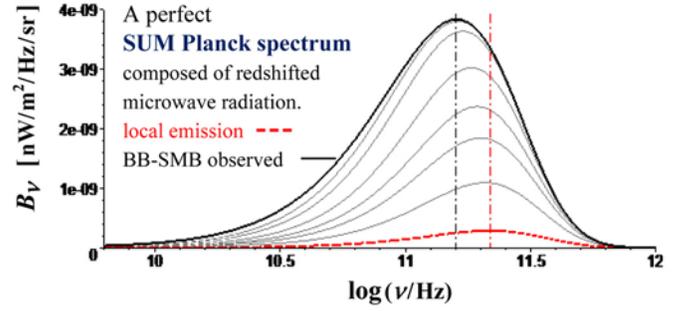

FIG. 9. – Demonstration of the SUM chance to have the CMB as a perfect BB spectrum composed of redshifted Microwave Background Radiation (MBR) within a stationary universe. The underlying solution is derived in Section 2.8 leading to the relations (86), (87) with $\kappa = 2$. The black line shows the CMB spectrum as actually observed, while the red broken line shows the emission of the here assumed universal MBR from DM (possibly in combination with microwave SR) in a representative local shell of 100 Mpc for example. In contrast, the grey lines between show only respective parts of the BB-SMB radiation coming from within z < 0.1, 0.2, 0.3, 0.5, 1.0, 1.7 according to (86) with the upper limit of the integral replaced correspondingly there.

The idealized local 'dark' emissivity $\beta(\nu_E)$ found theoretically in (87) tends to the linear expression $h\nu_E/(k\Theta_{DM})$ for frequencies $\nu_E \to \infty$. This behavior does not necessarily hold over the full frequency range, of course. Thus, in Figure 9 there is used a cut-off according to $\beta(\nu_E < 10^{12}$ Hz) in (87), otherwise $\beta=0$, without visible deviations from a perfect Planck spectrum in the observable frequency distribution shown in Figure 9 as black solid line.

Since in view of straight SUM 'dark' matter gets rid of its mysterious lack of physical interaction – the latter otherwise rather a 'matter of course' – acoustic DM oscillations may be caused in particular by the interplay of gravitational attraction and the DM radiation pressure.

Thus it might even turn out that the CCM explanation of anisotropies can be mathematically transferred into straight SUM, though with several physical modifications and on different scales of space and time, but leaving the observed angular distributions unchanged. Even local DM oscillations within our Milky Way might leave their imprints, while there should be – not exactly, though – the same order-of-magnitude temperature in each halo. Besides acoustic DM oscillations, also all other statistically isotropic unresolved deviations from DM homogeneity as in particular due to halos should appear as BB-SMB anisotropies.

Independent of any respective cosmological model, the CMB anisotropies – excellently measured in WMAP from [Bennett et al. 2003] to [Jarosik N. et al. 2011] or by the Planck Collaboration [Ade et al. 2011] – are clearly related to the distribution not only of visible luminous matter, but in particular to that of 'dark' matter, too. Therefore, taking into account an appropriately chosen transfer function there might always be a SUM 'explanation'.

Correspondingly, such a transfer function linking actual CMB observations to the universal matter distribution, has to describe how the various components of Extragalactic Background Light (EBL) coming from more or less diffuse sources

– more or less attenuated, obscured, absorbed or partially hidden behind completely opaque universal objects – may be changed by (local) re-thermalization before observed at last.

The middle grey lines of Figure 9 show that according to the simplest SUM solution (86), (87) of Section 2.8, by far most of the BB-SMB radiation reaching telescopes would have been emitted within $z < 1$. The other way round, the remaining parts of BB-SMB seem decreasing with distance. Such a decrease would depend on whether extra-cluster DM is distributed in additional small halos, for example, or more homogeneously instead.

Therefore the direct SUM counterpart to the well-known [Sunyaev & Zeldovich 1970] effect (SZ) should appear increasingly reduced at high redshifts and with some frequency-shifting modifications, too. The latter are limited by the vertical broken lines of Figure 9 corresponding to a redshift interval of $\Delta z \approx 0.4$ equivalent to a universal distance $\Delta r^* \approx 2$ Gpc. These vertical lines mark the peak positions of the locally emitted and the altogether observed BB-SMB radiation constituting the CMB.

It is in the nature of any 'particle' distribution that given a statistical radiation equilibrium, there might only approximately be found the same DM temperature in different clusters. Thus, small temperature differences might cause deviations in order of magnitude of those expected from the SZ effect itself or even more than that. Thus, inhomogeneities of the temperature distribution from rather homogeneously distributed parts of dark matter may be playing together with those caused by resolvable halos.

According to [Lieu, Mittaz, & Zhang 2006] the (thermal) tSZ effect is not clearly observed at its expected amounts. In addition, as reported in [Kashlinsky, Atrio-Barandela, & Ebeling 2012] (with references therein), there seems to be a 'Dark Flow' (DF) measured from the (kinematic) kSZ of galaxy clusters, what would mean another problem for the CCM. The discussion below will come back to both difficulties.

Within the SUM framework, the whole SZ effect – though undoubtedly present, in case of e.g. Coma with the same result as otherwise expected and observed – needs a thorough revision going beyond this preliminary approach.

Some preliminary ideas in this section may be considered working hypotheses to check the chance for a more elaborated concept of straight SUM, what certainly requires further efforts. Here it has been a first step in particular to show some extended parameter space to describe the CMB phenomenon in principle, after all.

*5.4 The BB-SMB as Only a Special Part of Universal Radiation*

As shown by the existence of a CIB as part of the EBL, the special fraction of radiation called CMB is – though highly dominating – not the only homogeneous-isotropic microwave radiation actually observed even in the mm-range, see [Hauser & Dwek 2001], [Kashlinsky 2005], [Ade et al. 2011] with references therein. Therefore in view of the SUM it is a natural question whether these contributions might be of same nature, only distinguished by the almost perfect (CMB) in contrast to the clearly deviating (CIB) black-body behavior of the observed spectral components.

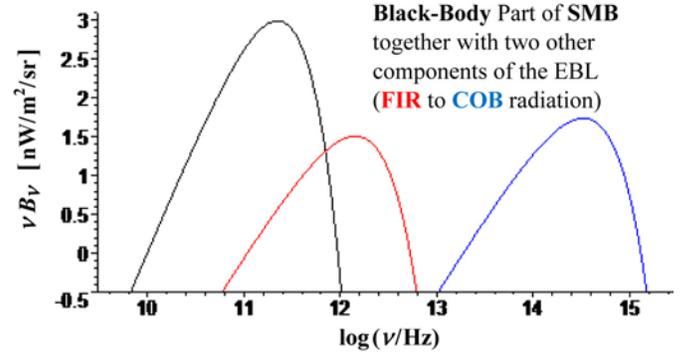

FIG. 10. – Schematic diagram of the BB part of the SMB (according to Fig. 9) together with strictly simplified parts of the CIB from dust clouds and the COB from stars as three main contributions to the approximately isotropic EBL (cf. Fig. 5 in [Hauser & Dwek 2001]). The black solid line on the left would represent the BB-SMB as composed according to the derivation in Section 2.8. The blue and the red line, respectively corresponding to redshifted starlight or radiation from partially unresolvable dust clouds, are calculated as $\nu B_\nu$ from (84) preliminarily using on trial source temperatures $\Theta_{Stars} = 6000\,K$ and $\Theta_{Dust} = 26\,K$ with absorption coefficients $\kappa_{Stars} = 6 \cdot 10^{-15}$ and $\kappa_{Dust} = 1 \cdot 10^{-5}$ assumed according to (77), where same coefficients $\kappa_U$ may result from various combinations of the respective values $\Omega_U$, $R_U$, $M_U$, $\alpha_U$, thus leaving a reasonable parameter space for necessary adjustments.

Consequently in view of straight SUM, the CIB might be only a small non-BB part of the SMB, while the predominant BB-SMB part would account for the CMB.

Furthermore there is also non-isotropic radiation in that frequency range coming from resolvable distant sources and already separated from the diffuse parts of the CIB, where in addition to the CMB the former is found an isotropic background including Far-Infrared Radiation (FIR) from dust.

According to a gross first-step simplification, the complete amount $B_R^*$ of universal radiation may be regarded as essentially composed of that from luminous matter in visible stars, gas, or dust clouds on the one hand, as well as, on the other hand here, that from 'dark' matter possibly in combination with a corresponding microwave SR. Both parts might altogether emit the universal background of $\gamma$-ray (C$\gamma$B), X-ray (CXB), optical (COB), infrared (CIB) down to microwave (CMB) or at last radio frequencies (CRB) as may be summarized according to

$$B_R^* = \sum_A B_A^* . \qquad (147)$$

where the index 'A' stands for (S)tars, (D)ust, (G)as, or dark (M)atter. In a stationary universe any relic-radiation, even that from the 'big bang', would have to be regarded as originated from 'local' sources. Taking into consideration, that 'dark' matter seems the predominating component of (147), the CMB should be the predominant component of extragalactic radiation $B_R^* \approx B_M^*$, too. Correspondingly, in Figure 10 a simple model of the BB-SMB partially overlapping the CIB, the latter reduced to one mean dust-temperature only, and in addition an equally simplified component of the COB are taken into account.

Now in the SUM framework, this figure illustrates the well-known fact that in contrast to the CMB as the DM microwave background assumed here, the energy densities of other types of EBL are comparably small. The red line stands for FIR, and the blue line for the COB, both parts coming from sufficiently distant (unresolvable) extragalactic sources according to relation (84) of Section 2.8, while the BB-SMB as derived in (86), (87) – otherwise already shown in Figure 9 – is represented by the black line here.

Besides $\Theta_{DM}$ = 2.7 K, tentative source temperatures $\Theta_{Stars}$ = 6000 K and $\Theta_{Dust}$ = 26 K are used there. The respective absorption coefficient as the remaining free parameter of relation (84) is preliminarily chosen as $\kappa_{Stars} = 6 \cdot 10^{-15}$ and $\kappa_{Dust} = 1 \cdot 10^{-5}$ assumed according to (77), where a same absorption coefficient may result from various combinations of the four respective values $\Omega_U$, $R_U$, $M_U$, $\alpha_U$, thus leaving a reasonable parameter space for necessary adjustments. Tentative values would be e.g. $\Omega_{Stars}$ = 1/144, $R_{Stars} = R_\odot$, $M_{Stars} = M_\odot$, $\alpha_{Stars}$ = 1 or only fictive assignments $\Omega_{Dust} \approx \Omega_{Stars}$, $R_{Dust}$ = 1 kpc, $M_{Dust} = 1 \cdot 10^{11} M_\odot$, $\alpha_{Dust}$ = 1/10 for illustration.

Therefore, taking into account that universal dust as well as luminous matter each actually exists at a broad range of various temperatures and not at all exclusively in form of spherical object, it should be possible to fit Figure 10 to the observed SED of the universal background radiation shown e.g. in Fig. 5 of [Hauser & Dwek 2001] or in addition at E. L. Wright's 2001 webpage [atro.ucla.edu/~wright/CIBR].

Quite naturally, the first straight SUM approach to an alternative BB-SMB explanation of the CMB radiation is necessarily incomplete and the simple ansatz (86), (87) may need essential improvements. It does not yet aim to 'predict' any definite details as otherwise excellently provided by the conventional 'relic'-radiation approach within the CCM framework. The challenge is to learn from the wealth of actual observations, how to reach compatibility of a deductive model based on proven physics as far as possible.

## III. DISCUSSION AND CONCLUSION

Though straight SUM would be in fundamental conflict with one single 'big-bang' universe, there might be 'local-bang' cosmoses within the stationary background instead. Adapting the whole CCM-cosmos into open SUM approximately, however, would demand appropriate modifications.

### 6. SOME RIVALING ASPECTS

1) *Underlying concepts:* The CCM concept has proved exceptionally fruitful, but there are several well-known problems concerning the singularity, inflation, horizons, coincidences, and fine tuning.

In particular a cosmological constant $\Lambda \equiv \Omega_\Lambda \cdot 3H_0^2/c^2$ with $\Omega_\Lambda \approx 0.73$ according to the CCM would mean nothing but $T_0 \approx 1/H_0$, as was shown in Section 3.2. This relation, however, could neither have been – nor would be again – ever fulfilled at any other time than only 'today'. Though the CCM is extremely successful, this anthropic coincidence – 'geocentric' in time, and hardly more convincing than 'geocentric' in space – may be the most striking of some strange features, which make that model casting serious doubts if applied to the entire universe. Whether justified or not, there are spontaneous associations with the numerical power of the Ptolemaic system long time ago (it may be added in this context, however, that there has been anticipated a kind of harmonic analysis many centuries before Fourier's work).

Within the CCM framework of a ΛCDM 'big bang' model, an inflation scenario is needed to solve problems like among others (a) a flat universe, (b) Gaussian random phase fluctuations, (c) 'superhorizon' scales – maybe most of those features (listed from [Bennett et al. 2003]) one simply would start from, given a stationary background universe.

While the CCM is dealing with a cosmological constant many orders of magnitude smaller than otherwise assumed, the SUM is dealing with a true Hubble constant given by the obvious relation $H^2 = 8/3\pi G \rho_c$ which may be concluded from the Schwarzschild radius $2GM_H/c^2$ of the 'Hubble mass', as reasoned in Section 2.13.

Even in a stationary background universe, however, if our evolutionary cosmos as observed by astronomers had a singular beginning approximately 13.7 Gys in the past, then this cosmos may be only the part known to us today. An infinite number of many such cosmic areas might arise and pass, again and again, just like the stars and spiral nebulae therein.

In this view, what do all those important achievements mean, explaining the cosmic evolution of matter, radiation and other more exotic components of the universe? Most, if not all, of the questions mentioned above appear in a different light. This because any peculiar features – if unacceptable for the universe as a whole – do not need justifications, if understood as those of our 'local' cosmos only. Therefore a suggestive solution seems to adopt today's cosmology embedded in an open SUM, the latter describing the natural background where the CCM 'big bang' once had took place.

Since in Section 3.2 approximate CCM parameters $\Omega_M$ = 0.263, $\Omega_\Lambda$ = 0.737 have been derived from a coincidence with the SUM's 'boundaries', particularly its FLRW-form (95) is suggesting an attempt to get the CCM cosmos with local evolution fitted therein. In contrast to the significant Hubble constant $H_{s\text{-SUM}} \equiv H$ (s. Sect. 2.5) most free parameters would have to describe our peculiar environment. The question is whether an open SUM might be an alternative, including pre-inflation as well as post-inflation scenarios, too.

According to the CCM, a limited region of observability within $r^* \approx 3.4 R_{H_0}$ is derived from this 'flat space' model. In contrast to the presupposed homogeneity, however, it may be more realistic to describe our evolutionary cosmos by basically inhomogeneous approaches as in particular by that of [Buchert 2000/01], [Wiltshire et al. 2007], [Coley 2010a/b] where effects of 'backreaction' are taken into account.

2) *'Primordial' nucleosynthesis:* The original big-bang attempt to an all-embracing primordial nucleosynthesis failed to explain the existence of heavier elements. In the SST framework, however, Hoyle's otherwise well-accepted clarification of the synthesis in stars suffers from the claim of an ongoing new creation of protons, there 'steadily' increasing their number. Both obstacles are not present in case of the SUM. Completely open for the synthesis of heavy elements in stars as well as for light elements from plasma sources or from spallation processes caused by cosmic rays, too, the relevant condition would be that there is an effective proton re-creation at e.g. SGCs. Also a disconcerting CCM overestimation of the Lithium abundance might find a solution here.

3) *CMB radiation and anisotropies:* A non-thermal universal microwave SR from the cm to the mm wavelength range as mentioned in the previous section is particularly observed for example in quasar SEDs as well as in the WMAP haze. While [Hooper, Finkbeiner, & Dobler 2007] previously claimed "evidence of dark matter annihilations", recently [Ade et al. 2012] stated "that the microwave haze is a separate component and not merely a variation in the spectral index of the synchrotron emission".

Regarding the CCM, the actual existence of some additional isotropic microwave radiation means that there are two essentially different contributions, where the CIB part is discerned by the theoretical presupposition of a 'big-bang' relic radiation based on phases of 'inflation', 're-combination', and 're-ionization'. In view of straight SUM, however, what is called CMB on the one hand and the mm-parts of CIB radiation on the other hand, should be of closely related origin. The latter seem only subsequently defined to be what remains after subtraction of a presupposed BB fraction from the actually measured extragalactic microwave background.

A main problem for straight SUM is the lack of a detailed explanation for the CMB anisotropies so far. These measured with still increasing precision from COBE to WMAP (and the upcoming *Planck* results) each an excellent support to the CCM, however, do not yet exclude a stationary background.

According to the CCM, the physics of the acoustic peaks – which seems understood as the physics of a hot plasma responding to fluctuations in the distribution of dark matter – is given in terms of the fraction of baryonic matter, the fraction of dark matter, the optical depth at 're-ionization', and the sound horizon at decoupling. These cosmological parameters are not directly measured, but concluded from observations using several unproved assumptions [Durrer 2008], as for example the paradigm of a temporary universal inflation due to a 'slow rolling' scalar field completely unknown to proven physics so far. Now, in addition, there seems to appear the need for a 're-construction' phase to low $^7$Li abundance, if not other adjustments solve this problem (see e.g. [Fields 2012]).

In the CCM, the sound horizon size at the time of last scattering is representing a fiducial peculiar length. Other appropriately chosen initial conditions of 'big bang' and various features of 'inflation' are effectively working as – or providing for – free parameters in modeling the fluctuation spectrum of the CMB. A transfer function contains appropriate information including a set of quite a few adjustable parameters to explain how that CMB observed today is related to the scenario around the phase of decoupling about $10^{10}$ years ago.

In view of straight SUM, however, there is no horizon concerning the infinite universe. That, instead, a mean Jeans length, an average diameter of voids, or other observable features may include some fiducial lengths to explain the CMB anisotropies, will need detailed investigations. A straight SUM scenario would probably also imply oscillating distributions of 'dark' matter. Since there may be additional DM halos besides those of galaxies or clusters, it seems necessary to consider another 'transfer' function as already mentioned above, relating the distributions of luminous and 'dark' matter in a straight SUM framework, too. The chance for an explanation of the CMB anisotropies seems almost evident by taking a glance at Figure 14-e of [Sharp 1986] if compared to Figure 7 of [Bennett 2003] for example (though the first is in Aitoff equal-area projection and the latter in Mollweide projection).

According to straight SUM, light elements should be re-created in a form of a 'primordial' nucleosynthesis all over the stationary universe. Here the distribution of heavier elements like in particular Fe as the heaviest of stable elements might be much higher than expected in the CCM framework.

Therefore even an idea [Hoyle, Burbidge, & Narlikar 2000], [Wickramasinghe 1992] that e.g. iron-whiskers – possibly a special sort of 'dark' matter – might be involved to explain the observed BB-SMB is debatable again. In contrast to effectively 'invisible' iron in extremely cold stars ('black dwarfs'), a much smaller amount in form of a quasi-uniform distribution of excitable iron dust could be sufficient to emit relevant parts of the DM microwave radiation assumed here.

Altogether, in the framework of straight SUM, the phenomenon of the CMB may stay an open question until existence, nature, and temperature of dark matter are definitely clarified.

In view of an open SUM, however, even a 'big bang' origin of the CMB and its anisotropies would not apply to the entire background universe. A correspondingly extended SUM would try to embed the CCM cosmos, together with the CMB as a relic from its origin. Regarding a infinite universe of unlimited possibilities there may always be an 'explanation', as long as one is ready to use some additional hypothetical assumptions, if necessary. This would be the same way, how 'big-bang' cosmology has been developed historically.

4) *'Dark' matter:* Before the CCM modified several principles of proven physics, it seemed a commonly accepted fact that "every physical body spontaneously and continuously emits electromagnetic radiation" (Wikipedia on "Planck's law" 01/02/13, for example). In this view, either DM would be no physical quantity – then any physical theory had to exclude it – or DM does emit electromagnetic radiation. Therefore a reasonable question is: What may b the temperature of that DM, which substance seems necessary to explain the unexpected rotation curves in galaxies [Rubin & Ford 1970] or the puzzling peculiar velocities in clusters [Zwicky 1933] as well as gravitational lensing far from visible objects?

Based on an approach corresponding to the argument for the existence of DM in our Milky Way, a calculation like in particular that of a pure Emden sphere may unpretentiously explain the essential feature of approximately constant velocities. On the assumption that pressure, volume, and temperature of simplified hypothetical DM distributions are related in the same way as in regular gases, there appear roughly similar rotation curves as actually observed if only the temperature of this dark matter in each galaxy took a respective constant value. Whether it can be statistically the same temperature for all of them would depend on masses and proportions of corresponding 'particles'. Anyway, this seems an unexpected indication for a universal BB radiation.

The idea that dark matter might consist of 'thermal' massive neutrinos is commonly considered to have been disproved. But this disprove may be weak, as far as it – like many other conclusions from cosmological observations – is based on the presupposition of one all-embracing 'Big Bang' of the entire universe.

From the non-zero rest masses of three types of neutrinos it follows that, despite propagating at approximately the speed of light after their release, they will be slowed down by the gravitational field of an infinite universe (s. Sect. 2.3). At thermal velocities they might show unexpected features. Dark matter of weakly interacting 'particles' – possibly with various admixtures of the three types of neutrinos in each galaxy – could therefore be at least partially responsible for the observed cosmic background radiation.

It might even be possible to estimate a mean mass and cross section of DM particles in order of magnitude. From the assumption of an isothermal DM distribution leading to the observed rotation curves in galaxies it would follow a particle mass of about 1/1000 the mass of the electron. Then corresponding to (77) the particle radius might be of order $R_{\text{cross}} \approx 10^{-15}$ cm, though this is only a doubtable speculation.

The problem of a cosmic missing mass-energy density might essentially exist as a result of intrinsic CCM limitations only, while in view of straight SUM an additional density of rather homogeneously distributed 'dark' matter has been considered in Section 2.7. Therefore, according to the measurements apparently confirming a CCM density parameter $\Omega_M \approx$ 1/4, a corresponding SUM conclusion may be to ascribe this parameter to that part which is gravitationally bound in systems like galaxies or clusters, whereas a fraction of order $\Omega_\Lambda \approx 3/4$ would be more homogeneously distributed at much lower densities, possibly in voids. Because of largely smoothed inhomogeneities it would be hard to detect the latter part by gravitational lensing effects.

Correspondingly, such a distribution of 'dark' matter would imply a fraction of about 1/4 reaching observers from 'point-like' galaxies or clusters, while about 3/4 would arrive from an effectively homogeneous part of DM. But this is not yet exactly what should be observed. Because of the Milky Way's halo, only a slightly smaller fraction would reach astronomers' instruments completely without local interaction.

By far most of the SMB radiation, however, is coming from 'individual' last scattering processes far outside the Local Group. Most of the point-source information as well as SZ deviations may be afterwards separated appropriately and then ascribed to anisotropies of the CMB or the CIB.

5) *Olbers' paradox:* As is well-known and explicitly shown in the context of (85) above, the very most of energy directly bound to starlight is effectively 'lost' on its way, otherwise the night sky would be approximately of same radiance as the surface of the sun. According to this relation, besides the thermal radiation absorbed in the conventional sense, there has been concluded the possibility of an additional amount either directly absorbed, or transported by thermal conduction or by thermal convection in unknown processes. Considering the solution $\kappa_U = \kappa = 2$ of relation (85), and taken constant mean values $\overline{\alpha}_U = \overline{\beta}_U/2$ for a special kind of non-greybody sources, one might find an energy equilibrium again, while the factor 1/2 in $\overline{\alpha}_U$ corresponds to the first summand "2" in (90a), (90b) again.

Altogether the attenuation concluded there, seems an effect composed of re-cycling and relative time dilation – quite different from a mere 'tiredness' of light or other past attempts which failed in denying the time dilation according to (19), a universal feature though applying locally only. While this is undoubtedly observed as well as the energy loss (55) of redshifted photons – both effects directly confirmed by the SNe-Ia measurements – the latter might be accompanied by an unexplained extinction yet to re-establish universal energy balance against the overall emission of starlight. In contrast to ordinary gravitational redshift, the photons propagating through the DM background field might intermediately store a corresponding amount of unknown potential energy on their way before they are actually absorbed.

The difference in the respective history of starlight compared to that of BB-SMB radiation may be that the first would be finally absorbed at a distance before returning in form of warm baryonic matter to e.g. SMOs, while the second may be respectively attenuated in DM at place and time where corresponding emission occurs.

6) *Sunyaev-Zel(')dovich effect:* Of all claimed observational evidence for the CMB as relic radiation from a 'big bang', the SZ effect seems a strong proof. It is just this effect in its standard interpretation, however, which is also raising some doubts. Remarkably, [Lieu, Mittaz, & Zhang 2006] have found anything but a clear tSZ effect in the WMAP data, where several clusters seem even to show rather the contrary. In another context, [Efstathiou & Migliaccio 2012] stated that „Early expectations that measurements of the tSZ effect … could be used for precision cosmology now seem naive."

Furthermore, as reported in [Kashlinsky, Atrio-Barandela, & Ebeling 2012] with references therein, some 'dark flow' measured by evaluation of the kSZ of galaxy clusters seems to agree in its axis and its absolute value roughly with the CMB dipole corrected for the Local Group (LG). This would apply after taking into account a probable 20%-30% overestimation of the derived cluster velocities.

In another context, averaging the Hubble constant in spherical radial shells, [Wiltshire et al. 2012] found the LG rest frame preferred rather than that of the CMB. Therefore:

a) On the one hand, the uniform DF velocity seems to correspond in its absolute value and in its axis approximately to the relative velocity between the local CMB frame and the LG.

b) On the other hand, in contrast to single galaxies gravitationally bound to clusters, the Local Group may be approximately at rest with respect to the universal 'comoving' coordinates as suggested by the evaluation of Wiltshire et al. quoted above.

Taken together, this means that with respect to the LG restframe there may be no DF of clusters at all. Just the other way round, such a 'dark flow' actually ascribed to distant clusters may turn out to apply to 'our' locally affected CMB instead.

In particular, besides other additional effects like intrinsic local temperature fluctuations, even inhomogeneities in the dark matter density distributions of same temperature might produce dipoles in local CMB spectra because of different amounts of redshifted radiation coming from within the respective emission regions. Such effects may be mixed with the dipoles due to peculiar velocities, of course.

In any case, a 'dark flow' would be another indication for the straight SUM concept, since in view of today's cosmology, this DF seemed to prove the existence of at least two different universal rest frames. The CCM, however, could not explain such a result without contradiction to its own presupposition of ultra-large scale isotropy.

Therefore the whole Sunyaev-Zel(')dovich effect, as in particular the tSZ, see e.g. [Lieu, Mittaz, & Zhang 2006], [Bonamente et al. 2006], and the kSZ, see e.g. [Hand et al. 2012], [Kashlinsky, Atrio-Barandela, & Ebeling 2012], should be reviewed in the SUM framework now, even if only to prove the CCM again. Future measurements will decide this issue, too.

7) *CMB-temperature evolution:* In addition to many other observational facts, also the CMB-temperature measurements in distant gas clouds reported in [Noterdaeme et al. 2011], s. also references therein, clearly support the CCM. With regard to the latter, however, it has to be taken into consideration that, on the one hand, it is only about a dozen measurements in the high redshift range so far, and that, on the other hand, it is no contradiction to straight SUM to find gas clouds of different temperatures there, as long as unexpected results are not statistically excluded.

Thus, it will still remain important to verify beyond all doubt whether or not the local CMB temperature fulfills the proportionality $\Theta_{CMB} \sim 1+z$ without any exception. Absorption lines originated at high redshifts from transitions between excitation states of suitable molecules, however, seem be preselected in view of energy differences close to assumed values $k\Theta_{CMB}$. Though recently [de Martino et al. 2012] have used the tSZ anisotropy induced by clusters of galaxies to confirm the $\Theta_{CMB} \sim 1+z$ proportionality, it may be of interest – at least as a completion – to search the other way round for indications to an approximately constant mean CMB temperature, too.

In case of an actual cosmic evolution as a whole, if once definitely confirmed by further CMB temperature measurements, though, this should not have begun earlier than from the modified Planck time $T' \approx T_\alpha = T_H e^{-T/\alpha+1} \approx 10^{-59} T_H$ for example, which process might correspond to a cosmic event of chaotic inflation [Linde 1983, 2005] considered from an initial 'fluctuation' [Mukhanov & Chibisov 1981] and apparently grown to Hubble length today. Then such a very event, however, should have happened in a stationary background universe largely equivalent to a pre/post-inflation scenario possibly described by an open SUM.

8) *Gravitational or Doppler redshift:* All the different versions of big-bang cosmology from [Lemaître 1927/31] up to today's CCM are based on the interpretation of cosmic redshift as a proof for a universal expansion. Since such a conclusion contradicts the concept of SUM essentially, this feature may be also addressed from another point of view.

According to the CCM, the expansion of space forces to accept that a receding distribution of matter or energy had reached superluminal velocities while inflation. If this scenario shall not contradict all experimental physics, one has always to distinguish two kinds of velocity, one of them related to kinetic energy, the other without. What may be the sense to talk about a velocity of receding galaxies without kinetic energy if there is no need for splitting those basic concepts, implying a schism of consistent physics.

Now given two legitimate interpretations of GR (s. Sect. 1.1), there evidently must not be concluded different physics from the same equations. As has been shown in Section 2.5, there is no need to understand the redshift as (completely) caused by an actual Doppler velocity of galaxies.

In physics any measure should be based at least in principle on a thinkable operationalization. [Weinberg's 1972] excellent '*Gravitation and Cosmology*' addressed the expansion feature of the 'big bang' approach explicitly, though with the reservation that "… proper distance is not very relevant to observational cosmology". A motion of any distant galaxy G further apart is reasoned there using a fictive chain of 'fleeing' galaxies lying close together on a line towards G, where the distance between each two of them is assumed to increase. But neither does the distance to, for example, Andromeda increase (this feature explained by its peculiar velocity), nor does exist any rigid 'proper Mpc-stick' to be compared with. In contrast to universal gravitation, a perfect meter-stick is rigid by local quantum mechanics. Calculating all relations in the sections above including those of the apparent SNe-Ia luminosities, there has been nowhere a need for a concept of proper distance to apply non-locally.

9) *Additional remarks:* The SUM as a model of unique mathematical simplicity – there may be also some elegance in its line element (17) – is considerably close to the SNe-Ia observational facts on scales $z > 0.1$ without referring to any peculiarities concerning the universe in its entirety. Taking into account that the presupposed features of isotropy and homogeneity are hardly justified below scales of about 400 Mpc, for example, a corresponding Hubble contrast could make this agreement complete over the full redshift range $0.01 < z < 1.8$ available so far. The other way round, it might be rather a surprise if neglecting any huge inhomogeneities or anisotropies within this range, there was no local deviation at all.

The SNe-Ia data – which besides the COBE [Mather et al. 1990], WMAP [Bennett et al. 2003], HST Key Project [Freedman et al. 2001], *HST* Calibration Program [Sandage et al. 2006], and SDSS of e.g. [Kessler et al. 2009], [Schneider et al. 2010] measurements are of exceptional importance – may represent the most valuable cosmological breakthrough of the last decades because their confrontation with competing theories probably requires the least input of unproven hypotheses about the universe. In spite of these magnificent discoveries, of course, there is still the need for more observational facts.

Given structures up to several hundred megaparsecs of the Sloan Great Wall, for example, the best fit accordance of the largely homogeneous CCM approach with the SNe-Ia data might possibly show even too much 'perfection'. As considered in 4.3, it is difficult to distinguish numerically between thinkable effects caused by large scale inhomogeneities of our cosmic environment on the one hand and a strange acceleration/deceleration performance of the entire universe on the other hand, the latter adhering to a strict presupposition of undisturbed homogeneity instead.

Without considering various hints here in detail, there might be more large-scale structure at redshifts $0.01 < z < 0.1$ than the CCM can actually explain.

Even if the first tentative explanation of the CMB according to Section 5.3 failed in detail, it might be worth to take straight SUM further on into consideration. Though such a failure would mean a fundamental missing link for this model, there are some missing links not less fundamental with regard to the CCM as e.g. (i) a fictive inflation of a slow-rolling scalar field, (ii) an implausible initial fine-tuning of a fictive cosmological constant (iii) either one fictive 'big bang' according to laws of nature where allegedly no nature has been before, or (iv) an elusion to disconnected fictive 'parallel universes' where the laws of nature are assumed to be different but respectively strictly valid again.

In the framework of straight SUM, all of these problems do not need any solution, since they do not exist. This approach may offer a chance of keeping an adapted CCM for our cosmos without having to assign strange features to the entire universe.

It was effectively forced to conceptualize such a provisional SUM scenario to show that there might be a viable stationary alternative to the CCM instead of the SST. As a new concept, it seems another chance for a stationary background cosmology on basis of Einstein's equations.

## 7. STRAIGHT SUM INSTEAD OF THE SST AS AN ARGUABLE ALTERNATIVE TO THE CCM

The idea that no universal horizons must limit physical reality has led to the SUM as a stationary cosmological solution of Einstein's equations. If at all, any well-known horizons in the framework of GRT are understood here to indicate, where in the interplay with gravitation, quantum mechanics may materialize its creative potential instead.

Confronting the SUM and the CCM with the SNe-Ia data as the fundamental criterion of instrumental cosmology, this direct comparison turned out to result in an ambivalent indication (Sect.s 2.6, 4.1 - 4.4).

With either the pure CCM or the straight SUM there seem to be two alternatives of relativistic cosmology if not both are combined to a model of our evolutionary cosmos within one stationary background universe. Several features are compared in Table 1. One may be left with the question, whether cosmic evolution affects the universe as a whole (according to the CCM) or 'locally' only (according to straight SUM).

Two simple postulates have been used to deduce the stationary universal line element $d\sigma^*_{SUM} = e^{Ht^*} d\sigma^*_{SRT}$ implying a constant coordinate speed of light $c^* = c$ (Sect. 2.1).

The SUM may describe the universal background on ultra-large scales in contrast to our evolutionary cosmos therein. Depending on one macroscopic constant $H$ in addition to $c$, $G$ only, its various aspects of stationarity have been shown.

In particular for galaxies statistically at rest with respect to universal coordinates – these otherwise named 'comoving' or 'conformal' – the SUM stands out from all other flat space solutions with redshift parameters $z = e^{Hl^*/c} - 1$ independent of time (Sect. 2.5).

The direct calculation of the SUM redshift is confirmed by an alternative calculation based on its FLRW-form. Thereby a general discrepancy has been clarified between the conventional Hubble parameter $H_c$ and a significant Hubble parameter $H_s$, the latter in case of SUM a constant $H_{s-SUM} \equiv H$. Except for Hubble's linear approximation, however, the former is shown to be misleading since it is unfortunately coupled to the 'proper'-distance concept, while the redshift of galaxies at rest is related by presupposition to universal ('comoving') coordinates instead (Sect. 2.9).

Now, in addition to local proper length, any such universal distance is a measurable physical quantity. Except for peculiar motions, it is unambiguously displayed in its values of redshift according to the actual SUM extension (53) of Hubble's linear approximate law.

The new model requires the existence of a 'dark' gravitational pressure of one third the critical energy density. This pressure has been found here to be necessarily negative, and it might correspond to something like a stationarily changing cosmological 'constant' (Sect. 2.4).

The SUM does not only prove stationary, but shows several essential features which have been caught in the theoretical framework of the CCM rather speculatively. Free of horizons concerning the universe it might turn out as an alternative to inflation, otherwise needed to arrive with e.g. an approximate flatness, 'superhorizon' scales, or some more features one simply would start from, given a stationary background universe. Not least, it may answer the question which line element of GRT would have governed an overall chaotic inflation in the background of the assumed 'big bang' (Sect. 6).

From the SUM, a heuristic approach has lead to CCM parameters $\Omega_M = 0.263$, $\Omega_\Lambda = 0.737$ which otherwise seem purely coincidental (Fig. 1). Thus the CCM scale factor effectively fulfills approximate SUM 'boundary' conditions at $Ht' = 0$ and $Ht' = -1$ (Sect. 3.2).

In accordance to straight SUM as the cosmological model of general *and* special relativity theory, there would be alternating processes of evolution and revolution all over the universal background, the latter possibly in quasars, 'black holes', SMOs and AGNi, hot core structures blown up to bubbles or even 'local-bang' cosmoses, if indeed.

There are reasons that according to SUM the physical actuality – anything but static – seems to be a lively interplay where special relativity theory represents quantum mechanics while, in contrast, general relativity represents gravitation. Consequently, the concept of an *infinite* stationary universe turns out to imply clear indication that individual cosmic structures have *finite* dimensions in space and time. This result is derived from the intrinsic limitations of proper length and proper time (Sect. 2.10) according to a self-restoring validity of SRT (Sect. 2.2). Proper quantities are found to be 'local' concepts only.

What otherwise is called 'age of the universe', now in view of the SUM turns out to be rather the maximum age of macroscopic structures. A re-creation of light elements more or less corresponding to the CCM's 'primordial nucleosynthesis' – as well as relevant physical knowledge and models which at present are ascribed to one hot 'big bang' – might apply to local processes instead. Lemaître's 'primeval atom' [Lemaître 1931c] would have been in a multitudinous universal community, though not as the mere singularity assumed only later. Several GRT theorems do not apply when taking QM into account. In view of straight SUM, any 'local-bang' events should have taken place within the one background universe.

Ultra-large scale stationarity, however, would demand local space-time areas of decreasing entropy. No laboratory experience would ever contradict a restriction of the natural increase to evolutionary scenarios only, while in the cores of SGCs for example – where any process of ordinary diffusion is overcome by gravitation – an unrestricted law of entropy may break down (Sect. 5.1).

Even in a stationary universe, empty space might appear relatively 'expanding' though only with respect to temporarily shrinking local proper-units. There would be a struggle of ultra-large scale entropic balance against gravitational re-creation, just as the other way round there is the well-known struggle of all structures against decline and decay. Here is no need for a physical beginning of space and time (Sect. 2.12).

In view of the SUM it remains the question, how far the limits of our evolutionary cosmos actually reach out. Where and when does the realm of our physical evolution actually merge into the infinite universe?

The stationary model is thought to describe the universe on ultra-large scales where the underlying assumptions of isotropy and homogeneity seem actually justified. According to this objective a comparison with the SNe-Ia data, the CCM, and its 'parents' EdS and SST, revealed straight SUM accordance with the [Riess et al. 2004/07] 'gold' sample as well as with 'the world's supernova distance-redshift data' [Kowalski et al. 2008] on scales $z > 0.1$ (Sect. 4.2).

In this view, according to Figures 3 and 4, there is serious indication that ongoing re-creation processes or something

**TABLE 1**
The Stationary Universe Model in comparison with the current Cosmological Concordance Model

| Some characteristic FEATURES (list extensible) | Model of a STATIONARY BACKGROUND UNIVERSE (SUM) | Concordance/Consensus Model of OUR COSMOS (CCM) |
|---|---|---|
| line element --- scale factor ($\Omega_R^{CCM}=0$) | $d\sigma_{SUM}^{*} = e^{Ht^*} d\sigma_{SRT}^{*}$ --- $a_{SUM} = 1 + Ht'$ | $d\sigma_{CCM}^2 = c^2 dt'^2 - a_{CCM}^2 dl^{*2}$ $a_{CCM}(t') = \left\{\left(\frac{1}{\Omega_\Lambda}-1\right)\sinh^2\left[\frac{1}{2}\ln\left(\frac{1-\sqrt{\Omega_\Lambda}}{1+\sqrt{\Omega_\Lambda}}\right) - \frac{3}{2}\sqrt{\Omega_\Lambda}Ht'\right]\right\}^{1/3}$ |
| $t^*, l^*$ | universal time, universal space | conformal time, comoving space |
| model parameters | natural constants $c, G, H$ | independent parameters $T_0, H_0, q_0, \Omega_0, \Omega_M, \Omega_\Lambda$, (several additional parameters of inflation) |
| cosmological constant (dark energy) | – none – (homogeneous dark matter background) | $\Omega_\Lambda \approx 0.73$, value coincidental if not determined by SUM 'boundaries' |
| redshift of starlight from sources at rest (in 'comoving' coordinates) | $z = e^{Hl^*}$, where $l^* = $ constant$\vert_{t^*}$, independent of time, directly showing stationarity*) | $z = z(t', l^*)$, dependent on time, as well as all functions of $z$, e.g. $H(z), q(z), \ldots$ |
| $H$ | galaxies at rest in the universal ('comoving') frame imply $H_s \equiv \dot{a}$ with the constant $H_{s\text{-SUM}} \equiv H$ as the *significant* Hubble parameter | both the significant parameter $H_{s\text{-CCM}}$ as well as the *conventional* Hubble parameter $H_{c\text{-CCM}}(t') \equiv \dot{a}/a$ depend on time |
| limits of space and time | $T_H, R_H$ maximum age/radius of structures subject to any SRT concepts | $T_0 \approx T_{Ho}$ the age of 'the universe', $R_0^* \approx 3.4 R_{Ho}$ its radius today |
| $H_0 T_0$ | $HT \equiv 1$ due to stationary values $H_0 \equiv H$ and $T_0 \equiv 1/H$ | $H_0 T_0 \approx 1$ coincidentally today (a temporary value) |
| 'deceleration' parameter $q \equiv -a(d^2a/dt^2)/(da/dt)^2$ | according to postulate I of stationarity: $q \equiv 0$ | assumedly *positive* after 'big bang', *negative* while inflation, then *positive* for some $10^9$ years, *negative* today – uncertain for the future |
| initial singularity | none with respect to universal coordinates, local pseudo-singularities instead, modified by quantum mechanics (breakdown of proper length and time) | unexplained origin in a 'big bang' concerning the entire universe, or a 'big bang' from e.g. chaotic inflation within a background unexplained in GR |
| spatial flatness | deduced from postulate II of a constant universal speed of light $c^* = c$ | approximately after a phase of 'superluminal inflation' |
| horizon problems | none | overcome by 'superluminal inflation' |
| law of entropy | restricted to *any* evolutionary processes | restricted to *one* evolution after 'big bang' |
| 'black holes' | supermassive (active) gravitational centers, 'bright sources', QM retains matter from vanishing forever | boundary of phenomenological GR-applicability, the Schwarzschild-radius accepted to limit physical reality |
| CMB | SMB from 'dark'-matter (possibly plus SR), natural anisotropies (incl. acoustic oscillations) | 'big bang' relic radiation, anisotropies caused by acoustic oscillations (ad-hoc fitted inflation) |
| n-bamg nucleosynthesis | ongoing re-creation with 'local bang' events | from one 'big bang' of the entire universe |
| straight-off compatibility with the SNe-Ia data (e.g. *local Hubble contrast*) | $0.1 < z < 1.8$ excluding the local region $z < 0.1$ (e.g. $0.01 < z < 1.8$) | $0.01 < z < 1.8$ the full range of observational data (e.g. $0.025 < z < 1.8$) |

NOTE. – *) The new stationary model SUM should not be confused with the 'Steady-state' Theory (SST) whose e.g. redshift parameters depend on time.

like 'local bangs' might affect scales at most of similar dimensions only. Full scale SUM compatibility with the SNe-Ia data has been obtained taking into account possible peculiar features of our 'local' cosmic environment like a Hubble contrast within the range of measured values (Sect. 4.3).

Given that such a local Hubble contrast was caused by inhomogeneities of matter and energy, then these might hardly affect the observations isotropically, however, if observers are not situated right in the middle of a void for example. More measurements of local angular distributions are certainly needed. In addition, it might prove unreasonable to take into consideration a large-scale temporal variability of the universe on the one hand, but no large-scale spatial inhomogeneity on the other hand.

That not only the old SST but also the EdS cosmology is disproved by the SNe-Ia observations, seems a safe conclusion. In contrast, the CCM represents a combination of both.

The observations of the last two decades may be seen to support a double mean SUM zero: $k \equiv 0$, $\bar{q} \equiv 0$. On the other hand, in the CCM framework the 'deceleration parameter' is claimed to be $q<0$ today, after it should have been $q>0$ in the past, though only after before it had been $q < 0$ while inflation (Sect. 2.13).

In view of straight SUM, all radiation going out from sufficiently large gravitationally bound objects might be near an energy equilibrium at its 'surface' with all radiation coming in. In particular, however, Figure 8 seems to support an assumption that for any special sort of universal objects U with low ratio ($F_U^*/I_U$), this ratio may be roughly proportional to its density parameter $\Omega_U$, thus leading to an unexpected cosmological mass-to-radius relation (Sect. 5.2).

Several considerations show not only the possibility but do even suggest the existence of a DM black-body background radiation as a predominant part of the BB-SMB stationarily emitted within the universe (Sect. 2.8). A tentative straight SUM approach to the CMB assumes that such a microwave radiation originates essentially from the approximately homogeneous fraction of 'dark'-matter distributed in voids as well as from the much more inhomogeneous fraction in halos like those of galaxies or clusters. In particular this preliminary attempt, though, clearly needs future elaboration (Sect. 5.3).

In spite of the probability that some tentative assumptions would fail, it might be reasonable, temporarily to accept even those 'missing links' of the straight-SUM concept as, for example, concerning the CMB anisotropies. These may be caused by DM oscillations or other inhomogeneities due to DM halos, but are not yet explicitly explained here.

Concerning both SZ effects in the straight SUM framework with numerical modifications primarily in the high-$z$ range, the results of [Lieu, Mittaz, & Zhang 2006], or the 'dark flow' stated by [Kashlinsky, Atrio-Barandela, & Ebeling 2012], both with references therein, have raised doubts in the 'big bang' origin of the CMB, which is assumed here as only a special part of EBL (Sect. 5.4).

The case for a necessity to adapt a CCM cosmos into an open SUM background, would be to verify beyond all doubt – from a sufficiently large sample at sufficiently large distances, though – whether or not the local CMB temperature fulfills without any exception the CCM proportionality $\Theta_{CMB} \sim 1+z$, which in several observations seems already confirmed by corresponding results of [Noterdaeme et al. 2011] and papers referenced therein. The other way round, in case of some $\Theta_{CMB}(z) \approx$ constant $\approx$ 3 K measurements at high redshifts this would falsify the CCM after all (Sect. 6).

In view of straight SUM, several important CCM features resorting to peculiar phases in the assumed history of the universe have to be alternatively explained by e.g. selection effects including various forms of attenuation, or by 'local' cosmic evolution implying e.g. peculiar flows. These chances cannot be excluded, quite the contrary. Essential processes in the lively universe may be far from being understood. The SNe-Ia breakthrough at the turn to the 21st century will not remain the last unexpected cosmological discovery forever.

In view of Occam's razor – also regarding simplicity and symmetry – the SUM might be appropriate to describe the entire universe, whereas the CCM would rather describe our evolutionary cosmos. Even the criterion of esthetics is of some practical importance, since for the sake of an unbiased comparability it could be reasonable, to refer to the simplest model which allows a systematic classification of observational data. In this view, the SUM seems predestined as a reference model because of its unique mathematical simplicity.

Furthermore, if one takes Einstein's original equations *without* cosmological constant, then the SNe-Ia measurements would have obviously confirmed straight SUM.

Several fundamental facts well-known as its main pillars seem to prove 'big bang' cosmology beyond all doubt. Almost as strong as these pillars, however, as weak seems some ground. In addition to (a) the inflation paradigm dealing with a scalar inflaton field never observed, these are among others (b) the baryon asymmetry, (c) a singular origin from blind chaos leading to fixed physical laws, d) 'dark energy', in contrast to 'dark' matter irritating fundamentally, e) an *imperfect* cosmological principle excluding time from universal symmetry, and on top f) the strange coincidental 'age of the universe' equaling the Hubble time just only today. These unexplained features of the current ΛCDM 'big bang' cosmology may have been widely accepted not least in view of apparently no arguable alternative so far.

Now in the SUM framework, several problems might disappear or may be at least seen in different light. In particular, it has been considered, that …

1. … the universal redshift is deducible as gravitational effect without the need for a corresponding Doppler motion, thereby introducing a significant Hubble parameter in contrast to the conventional one.

2. … the law of entropy may be restricted – without contradiction to any laboratory experience – to evolutionary processes, thus allowing for 'primordial' nucleosynthesis in re-creation processes of AGNi or in 'local bangs'.

3. … there may be a nearly homogeneous, optically almost transparent, non-lensing distribution of 'dark matter' filling the gap between observable matter and the critical density, thus accounting for what is ascribed to 'dark energy' today.

4. … most of the CMB might be essentially a thermal radiation from 'dark matter', where its inhomogeneities apparently seem also to reflect acoustic oscillations at a statistical mean universal temperature.

Summarizing that several fundamental observational facts are in approximate accordance or at least not definitely in conflict with the results derived in the deductive part of this paper, it seems a legitimate conclusion that relativistic cosmology may have gained future scope between two extreme alternatives: a singular CCM origin of the universe including space and time on the one hand, and a straight SUM background including local quasi-bang events on the other hand. This paper claims that, in any case, what modern cosmology describes as an unevenly evolving evolutionary cosmos, has not necessarily to be the entire universe.

Even if a straight SUM was able to describe the universe on its own, however, it could not be expected to explain the plenty of cosmological observations at once. Many open questions should be answered whether positive or negative in the new context, as this has happened with many open questions of the 'big bang' cosmology in the past. Correspondingly several unbiased attempts may be needed to provide future clarification.

APPENDIX: ON THE *PLANCK* 2013 RESULTS

Recently in a series of pre-prints the *Planck* 2013 results were published online, s. [Planck Collaboration I 2013] and references therein.

Though now several important aspects give reason to add this brief appendix, the SUM concept developed above is not affected itself. In contrast, on the other hand, while the CCM density parameters are significantly changed, several WMAP anomalies in form of 'intriguing' asymmetries are confirmed or even strengthened by the results in [Planck Collaboration XXIII 2013].

Some minor modifications in comparison with the text do not change the conclusions of this paper on hand essentially. Taking into account a *Planck* value $\Omega_\Lambda \approx 0.69$, now in Figure 6 the respective gradients of the straight lines in the c), e) panels representing the mean quadratic CCM deviations would appear improved from $-0.07, -0.08$ to $-0.03, -0.04$. In Figure 4 showing the most relevant fits at universal scales $z > 0.1$, the same value would lead from $-0.16, -0.18$ to $-0.12, -0.14$ (while the corresponding SUM gradients stay at $+0.02, +0.03$ and $-0.03, -0.02$).

In view of a thinkable attempt to embed the CCM cosmos into an open SUM background as also considered above, it is of more importance that taking into account the new [Planck Collaboration XVI 2013] high-precision results $H_0 = 67.3$ km s$^{-1}$ Mpc$^{-1}$, $\Omega_m = 0.315$, "… these values are in tension with recent direct measurements". The newly emerging problems concerning the conventional parameter $H_0$ again, give in view of straight SUM reasonable indication for a local Hubble contrast as concluded in Section 4.3.

It might become increasingly difficult to maintain the claim of a 'Cosmological *Concordance* Model' describing the commonly presupposed 'big bang' at all. On the other hand, in view of straight SUM it should not bear serious difficulties to explain in principle any asymmetries of the BB-SMB (CMB) from 'local' inhomogeneities.

According to Section 5.3, by far most of this radiation would have been emitted within $z < 1$ instead at from a surface of last scattering at $z \approx 1100$. With regard to the increasingly reduced SZ effect considered in the context of Figure 9, it seems remarkable that the [Planck Collaboration XXIX 2013] SZ catalogue as the deepest all-sky cluster catalogue ever, reports corresponding redshifts up to about $z \approx 1$ only.

In the CCM framework, a process of 're'-ionization must have followed the phase of 're'-combination. But in particular the former remains unclear because of "unexplained residuals", at least until further *Planck* results will be released.

What has been measured by the Planck collaboration to unprecedented resolution are frequency-dependent angular distributions, which in a straight SUM framework obviously would have to be scaled up from an inflationary 'big bang' model to a stationary background universe.

This means the need for a future attempt of a parameter transfer to straight SUM: instead of the CCM's 'strange recipe', now one may take primarily six 'acoustic' parameters to fit the acoustic oscillations into the SUM framework.

A fiducial peculiar CCM length is tightly related to the first acoustic peak. Any structure at a universal distance one hundred times its diameter, however, would appear at about an angle of $0.6°$ on the sky, as might do voids at Hubble distance $R_H$, as clusters at the transition scale to universal homogeneity at $z \approx 0.1$ (s. Sect. 4.2), or possibly even as galaxies at some mean distance between them, for example.

Now the [Planck Collaboration XVI 2013] measurements state a density parameter $\Omega_M \approx 1/3$ (instead of 1/4), straight SUM would ascribe this to only the part of universal ('dark') matter gravitationally bound in galaxies or clusters, whereas a fraction $\Omega_\Lambda \approx 2/3$ (instead of 3/4) seems to be much more homogeneously distributed at lower densities, as primarily in voids. In contrast to the first, because of smoothed inhomogeneities the latter might show no gravitational lensing effects. Such a distribution of 'dark' matter should imply corresponding fractions of the BB-SMB reaching an observer from point-like halos or from homogeneous DM respectively.

In view of straight SUM there may appear some 'local' CMB asymmetries from the universal distribution of 'dark' matter primarily emitting the BB-SMB. That six CCM parameters for acoustic oscillations could be not enough to model the CMB anisotropies together with the various WMAP anomalies, now seems confirmed by the *Planck* results.

Any thinkable attempt to embed a modified CCM cosmos into an open SUM would probably presuppose a special location, then justified by an anthropic principle related to that one explaining the approximate CCM coincidence of the Hubble time with the assumed 'age of the universe' stated in Section 3.2. In a suggestive historical analogy such an attempt might correspond to a Brahe-model, which once tried to reconcile the old geocentric with the newly rediscovered solar-centric view. Now with a *Planck* density parameter $\Omega_\Lambda \approx 0.69$ (instead of 0.73) the deviation from a perfect accordance of both periods above is increased from $< 1\%$ to $\sim 5\%$, thus weakening that coincidental argument for a direct embedding.

Facing the "intriguing inconsistencies" implied in the results of [Planck Collaboration XXIII 2013], and these taken together with several CCM problems already well-known before, it might be worthwhile to focus on the straight SUM framework in this context instead.

Given Einstein's original equations without cosmological constant, the SNe-Ia measurements would obviously confirm straight SUM directly. Instead of even more sophisticated 'new physics' – occasionally addressed as a possible consequence if necessary – a shift of two paradigms with no loss of proven GR applicability may be sufficient. At first, the concept of SRT 'proper' quantities must not be overstrained to universal scales (s. Sect. 2.10), which questionable practice is manifest from a strict preference for the FLRW form so far (it has been shown in Section 2.5 that this concept implies an unfortunately misleading conventional Hubble parameter). At second, physicists might accept genius Einstein's famous assessment of the cosmological constant as his biggest blunder (*'größte Eselei'*) after all. Essential aspects resulting from both these claims are summarized in Table 1 above.

REFERENCES TO THIS APPENDIX